\documentclass[iop,apj, numberedappendix]{emulateapj}
\usepackage[caption=false]{subfig}
\usepackage{times}
\usepackage{amsmath, amssymb}
\usepackage{longtable}

\usepackage{multirow}
\usepackage[breaklinks,colorlinks,citecolor=blue,linkcolor=magenta]{hyperref} 

\usepackage[all]{hypcap} %Links go to figures; breaks on deluxetables (use \capstartfalse \capstarttrue to fix it)

\shorttitle{The PS-ELQS}

\shortauthors{Schindler et al.}

\begin{document}
 
%  Title
\title{The Extremely Luminous Quasar Survey in the Pan-STARRS 1 footprint (PS-ELQS)} 

% Authors 
\author{Jan-Torge Schindler\altaffilmark{1,2}}
\author{Xiaohui Fan\altaffilmark{1}}

% \author{Ian D. McGreer\altaffilmark{1}}
% \author{Qian Yang\altaffilmark{2,3}}
% \author{Jin Wu\altaffilmark{2,3}}
% \author{Linhua Jiang\altaffilmark{2}}
% \author{Richard Green\altaffilmark{1}}
\author{Yun-Hsin Huang\altaffilmark{1}}
% https://orcid.org/0000-0003-4955-5632
\author{Minghao Yue\altaffilmark{1}}
%0000-0002-5367-8021
\author{Jinyi Yang\altaffilmark{1}}
\author{Patrick B. Hall\altaffilmark{3}}
%0000-0002-1763-5825
\author{Lukas Wenzl\altaffilmark{2}}
%https://orcid.org/0000-0001-5245-2058
\author{Allison Hughes\altaffilmark{1}}
%https://orcid.org/0000-0002-1718-0402 
\author{Katrina C. Litke\altaffilmark{1}}
%0000-0002-4208-3532
\author{Jon M. Rees\altaffilmark{4}}
%0000-0002-5376-3883

%\author{Friends}

\altaffiltext{1}{Steward Observatory, University of Arizona, 933 North Cherry Avenue, Tucson, AZ 85721,USA}
\altaffiltext{2}{Max Planck Institute for Astronomy, Königstuhl 17, 69117 Heidelberg, Germany}
\altaffiltext{3}{Department of Physics and Astronomy, York University, Toronto, ON M3J 1P3, Canada}
\altaffiltext{4}{Center for Astrophysics and Space Science, University of California San Diego, La Jolla, CA 92093, USA}
% \altaffiltext{2}{Kavli Institute for Astronomy and Astrophysics, Peking University, Beijing 100871, China}
% \altaffiltext{3}{Department of Astronomy, School of Physics, Peking University, Beijing 100871, China}
% 

\defcitealias{Schindler2017}{ELQS1}
\defcitealias{Schindler2018}{ELQS2}

\begin{abstract}

We present the results of the Extremely Luminous Quasar Survey in the $3\pi$ survey of the Panoramic Survey Telescope and Rapid Response System (Pan-STARRS; PS1). This effort applies the successful quasar selection strategy of the Extremely Luminous Survey in the Sloan Digital Sky Survey footprint ($\sim12,000\,\rm{deg}^2$) to a much larger area ($\sim\rm{21486}\,\rm{deg}^2$). 
This spectroscopic survey targets the most luminous quasars ($M_{1450}\le-26.5$; $m_{i}\le18.5$) at intermediate redshifts ($z\ge2.8$). Candidates are selected based on a near-infrared JKW2 color cut using WISE AllWISE and 2MASS photometry to mainly reject stellar contaminants. Photometric redshifts ($z_{\rm{reg}}$) and star-quasar classifications for each candidate are calculated from near-infrared and optical photometry using the supervised machine learning technique random forests. 
We select 806 quasar candidates at $z_{\rm{reg}}\ge2.8$ from a parent sample of 74318 sources. After exclusion of known sources and rejection of candidates with unreliable photometry, we have taken optical identification spectra for 290 of our 334 good PS-ELQS candidates. 
We report the discovery of 190 new $z\ge2.8$ quasars and an additional 28 quasars at lower redshifts. A total of 44 good PS-ELQS candidates remain unobserved. Including all known quasars at $z\ge2.8$, our quasar selection method has a selection efficiency of at least $77\%$.
At lower declinations $-30\le\rm{Decl.}\le0$ we approximately triple the known population of extremely luminous quasars.
We provide the PS-ELQS quasar catalog with a total of 592 luminous quasars ($m_{i}\le18.5$, $z\ge2.8$).
This unique sample will not only be able to provide constraints on the volume density and quasar clustering of extremely luminous quasars, but also offers valuable targets for studies of the intergalactic medium.

\end{abstract}


\section{Introduction}

Quasars are excellent tracers of the formation and evolution of highly accreting supermassive black holes (SMBHs) across cosmic time. Their large luminosities not only allow us to detect and study them within the first billion years of the universe \citep[e.g.][]{Fan2000}, but further provide strong background sources with which one can probe the large-scale structure formation of the universe and the nature of the intergalactic medium \citep[e.g.][]{Simcoe2004, Prochaska2005}.
The highest redshift quasars at $z\ge7$ provide strong constraints on the re-ionization of the universe \citep{Mortlock2011, Banados2018, Wang2018, Matsuoka2019} and on models of SMBH formation \citep{Volonteri2012}.

Large quasar surveys provide the necessary number statistics to study the evolution of active SMBHs. The Sloan Digital Sky Survey \citep[SDSS;][]{York2000}, the Baryon Oscillation Spectroscopic Survey \citep[BOSS;][]{Eisenstein2011, Dawson2013} and the extended BOSS \citep[eBOSS;][]{Dawson2016} have identified over 500,000 quasars at $z\lesssim6$ and dozens of quasars at $z>6$ \citep{Fan2001b, Fan2003, Fan2004, Fan2006, Jiang2008, Jiang2009, Jiang2016}

Efforts at higher redshifts have also utilized other large surveys like the CFHQS\citep[e.g.][]{Willott2007, Willott2010a}, UKIDSS\citep[e.g.][]{Venemans2007, Mortlock2011}, VIKING \citep{Venemans2013}, VST-ATLAS \citep{Carnall2015, Chehade2018}, DES \citep{Reed2015, Reed2017,Yang2018c_arxiv, Reed2019},  Pan-STARRS1 \citep{Morganson2012, Banados2014, Banados2016, Mazzucchelli2017, Pons2019} and the DESI Legacy Imaging Surveys \citep{Wang2018b_arxiv}.
The Hyper Surprime-Cam Subaru Strategic program \citep{Aihara2018} allowed the exploration of the fainter quasar population at intermediate \citep{Akiyama2018} and high redshifts \citep{Kashikawa2015, Matsuoka2016, Matsuoka2018, Matsuoka2018b}.

While the recent efforts have mainly focused on the high redshift quasar regime, surveys to identify intermediate redshift quasars outside the SDSS footprint have been scarce. 
Especially the $3\pi$ wide area coverage of the Pan-STARRS1 \citep[Panoramic Survey Telescope and Rapid Response System,][]{Kaiser2002, Kaiser2010} survey \citep[PS1;][]{Chambers2016} provides an excellent opportunity to explore the extremely luminous quasar population ($M_{1450}\lesssim-28$). 

While these extremely luminous quasars at intermediate redshift are similarly rare as high redshift quasars ($10^{-9}\,\rm{Mpc}^{-3}\rm{mag}^{-1}$), see e.g. \citet{Ross2013}, they are valuable sources to study the He-reionization of the universe \citep{Worseck2011, Worseck2016}, to explore the ionization state of the IGM \citep{Schmidt2018b_arxiv},  to investigate quasar clustering \citep[e.g.][]{Myers2006} and to constrain the evolution of the bright-end of the quasar population \citep{Schindler2018}.

In this work we build on the Extremely Luminous Quasar Survey in the SDSS footprint \citep[ELQS; ][hereafter ELQS1]{Schindler2017} to discover $z=2.8-5$ quasars with $m_{i}\leq18.5$ in $\sim21486\,\rm{deg}^2$ of the PS1 $3\pi$ footprint. 
 
We first describe the photometry that the quasar selection is based on (Section\,\ref{sec_photometry}) and give an overview over quasar catalogs in the literature that we use (Section\,\ref{sec_qso_catalogs}). We subsequently present our quasar selection strategy in Section\,\ref{sec_qso_selection} and the construction of the PS-ELQS candidate catalog in Section\,\ref{sec_pselqs_cand_cat}. Section\,\ref{sec_observations} discusses the spectroscopic observations and the data reduction, before we present the PS-ELQS quasar catalog in Section\,\ref{sec_qso_catalog}. We discuss our results in Section\,\ref{sec_discussion} and provide a summary in Section\,\ref{sec_conclusion}. Discovery spectra and tables detailing properties of the newly discovered quasars are available in the appendix.

We present magnitudes in the AB system \citep{Oke1983}, which are corrected for galactic extinction \citep{Schlegel1998}. All optical passbands refer to PS1, unless otherwise noted. Extinction corrected magnitudes are denoted by $m_x$, where $x$ refers to the photometric band, as opposed to extinction uncorrected magnitudes $x$. We employ a standard $\Lambda$CDM cosmology with $H_0=70\,\rm{km}\,\rm{s}^{-1}\,\rm{Mpc}^{-1}$, $\Omega_m = 0.3$ and $\Omega_{\Lambda}=0.7$, generally consistent with recent measurements \citep{PlanckCollaboration2016}.

\section{Photometry}\label{sec_photometry}

\subsection{The Wide-field Infrared Survey Explorer (WISE)}
Our quasar selection takes advantage of the WISE AllWISE data release providing infrared photometry over the entire sky at 3.4, 4.6, 12, and 22$\,\mu\rm{m}$ (W1, W2, W3, W4). AllWISE combines data from the original cryogenic mission and its post cryogenic extension \citep{Mainzer2011}\footnote{\url{http://irsa.ipac.caltech.edu/cgi-bin/Gator/nph-scan?submit=Select&projshort=WISE}}.
For our selection process we use the W1 ($3.4\,\mu{\rm{m}}$) and W2 ($4.6\,\mu{\rm{m}}$) photometry, for which the AllWISE source catalog achieved $95\%$ photometric completeness for all sources with limiting magnitudes brighter than $19.8$, $19.0$ (Vega: $17.1$, $15.7$), respectively. Vega magnitudes were converted to the AB magnitude system using  $W1_{\rm{AB}} = W1_{\rm{Vega}} + 2.699$ and $W2_{\rm{AB}} = W2_{\rm{Vega}} + 3.339$ and extinction corrected using  $A_{\rm{W1}}, A_{\rm{W2}} = 0.189, 0.146$.

\subsection{The Two Micron All Sky Survey (2MASS)}
We extend the WISE photometry to the near-infrared taking advantage of the 2MASS survey, which mapped the entire sky in the near-infrared bands J ($1.25\,\mu{\rm{m}}$), H ($1.65\,\mu{\rm{m}}$) and $\rm{K}_s$ ($2.17\,\mu{\rm{m}}$). 
The 2MASS point source catalog (PSC) includes all sources detected with a singal-to-noise ratio of 
$\rm{SNR}\geq7$ in one band or $\rm{SNR}\geq5$ detections in all three bands. 
Unfortunately, due to strong confusion of sources closer to the galactic plane, the photometric sensitivity is a strong function of galactic latitude. 
Generally, all sources brighter than $16.7$, $16.4$,  $16.1$ (Vega: $15.8, 15.0, 14.3$) in the J,H and $\rm{K}_s$ bands are detected with  $10\sigma$ photometric sensitivity. However, based on the on-line documentation\footnote{Figure\,7 on \url{https://www.ipac.caltech.edu/2mass/releases/allsky/doc/sec2_2.html}} we estimate the $10\sigma$ limiting magnitudes for higher latitudes to be $\rm{J}=17.7$, $\rm{H}=17.5$, $\rm{K}_s=17.1$.
Conveniently, the 2MASS PSC has been pre-matched to the WISE AllWISE source catalog.  The match corresponds to the closest 2MASS object within a $3\arcsec$ radius of the WISE position.
All 2MASS Vega magnitudes are converted to the AB system using  $J_{\rm{AB}} {=} J_{\rm{Vega}} + 0.894$, $H_{\rm{AB}} {=} H_{\rm{Vega}} + 1.374$, ${K_s}_{,\rm{AB}} {=} K_{s, \rm{Vega}} + 1.84$ and corrected for galactic extinction ($A_{\rm{J}}, A_{\rm{H}}, A_{\rm{K_s}} {=}  0.723, 0.460, 0.310$).

\subsection{The Panoramic Survey Telescope And Rapid Response System 1 (PS1)}
We combine the near-infrared/infrared photometry of 2MASS and WISE by DR1 optical photometry from the PS1 $3\pi$ survey \citep{Chambers2016}. PS1 delivers optical photometry in the $g$-, $r$-, $i$-, $z$-, and $y$-bands up to a depth of 23.3, 23.2, 23.1, 22.3 and 21.3 magnitudes ($5\sigma$, $3\pi$ stack) over $3\pi$ steradian of the sky ($\rm{Decl.}>-30$). Saturation only occurs at magnitudes of $\sim12-14$, depending on the seeing conditions. 
The PS1 photometry is nominally on the AB system. All magnitudes were corrected for galactic extinction ($A_{\rm{g}}, A_{\rm{r}}, A_{\rm{i}}, A_{\rm{z}}, A_{\rm{y}} {=}$ $3.172 , 2.271, 1.682, 1.322, 1.087$).

\section{Quasar Catalogs in the Literature}\label{sec_qso_catalogs}

To match promising candidates with known quasars from the literature, we make use of the large quasar samples discovered by SDSS I/II \citep{Abazajian2009}, BOSS, and eBOSS published in the SDSS DR7 \citep[DR7Q;][]{Schneider2010}, DR12 \citep[DR12Q;][]{Paris2012} and DR14 \citep[DR14Q;][]{Paris2018} quasar catalogs.

The quasar selection for the SDSS I/II spectroscopic survey is described in \citet{Richards2002} and selects quasars as outliers of the stellar locus in the ugri and griz color space. Inclusion regions are designed to include quasars in certain redshift ranges, that are highly contaminated with stellar sources. The resulting DR7Q includes 100,000 quasars over $9,380\,\rm{deg}^2$ region of the SDSS DR7 footprint. 
The BOSS quasar selection \citep{Bovy2011} was optimized to find quasars in the targeted redshift range of BOSS at $2.2 < z < 3.5$. The newly discovered quasars in BOSS were published in DR12Q.
The eBOSS quasar selection \citep{Myers2015} is based on the XDQSO method \citep{Bovy2011} and a mid-infrared color cut to provide a uniform quasar sample over $7500\,\rm{deg}^2$  with $g_{\rm{SDSS}}<22$ or $r_{\rm{SDSS}}<22$.

 The latest version of the SDSS quasar catalog (DR14Q) was then designed to include all quasars observed during any of the stages of SDSS. Therefore the DR14Q includes nearly all of the DR7Q and DR12Q quasars.
All in all, SDSS discovered more than 500,000 quasars in the northern hemisphere and makes up for the majority of the known quasars in the PS1 footprint. We use all three SDSS quasar catalogs mentioned above to match our candidate sample against known sources. Furthermore, the quasar training set for the random forest regression and classification is built from DR7Q and DR12Q quasars.

In addition to the SDSS quasar catalogs we also match our candidates against the Million Quasar Catalog \citep[MQC, version 5.7b;][]{Flesch2015}. The MQC is a compilation of type I and type II AGN from all the available literature, including a large fraction of quasar candidates. All quasars from the SDSS quasar catalogs can also be found within MQC.
For the cross-match to our candidate list we exclude all quasar candidates. 

Yang et al. (publication in preparation) are also working on a spectroscopic survey of bright quasars at intermediate redshifts similar to PS-ELQS. They are exploring two quasar selections \citep{Wu2010, Wu2012} targeted at $z\approx2-3$ and at $z\geq4$ to assess different selection criteria for the upcoming LAMOST quasar survey. The spectroscopic identification campaigns were carried out with the Lijiang telescope ($2.4\,\rm{m}$) and the Xinglong telescope ($2.16\,\rm{m}$). A number of PS-ELQS candidates were spectroscopically identified by their efforts.

\section{The PS-ELQS Quasar Selection}\label{sec_qso_selection}

The PS-ELQS quasar selection is in many aspects analogous to the original ELQS quasar selection \citepalias[see Section\,4-6, ][]{Schindler2017}. 
However, the optical photometry from PS1 does not provide u-band measurements like SDSS did. While our initial near-infrared JKW2 color cut selection is not affected by this, we loose information on quasars, where the broad emission lines transition from the u-band to the g-band. As a consequence our photometric redshift estimates for quasars at $z\lesssim3$ might be more unreliable.  
The additional y-band filter of PS1, which is essential for high redshift quasar selections, only adds little information for the search of intermediate redshift quasars.

The following subsections describe the PS-ELQS footprint as well as the individual steps of our quasar selection strategy in detail.

\subsection{PS-ELQS footprint} 

The ELQS aimed to discover extremely bright, intermediate redshift quasars, which are very rare. The survey used optical photometry provided by the SDSS over $\sim12.000\,\rm{deg}^2$, excluding the galactic plane. With the public data release of the PS1 $3\pi$ survey we can extend our previous efforts to a much larger area and into regions, which have not been included in previous quasar surveys. 
Therefore PS-ELQS is designed to cover the entire PS1 footprint except the galactic plane ($|b|\le20$). Compared to the ELQS footprint we cover an additional  $\sim9600\,\rm{deg}^2$, of which the majority ($\sim5600\,\rm{deg}^2$) lies at $-30\le\rm{Decl.}\le0$.

%2306.588760451589 5577.1062724

\begin{figure*}[htb]
\centering 
\includegraphics[width=1\textwidth]{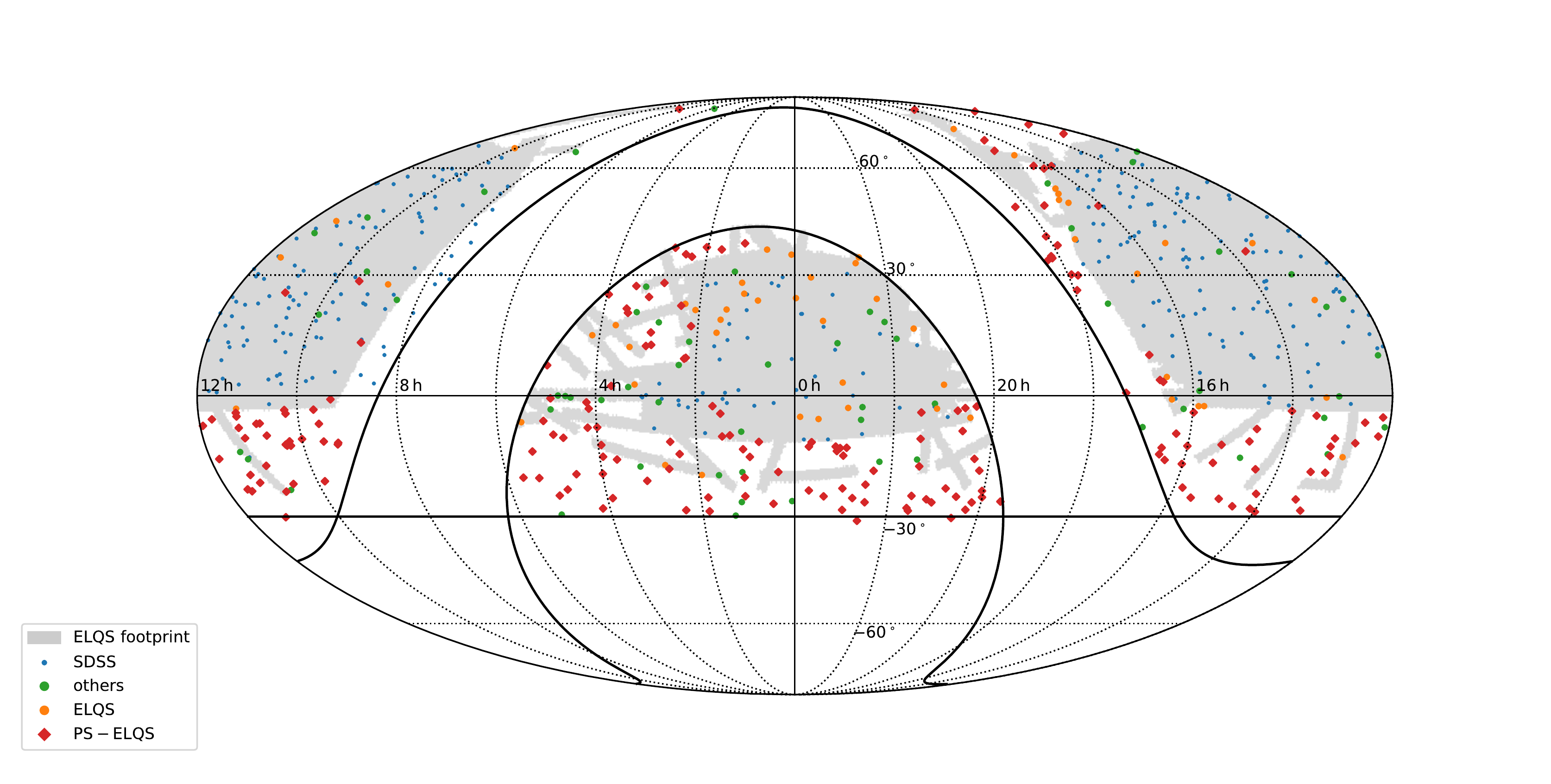}
\caption{Mollweide projection of the entire sky in equatorial coordinates. The original ELQS footprint is shown in grey. The PS-ELQS footprint covers all area above the thick solid line at  $\rm{Decl.}=-30$ and outside of the two lines outlining the galactic plane ($|b|\le20$). Data points show the position of all 592 quasars in the PS-ELQS quasar catalog (Section\,\ref{sec_qso_catalog}) colored by their reference.} 
\label{fig_coverage}
\end{figure*}

Figure\,\ref{fig_coverage} shows a Mollweide projection of the entire sky in the equatorial coordinate system. The coverage of the original ELQS is shown in grey. Two thick solid lines map the outline of the galactic plane ($|b|\le20$), which we exclude from our selection. A third solid line at $\rm{Decl.}=-30$, shows the southern border of the PS1 footprint. We effectively select the all the area above $\rm{Decl.}=-30$ and outside of $|b|\le20$ for the PS-ELQS survey. Colored data points show all 592 quasars from the PS-ELQS quasar catalog described in Section\,\ref{sec_qso_catalog}. The color refers to their source of identification.

We employ the Hierarchical Equal Area isoLatitude Pixelization \citep[HEALPix;][]{Gorski2005} to roughly estimate the area of the PS-ELQS footprint similar to \citet{Jiang2016}. 
HEALPix divides the sky into a grid of curvilinear, equal-sized quadrilaterals. At the lowest resolution the sky is represented by 12 pixels. To create higher resolution maps each pixel is subdivided into four pixels per resolution level. Therefore the total number of pixels follows $N_{\rm{pix}} = 12\cdot2^{\rm{lvl}}$.   

Our coverage estimate is based on $1.918.290$ sources selected by our photometric JKW2 color-cut selection (Section\,\ref{sec_phot_sel}), which were matched to PS1 according to our quality criteria but without enforcing the extended object rejection. We expect the PS-ELQS footprint to roughly cover $\sim20,000\,\rm{deg}^2$, resulting on average in one source per $\sim0.01\,\rm{deg}^{2}$. 
We choose resolutions with $\rm{lvl}=6, 7, 8, 9$ to calculate our coverage maps, resulting in a total number of pixels of $49152, 196608, 786432, 3145728$ with $\sim0.84, 0.02, 0.05, 0.01\,\rm{deg}^{2}$ per pixel. Our coverage estimates for these three resolution levels are $21697, 21487, 20676, 13501\,\rm{deg}^{2}$, respectively. 
There is a large decrease in coverage from $\rm{lvl}=8$ to $9$. At the highest resolution the pixel density is approaching the source density and we are effectively oversampling the area. 

We adopt a resolution of $\rm{lvl}=7$ with an effective area of $21486^{+210}_{-833}\,\rm{deg}^{2}$ for our final coverage estimate. The uncertainties reflect the differences to the coarser ($\rm{lvl}=6$) and finer  ($\rm{lvl}=8$) resolutions.

\subsection{Photometric Selection}\label{sec_phot_sel}

We begin our photometric selection with the WISE AllWISE catalog pre-matched to all sources from the 2MASS point source catalog (PSC).  An overview of the selection process is given in Section\,\ref{sec_pselqs_cand_cat} . The source selection is restricted to higher galactic latitudes ($ |b| \ge 20$) to exclude the galactic plane, where a high source density leads to significant source confusion. The selection further requires a signal-to-noise ratio of $S/N\ge5$ in the WISE W1 and W2 bands and J-band detections ($J>0$) for all objects. 
At the heart of the near-infrared selection is the JKW2 color cut \citepalias{Schindler2017},
\begin{equation}
K_{\rm{Vega}}-W2_{\rm{Vega}} \ge 1.8 - 0.848\cdot(J_{\rm{Vega}}-K_{\rm{Vega}}) \ ,
\end{equation}
which allows to clearly separate quasars at $z<5$ from the stellar locus in J-K-W2 color-space. 

We obtained 3,815,192 sources, which are then further matched to optical photometry from the Pan-STARRS PS1 catalog within a $3\farcs96$ aperture using the STSCI MAST casjobs interface\footnote{\url{http://mastweb.stsci.edu/ps1casjobs/}}. We adopt the flags outlined in \citet[][their Table\,6]{Banados2014} to ensure the selection of reliable photometry according to the Image Processing Pipeline \citep{Magnier2006, Magnier2007}. 
The full SQL query to retrieve the PS1 data is provided in Appendix\,\ref{app_ps1_sql}. We have included a loose criterion to reject extended sources (Section\,\ref{sec_ext_study}) in the query to reduce the download size of the data set.  The match to the Pan-STARRS PS1 catalog returned a total of 74318 sources.

The Pan-STARRS PS1 photometry has been matched with the AllWISE position. While the AllWISE PSF is larger than the 2MASS PSF, their average astrometric precision with respect to the U.S. Naval Observatory CCD Astrograph Catalog is similar (2MASS:$\sim 80\,\rm{mas}$, AllWISE:$\sim 87\,\rm{mas}$). However, the 2MASS PSC online documentation\footnote{\url{https://old.ipac.caltech.edu/2mass/releases/allsky/doc/sec2_2.html}} notes that stars fainter than $K_{\rm{Vega}}\sim14$ have worse position residuals, which indicates that the extraction uncertainties dominate rather than the uncertainties in the mapping into the IRCS reference frame. As our selection is limited by the depth of the 2MASS survey and the majority of the pre-matched  3,815,192 sources have $K_{\rm{Vega}}>14$, we preferred to use the AllWISE position over the 2MASS position for the cross-match to PanSTARRS PS1.

The optical and near-infrared photometry is extinction corrected using the python \texttt{dustmaps} module \citep{Green2018} with the values of \citet{Schlegel1998}.

\subsection{Rejection of extended sources}\label{sec_ext_study}

The JKW2 color cut is highly successful in rejecting stellar contaminants. However, as described in \citetalias{Schindler2017}, galaxies straddle the color cut and become our main contaminants once the majority of stars are excluded. 

We use the absolute value of the magnitude difference ($\Delta m$) between the PS1 mean PSF magnitudes (\texttt{iMeanPSFMag}) and the PS1 mean aperture magnitudes (\texttt{iMeanApMag}) as our main quantity to identify extended sources. 

\begin{figure}
\includegraphics[width=0.5\textwidth]{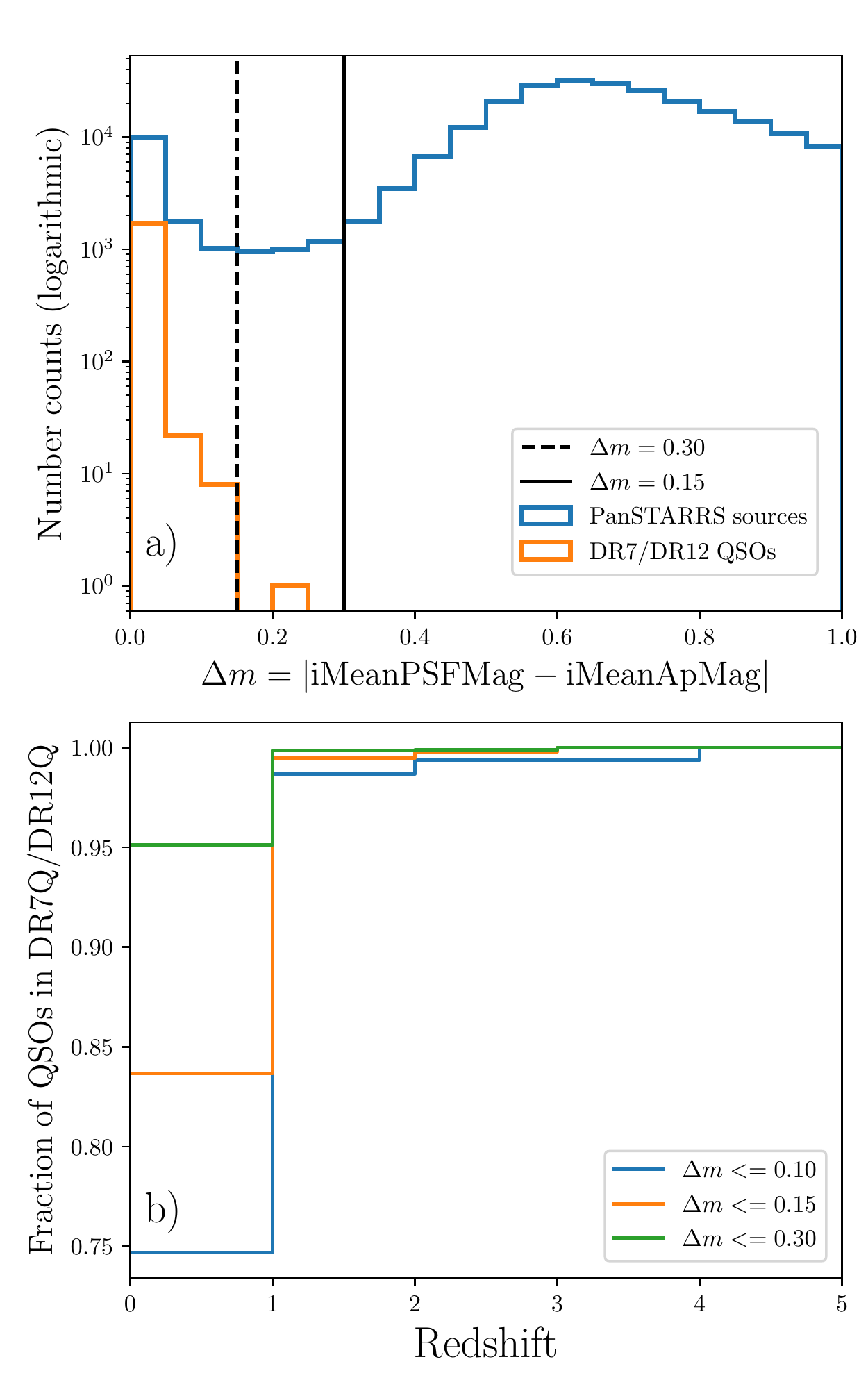}
\caption{(a) Distribution of PS1 sources (blue) and SDSS DR7Q/DR12Q quasars (orange) as a function of the absolute value of the difference ($\Delta m$) between the PS1 mean PSF magnitude and the PS mean aperture magnitude. The PS1 sources are selected by the photometric criteria discussed in Section\,\ref{sec_phot_sel} in a region of $b\le-20$ and $120 \le l \le 240$ (galactic coordinates). The SDSS quasars are restricted to $i \le 18.5$ and to redshifts $z\ge2.5$, as targeted in this study. (b) The fraction of bright ($i \le 18.5$) quasars in the DR7Q/DR12Q included in three different $\Delta m$ cuts as a function of redshift. At redshifts $z\ge1.0$ the majority of all quasars are unaffected by the $\Delta m$ cut.}
\label{fig_ext_study}
\end{figure}

In Figure\,\ref{fig_ext_study} (a) we display two data sets as a function of their magnitude difference, $\Delta m$. The first histogram (blue solid line) is calculated from all sources in a region of $\rm{b}\le-20$ $120\le\rm{l}\le240$ (galactic coordinates) that passed the JKW2 color cut and were matched to PS1 photometry according to our criteria above. This corresponds to roughly $2250\,\rm{deg}^2$ or $\sim10\%$ of the total survey area. There are 267,951 sources in this data set, of which 12,579 (15,696) have $\Delta m \le 0.15\ (0.3)$. The distribution of the sources as a function of $\Delta m$ has a minimum around $\Delta m \approx 0.2$, with the majority of sources in the data set exhibiting higher values of $\Delta m$. 
The second data set contains the quasars from the combined SDSS DR7 and SDSS DR12 quasar catalogs matched to PS1 photometry. In addition, we only include quasars with $i < 18.5$  and (visually inspected) $z>2.5$ to restrict the quasar sample to the same magnitude range and redshift range that we target with our selection.  All of the remaining 1736 quasars have $\Delta m < 0.3$ and all except one are even included within  $\Delta m < 0.15$.

In Figure\,\ref{fig_ext_study} (b) we display the fraction of quasars, which are restricted by various cuts on $\Delta m$, to all SDSS DR7/DR12 quasars with $m_{\rm{i}} < 18.5$. It becomes evident that any of the three restrictions on $\Delta m$ decreases the number of quasars at the lowest redshifts. At redshifts beyond $z\approx1.0$ the majority of quasars ($>99\%$) are included even when the stronger restriction of $\Delta m \le 0.15$ is applied.

For the PS-ELQS quasar selection we thus reject extended sources using $\Delta m < 0.15$ as our main criterion. Based on the SDSS DR7/DR12 quasar samples we estimate that this restriction has a completeness of $99.8\%$ for quasars at $z>2.5$. 
After applying the $m_{\rm{i}} < 18.5$  magnitude cut and the criterion to reject extended objects to our 74318 candidates, we retain 43430 sources.

\subsection{Random Forest redshift regression and classification}

Random Forests \citep{Breiman2001} are a supervised machine learning technique, that can be efficiently applied to multi-class classification or standard regression problems. The algorithm is non-parametric and avoids the problem of overfitting. 
In the past random forests have been successfully used on many astronomical data sets \citep{Dubath2011, Richards2011, Carliles2010, CarrascoKind2013}, including quasar classification \citep{Carrasco2015} and redshift estimation \citep{DIsanto2018}.

For any supervised machine learning technique, the results of the classifier (or regressor) are highly dependent on the training set. For our purposes we base our training set on quasars from the combined SDSS DR7 and DR12 quasar catalogs as well as on a spectroscopic sample of stars from SDSS DR13. These are essentially the same training sets used in \citepalias{Schindler2017}, matched to the PS1 source catalog within $3.96\arcsec$ to obtain PS1 DR1 photometry. Table\,\ref{tab_training_detail} provides the total numbers of stars and quasars in the different training sets used below. 
% PS1 stars 
% 121426+46998+44320+32866+22862 +30366+11740+11194+8172+5647
% PS1 quasars 
% 49283+41661+28319+3153 + 12394+10416+7040+754

% PS1+W1W2 stars
% 79439+41820+37894+14652+12704 + 19704+10532+9535+3711+3146
% PS1+W1W2 quasars
%40776+37492+26055+2557 + 10156+9358+6537+669

%PS1+TMASS+W1W2 stars
%55014+31518+30750+10402+7784 +  13780+7861+7680+2647+1899
%PS1+TMASS+W1W2 quasars
%3061+917+707+55 + 798+215+161+12

\begin{table}[t]
\caption{Overview over the number of quasars and stars in the training sets}
\begin{tabular}{ccc}
\tableline
\tableline
Data set & DR13 Stars & DR7/DR12 Quasars \\
\tableline
full catalogs & 383966 & 213781 \\
\tableline 
\multicolumn{3}{c}{Classification} \\
\tableline
PS1, $i_{\rm{SDSS}} < 21.5$ & 335591 & 153020 \\
PS1+W1W2, $i_{\rm{SDSS}} < 21.5$ &233137 &  133600\\
PS1+TMASS+W1W2, $i_{\rm{SDSS}} < 21.5$ & 169335& 5926 \\
\tableline 
\multicolumn{3}{c}{Regression} \\
\tableline
PS1+W1W2 & - & 134097\\
PS1+W1W2, $i_{\rm{SDSS}} < 18.5$ & - & 13119\\
\tableline
\end{tabular}
\label{tab_training_detail}
\end{table}

After the photometric selection and the extended object rejection our candidate sample is still contaminated by stars and low redshift ($z<2.8$) quasars. To enhance our efficiency we use the random forest classifier to reject obvious stellar contaminants and then estimate a photometric redshift with the random forest regressor to select only the $z\ge2.8$ quasars.

We use the \texttt{scikit-learn} \citep{scikit-learn} python implementation of the random forest classifier and regressor with its default parameters unless otherwise noted. 
%Each of the binary trees of the random forest is constructed using the Gini impurity to determine the best split at each step. 

\subsubsection{Classification}

To further enhance the efficiency of our selection we employ random forest classification. We classify our candidates into four different redshift classes ("vlowz" : $0< z\le 1.5$, "lowz" : $1.5< z\le 2.2$, "midz" : $2.2< z\le 3.5$, "highz" : $3.5< z$) and five different stellar spectral classes (A, F, G, K, M). The redshift classes are designed to split the quasars at redshifts where emission lines move from one passband into the next redder passband, introducing strong features in the corresponding flux ratios \citep[see also,][]{Richards2015}. For evaluation purposes we also summarize all stellar classes under the "STAR" label and all quasar classes under the "QSO" label, effectively resulting in a binary classification.

We combine the SDSS DR7Q/DR12Q quasars with the SDSS DR13 spectroscopic stars \citepalias[see][]{Schindler2017} matched to PS1  photometry to form the classification training set. Our only requirement is that all objects are brighter than $i_{\rm{SDSS}}<21.5$ to exclude very faint objects with substantial photometric uncertainties.

We test the performance of the classification for three subsets of the full training set with different features. The first uses the four adjacent flux ratios ($g/r$, $r/i$, $i/z$, $z/y$) of the five photometric PS1 bands and the PS1 $i$-band magnitude as features. For the second and third training set we first include the WISE W1 and W2 bands (PS1+W1W2) and for the third we also include all three 2MASS passbands (PS1+TMASS+W1W2). The feature set is expanded accordingly, when we include the WISE and 2MASS photometry, by adding the additional flux ratios ($+$$[y/W1$, $W1/W2]$; $+$$[y/J$, $J/H$, $H/K_s$, $K_s/W1$, $W1/W2]$) and the W1 and the J-band magnitude. We require each object in the subsets to have information in all used features (see constraints in Table\,\ref{tab_class_results}), resulting in varying number of sources per data set. 
For each subset we calculate the best combination of hyperparameters for the classifier on a grid of \texttt{n\_estimators}$=[200, 300, 400]$, \texttt{min\_samples\_splot}$=[2, 3, 4]$, and \texttt{max\_depth}$=[15, 20, 25]$. 
As in the case for the photometric redshift regression, we apply five-fold cross-validation on the full training set using $80\%$ of the sources for training and the remaining $20\%$ for validation. 

The best hyperparameters are evaluated using the $F_1$ score \citep{Bishop2006}. The $F_1$ score, which is also called the traditional F-measure or the balanced F-score, is the harmonic mean of the precision and the recall of the classification. 
\begin{equation}
 F_1 = 2 \cdot \frac{\rm{precision}\cdot\rm{recall}}{\rm{precision}+\rm{recall}} \ .
\end{equation}
Here precision  (p; or efficiency) is defined as the ratio of true positives to the sum of true and false positives and the recall (r; or completeness)  is defined as the ratio of true positives to the sum of true positives and false negatives.

Table\,\ref{tab_class_results} provides an overview of the best classification results for the three different subsets. 
The first three columns show the sizes of the training and validation sets for each subset, the constraints on the subset, and the features used. The last three columns provide the precision (p), recall (r) and $F_1$ measure for the "highz" quasar class as well as the binary classification between quasars ("QSO") and stars ("STAR"). We would like to stress that only 12 "highz" quasars were included in the PS1+W1W2+TMASS validation set (third row in  Table\,\ref{tab_class_results}) to determine the p, r and $F_1$ values for the "highz" class, introducing high stochastic uncertainties on those values.

The inclusion of flux ratios beyond the PS1 photometry leads to generally better classification results. On the downside, the sample sizes decrease with the addition of the WISE and 2MASS photometry, as we require all objects to have information in all features considered. When the training sets become too small, they will not be able to fully populate the available feature space and thus will lead to worse classifications. Additionally, small validation sets will introduce large errors on the classification metrics. 
As a result, the recall value and $F_1$ score of the "highz" class in the PS1+W1W2+TMASS set is worse than in the other two subsets, although more features are included.

Hence, we adopt the PS1+W1W2 subset as the best training and feature set for our classification problem (marked with $\star$ in Table\,\ref{tab_class_results}). It achieves the best classification results of all three subsets. The best hyperparameters for this set were evaluated to be  \texttt{n\_estimators}$=400$, \texttt{min\_samples\_splot}$=4$, and \texttt{max\_depth}$=25$.

A helpful visualization to understand the classification result is the confusion matrix. Each row of the matrix marks the true class of the objects within it, which are predicted to belong to different classes according to the columns of the matrix. The diagonal entries show the number of correctly classified objects, while all off-diagonal entries show the numbers of incorrectly classified objects. It provides a good overview over which classes are commonly confused and which can be separated easily with the features supplied to the classifier.

\begin{figure}
\centering 
\includegraphics[width=0.5\textwidth]{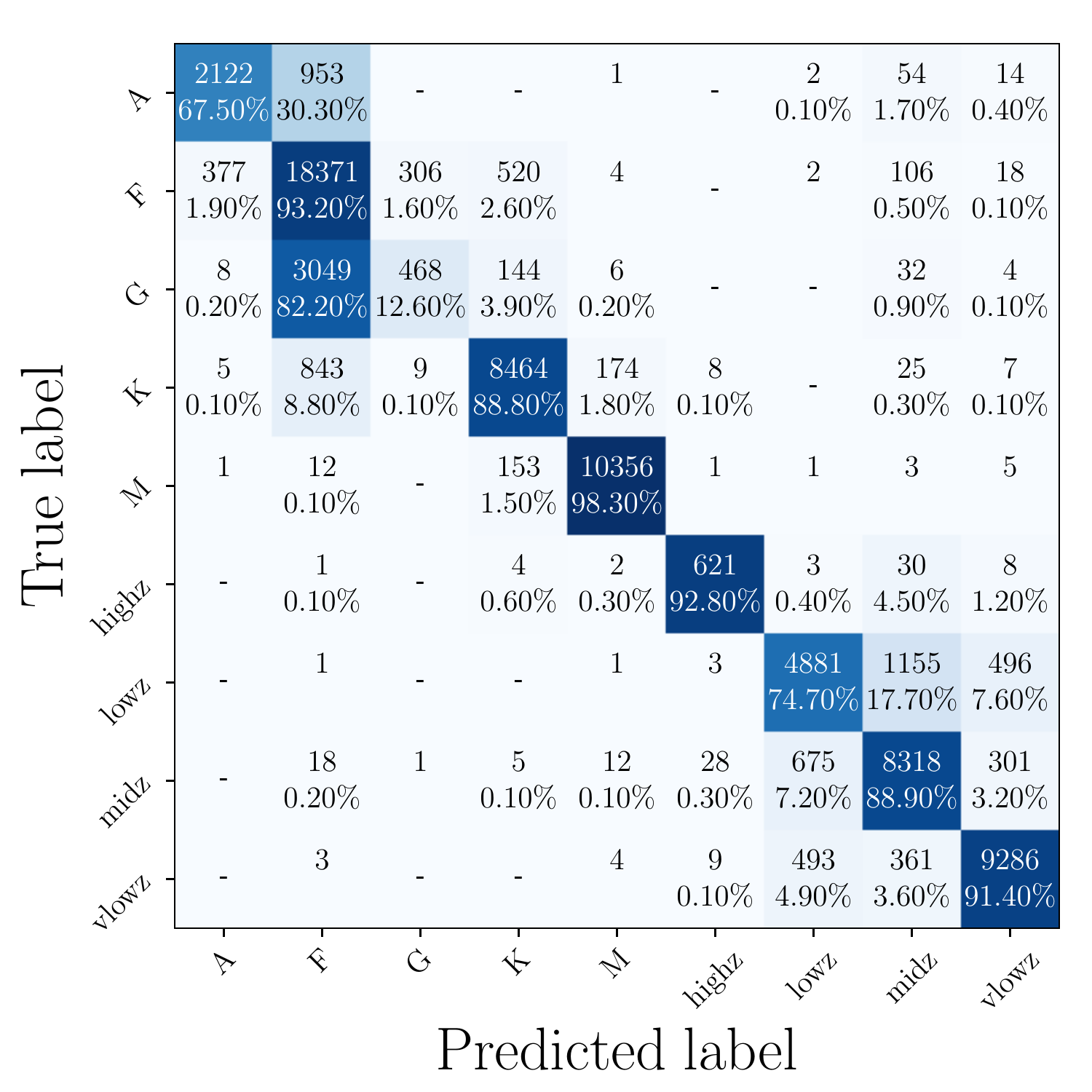}
\caption{The confusion matrix for the PS1+W1W2 subset ($\star$ in Table\,\ref{tab_class_results}). The rows show the true labels (classes) of the objects, whereas the columns indicate the predicted labels (classes). Each entry shows the total number of objects of the true row label classified as the predicted column label. The percentages show the fraction of objects in that entry to the total number of objects in the row. The entries are color coded to highlight the entries with the majority of objects in each row.}
\label{fig_cnf_mat}
\end{figure}

We display the entire confusion matrix for the PS1+W1W2 subset ($\star$) in Figure\,\ref{fig_cnf_mat}. The individual entries show the number of objects belonging to the row class (true label) and classified as the column class (predicted label). The percentages below each entry are with respect to the total number of objects in the true class (full row). Therefore the percentages of the diagonal entries show the completeness with respect to its row/column label.

The quasar and star classes are well separated. The largest number of stellar contaminants enter in the "midz" quasar class ($2.2\le z \le 3.5$). This is one redshift range, where the quasar distribution overlaps strongly with the stellar locus in optical color space. Indeed, the majority of quasars misclassified as stars also stem from this redshift class. Within the quasar classes "midz" and "lowz" quasars show the highest level of confusion with each other.

%Best model 0.844 (+/-0.001) for 'min\_samples\_split': 4, 'n\_estimators': 400, 'max\_depth': 25

\begin{table*}[t]
\caption{Results of the Random Forest classification on the full empirical training set}
\begin{tabular}{ccccccc}
 \tableline
 \tableline
 Training / Validation size\tablenotemark{a} & Constraints & Features\tablenotemark{b} & p / r / F1 (highz) & p / r / F1 (QSO) & p / r / F1 (STAR)  & \\
 \tableline
  390888 / 97723 &  PS1 fl.r., $i_{\rm{SDSS}}<21.5$ & PS1 & 0.87 / 0.85 / 0.86 & 0.93/0.91/0.92 & 0.96/0.97/0.97  & \\
  293389 / 73348 &  PS1+W1W2 fl.r., $i_{\rm{SDSS}}<21.5$ & PS1+W1W2 & 0.93 / 0.93 / 0.93 & 0.99/1.00/0.99 & 1.00/0.99/0.99  &$\star$ \\
  140208 / 35053 &  PS1+TMASS+W1W2 fl.r., $i_{\rm{SDSS}}<21.5$ & PS1+TMASS+W1W2 & 1.00 / 0.67 / 0.80 (12) & 1.00/0.98/0.99 & 1.00/1.00/1.00 & \\
 \tableline
\tablenotetext{1}{For the 5-fold cross validation the full data sets are split into a training (80\%) and validation (20\%) set . We provide the number of objects for each set in this column.}
\tablenotetext{2}{ We abbreviated flux ratios to "fl.r." in this column.}
\end{tabular}
\label{tab_class_results}
\vspace{0.5cm}
\end{table*}

\subsubsection{Photometric redshift regression}

Our selection process rejects stars with the JKW2 color cut and galaxies due to their large difference in mean PSF and aperture magnitudes. Therefore, quasars at $z<2.8$ become the dominant contaminants. We use random forest regression to calculate photometric redshifts, $z_{\rm{reg}}$, and then select quasar candidates with $z_{\rm{reg}}\ge 2.8$.

The training set includes all SDSS DR7/DR12 quasars with full PS1 and WISE W1 and W2 photometry. We also build a smaller subset, limiting the full quasar training set to $i<18.5$. 

The features used for the random forest regression are the 6  adjacent flux ratios ($g/r$, $r/i$, $i/z$, $z/y$, $y/W1$, $W1/W2$) from the five photometric bands of PS1 in addition to W1 and W2. We further add the PS1 $i$-band magnitude and the $W1$ magnitude to the feature set. 
As discussed in \citetalias{Schindler2017}, including 2MASS photometry dramatically reduces the number of training objects and therefore does not allow for sufficient training in the large feature space.

We perform grid searches on the full training set and the magnitude limited subsample to determine the hyperparameters of the best regression model. The grid of hyperparameters includes the number of binary trees (\texttt{n\_estimators}$=[200, 300, 500]$), the minimum number of samples to be split (\texttt{min\_samples\_splot}$=[2, 3, 4]$), and the maximum depth of the tree (\texttt{max\_depth}$=[15, 20, 25]$).
To test the hyperparameters we use five-fold cross-validation on the full training set using each time $80\%$ of the sources for training and the remaining $20\%$ for validation. 

The best hyperparameters are evaluated using the standard $R^2$ regression score,
\begin{equation}
 R^2 = 1- \frac{\sum_i \left(z_{\rm{spec}, i}-z_{\rm{reg}, i}\right)^2}{\sum_i\left(z_{\rm{spec}, i}-\bar{z}\right)^2 } \ ,
\end{equation} 
where $z_{\rm{spec}, i}$ are the true redshifts, $\bar{z}$ is the mean of all $z_{\rm{spec}, i}$, and the predicted redshift values are denoted by $z_{\rm{reg}, i}$.
The other common metric in the literature assesses the goodness of the redshift estimation with redshift normalized residuals  ($\delta z = \left| z_{\rm{reg}} - z_{\rm{spec}} \right| / (1+z_{\rm{spec}})$). 
Most studies report the fraction of quasars in the validation set with residuals smaller than a given threshold $e$. 
%The other common metric for photometric redshift estimation in the literature is the fraction of quasars with redshift normalized residuals smaller than a given residual threshold $e$ to the total number of quasars in the validation set $N_{\rm{tot}}$.
\begin{equation}
 \delta_{e} = \frac{N(\left| z_{\rm{reg}, i} -z_{\rm{spec}, i} \right| < e\cdot(1+z_{\rm{spec}, i}))}{N_{\rm{tot}}} \ ,
\end{equation}
where $N_{\rm{tot}}$ denotes the total number of quasars in the validation set. Residual thresholds of $e= 0.1, 0.2$, and $0.3$ are typically chosen in this context.

\begin{figure}
\centering 
\includegraphics[width=0.5\textwidth]{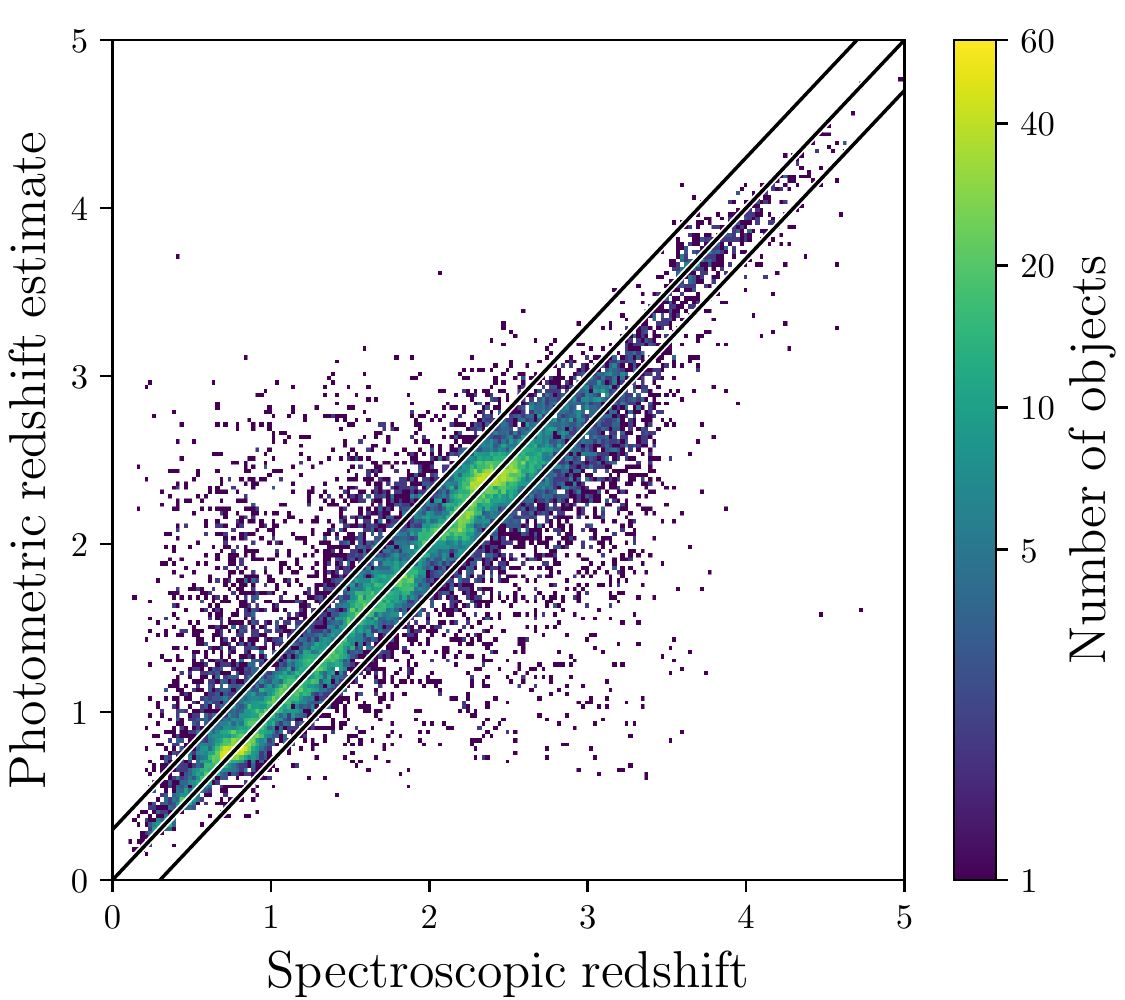}
\caption{Distribution of quasars in the validation set ($\star$ in Table\,\ref{tab_reg_results}) as a function of photometric redshift estimate (regression redshift) and measured spectroscopic redshift, color coded by the number of objects per bin. The three solid black lines illustrate the $\Delta z = |z_{\rm{spec}}-z_{\rm{reg}}| = 0$ diagonal and the $\Delta z \le 0.3$ region.}
\label{fig_reg_zz}
\end{figure}

We show the best results of the regression grid search for the full quasar training set and the magnitude limited subsample in Table\,\ref{tab_reg_results}. 
While the magnitude limited subset (second row) achieves slightly better results, the training set is reduced to $10\%$ of its full size. In addition, it will be additionally biased against higher redshift quasars, because they are generally fainter. Therefore we adopted the full quasar training set (marked with $\star$ in Table\,\ref{tab_reg_results}). The best hyperparameters for this case are \texttt{min\_samples\_split}$= 2$, \texttt{n\_estimators}$= 500$, and \texttt{max\_depth}$= 25$.
The resulting distribution of spectroscopic redshifts to random forest regression redshifts in the validation set are shown in Figure\,\ref{fig_reg_zz}. While the results at $z\gtrsim3$ are mostly within the region of $\Delta z  = |z_{\rm{spec}}-z_{\rm{reg}}| \le0.3$, a larger distribution of outliers persists at lower redshifts.

%Best (most inclusive) model 0.817 (+/-0.005) for 'min\_samples\_split': 2, 'n\_estimators': 500, 'max\_depth': 25
 
\begin{table*}[t]
\centering
 \caption{Results of the Photometric Redshift estimation methods}
\begin{tabular}{cccccccccc}
\tableline
\tableline
Data set & Training / Validation size\tablenotemark{a} & Constraints & Features  & $\delta_{0.3}$ & $ \delta_{0.2}$ & $\delta_{0.1}$ & $\sigma\tablenotemark{b}$ & $R^2$ & \\
\tableline
DR7DR12Q & 107277 / 26820 & PS1+W1W2 fl.r. & PS1+W1W2  & 0.95 & 0.91 & 0.79 & 0.352 & 0.817  &$\star$\\
DR7DR12Q & 10495 / 2624 & PS1+W1W2 fl.r., $i<18.5$ & PS1+W1W2  & 0.98 & 0.95 & 0.88 & 0.265 & 0.883 &\\
 \tableline
 \tablenotetext{1}{For the 5-fold cross validation the full data sets are split into a training (80\%) and validation (20\%) set . We provide the number of objects for each set in this column.}
 \tablenotetext{2} {Standard deviation of the residual of the photometric redshift estimate (regression redshift) and the measured spectroscopic redshift.}
\end{tabular}
\label{tab_reg_results}
\end{table*}

\section{Construction of the PS-ELQS quasar candidate catalog}\label{sec_pselqs_cand_cat}

\begin{figure}
\centering
\includegraphics[width=0.5\textwidth]{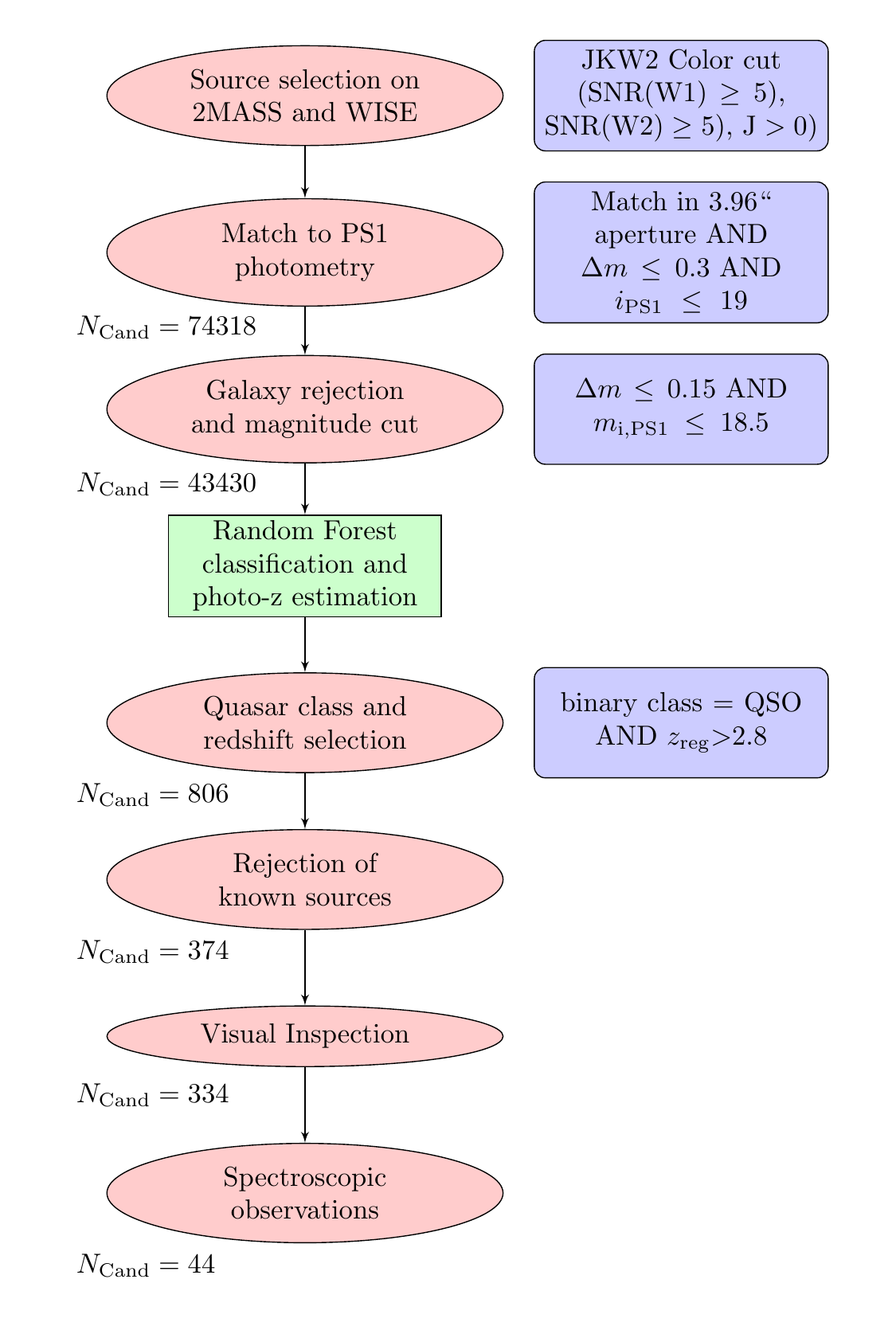}
\caption{Flowchart of the PS-ELQS quasar candidate selection.}
\label{fig_flowchart}
\end{figure}

We provide an overview over the candidate selection process in Figure\,\ref{fig_flowchart} and Table\,\ref{tab_cand_stat}. The selection process begins with the parent sample of 2MASS and WISE AllWISE sources, which pass the JKW2 color cut and have reliable photometry ($\rm{SNR(W1)}\geq5$, $\rm{SNR(W2)}\geq5$, $\rm{J}>0$). We then match these sources to the PS1 photometry within a  $3.96\arcsec$ aperture, requiring the objects to be brighter than $i=19.0$ and fulfill $\Delta m \le 0.3$. Furthermore all objects have to satisfy a range of quality flags. The full sql query is shown in Appendix\,\ref{app_ps1_sql}.
After we retrieved the PS1 photometry for all remaining sources (74318), we apply the more stringent rejection of extended sources ($\Delta m \le 0.15$) and restrict the sample to all sources with $i\le18.5$. 

At this point we run the random forest regression and classification. The regression provides us with a regression redshift $z_{\rm{reg}}$, our photometric redshift estimate. The classification determines the most likely class of the object (\texttt{rf\_mult\_class\_pred}) and the summed probability of the object belonging to any of the quasar classes (\texttt{rf\_qso\_prob}). All objects that are generally classified as quasars ("QSO"), according to the binary classification, and have regression redshifts of $z_{\rm{reg}}\ge 2.8$ form the PS-ELQS candidate sample.

The PS-ELQS candidate sample selects a total of 432 known quasars from the literature, including 70 sources observed as part of the ELQS, which are then excluded from the candidate sample along with all other known sources.
We visually inspect all unknown candidates for unreliable photometry. We reject all candidates, whose point spread function is blended with nearby sources, where image artifacts are evident or the source seems to be extended. After the exclusion of known quasars and the rejection of 40 candidates with unreliable photometry, the "Good" PS-ELQS candidate sample has a total of 334 objects. These candidates are then prioritized according to the criteria described in Table\,\ref{tab_priority}. 

Spectroscopic observations could successfully identify 290 good PS-ELQS candidates, of which 190 are quasars at $z\ge2.8$. A total of 44 good PS-ELQS candidates have not been observed yet, but are targeted in future observing campaigns. A list of the remaining PS-ELQS candidates is given in Table\,\ref{tab_cand} in Appendix\,\ref{app_candidates}.

\begin{table}
 \centering
 \caption{PS-ELQS candidate selection}
 \label{tab_cand_stat}
 \begin{tabular}{lc}
%data based on make_pselqs_catalog.py
 WISE+2MASS+PS1 parent sample &  \multirow{2}{*}{74318}\\
 (JKW2 color cut, $i < 19.0$, $|\Delta m|<0.3 $) & \\
 \tableline
 Photometric sample  & \multirow{2}{*}{43430}\\
 ($i < 18.5$, $|\Delta m|<0.15 $) & \\
 \tableline
 PS-ELQS candidate sample & \multirow{2}{*}{806} \\
 ($z_{\rm{reg}}\ge2.8$ and class=``QSO'')  & \\
 \tableline
 Known QSOs in the literature in the candidate sample & 432 \\
 Known $z\ge2.8$ QSOs in the literature in the candidate sample & 402 \\
 \tableline
 Observed ELQS sources in the candidate sample & 70 \\
 Known $z\ge2.8$ ELQS QSOs in the candidate sample & 54 \\
 \tableline
 Good PS-ELQS candidates  & \multirow{2}{*}{334} \\
(excluding bad photometry and known sources) & \\
 \tableline 
 Good PS-ELQS candidates observed & 290 \\
 Remaining good PS-ELQS candidates & 44 \\
 \tableline
 New PS-ELQS quasars& 218 \\
 New PS-ELQS quasars at $z\ge2.8$  & 190 \\
\tableline 
\end{tabular}
\end{table}

%crit_a = 'vis_id=="good" and rf_emp_photoz>=3.5 and PS_mag_i <=18.0'
%    df.loc[df.query(crit_a).index, 'priority'] = 1
%
%    crit_b = 'vis_id=="good" and 3.5>rf_emp_photoz>=3.0 and PS_mag_i <=18.0'
%    df.loc[df.query(crit_b).index, 'priority'] = 2
%
%    crit_c = 'vis_id=="good" and 3.0>rf_emp_photoz>=2.5 and PS_mag_i <=18.0'
%    df.loc[df.query(crit_c).index, 'priority'] = 3
%
%    crit_d = 'vis_id=="good" and rf_emp_photoz>=3.5 and 18.0<PS_mag_i <=18.5'
%    df.loc[df.query(crit_d).index, 'priority'] = 3
%
%
%    crit_e = '(vis_id=="ext" or vis_id=="good") and  3.5>rf_emp_photoz>=3.0 and 18.0<PS_mag_i <=18.5'
%    df.loc[df.query(crit_e).index, 'priority'] = 4
%
%    crit_e = '(vis_id=="ext" or vis_id=="good") and  3.0>rf_emp_photoz>=2.5 and 18.0<PS_mag_i <=18.5'
%    df.loc[df.query(crit_e).index, 'priority'] = 5

\begin{table}
\caption{PS-ELQS quasar candidates and their selection priorities}
\begin{tabular}{ccc}
\tableline
Priority & Criteria & Good PS-ELQS \\
 & & candidates \\
 & & (44 remaining)  \\
\tableline 
\tableline 
1 & $3.5 \le z_{\rm{reg}}$ and $m_i\le18.0$ & 53 (3)\\
2 & $3.0 \le z_{\rm{reg}}\le3.5$ and $m_i\le18.0$ & 49 (5)\\
3 & ($2.5 \le z_{\rm{reg}}\le3.0$ and $m_i\le18.0)$ & 122 (12)\\
 & OR ($3.5 \le z_{\rm{reg}}$ and $18.0 < m_i\le18.5)$ & \\
4 & $3.0 \le z_{\rm{reg}}\le3.5$ and $18.0 < m_i\le18.5$ & 59 (8)\\
5 & $2.5 \le z_{\rm{reg}}\le3.0$ and $18.0 < m_i\le18.5$& 46 (16)\\
10 & all the remaining candidates & 5 \\
\tableline
\end{tabular}
\label{tab_priority}
\end{table}

%All primary candidates to observe: 44
%5    16
%3    12
%4     8
%2     5
%1     3
%All PS-ELQS candidates observed, priorities: 
%3     110
%4      51
%1      50
%2      44
%5      30
%10      5

% 
% crit_a = 'vis_id=="good" and rf_emp_photoz>=3.5 and PS_mag_i <=18.0'
%     df.loc[df.query(crit_a).index, 'priority'] = 1
% 
%     crit_b = 'vis_id=="good" and 3.5>rf_emp_photoz>=3.0 and PS_mag_i <=18.0'
%     df.loc[df.query(crit_b).index, 'priority'] = 2
% 
%     crit_c = 'vis_id=="good" and 3.0>rf_emp_photoz>=2.5 and PS_mag_i <=18.0'
%     df.loc[df.query(crit_c).index, 'priority'] = 3
% 
%     crit_d = 'vis_id=="good" and rf_emp_photoz>=3.5 and 18.0<PS_mag_i <=18.5'
%     df.loc[df.query(crit_d).index, 'priority'] = 3
% 
%     crit_e = '(vis_id=="ext") and rf_emp_photoz>=3.0 and PS_mag_i <=18.5'
%     df.loc[df.query(crit_e).index, 'priority'] = 4
% 
%     crit_e = '(vis_id=="ext" or vis_id=="good") and  3.5>rf_emp_photoz>=3.0 and 18.0<PS_mag_i <=18.5'
%     df.loc[df.query(crit_e).index, 'priority'] = 4
% 
%     crit_e = '(vis_id=="ext" or vis_id=="good") and  3.0>rf_emp_photoz>=2.5 and 18.0<PS_mag_i <=18.5'
%     df.loc[df.query(crit_e).index, 'priority'] = 5

\section{Spectroscopic follow-up observations and data reduction}\label{sec_observations}

Dedicated observational campaigns for the PS-ELQS began in fall 2017 after the candidate selection was frozen in. Observations were completed in June 2018. In this Section we will describe the observational setups and the data reduction procedure.

\subsection{SOAR}
The focus of this quasar survey was the mostly unexplored area of the PS1 $3\pi$ survey between galactic latitudes of $-30<b<0$. Therefore the majority of our observations were carried out with the Goodman High Throughput Spectrograph \citep[Goodman HTS; ][]{Clemens2004} on the Southern Astrophysical Research (SOAR) Telescope ($4.1\,\rm{m}$). Spectra were taken in 2017 October 6-10 , 2018 January 22-24, 2018 April 4-6, 2018 June 2-4. We used the 400\,g/mm grating with central wavelengths of $6000\text{\AA}$ and $7300\text{\AA}$. The first setup utilized the GG-385 blocking filter, whereas we used the GG-495 blocking filter for the second setup. The spectra have a wavelength coverage of $\sim4000-8000\text{\AA}$ and $\sim5300-9300\text{\AA}$ for the two different central wavelengths. We chose slit widths of $1\farcs0$ or $1\farcs2$ dependent on the weather conditions, resulting in spectral resolutions of $R\approx830$ and $R\approx690$, respectively. Exposure times varied between $3\,\rm{min}$ and $15\,\rm{min}$ depending on the target magnitude and the atmospheric transparency.

%April 2018 4-6, October 2017 6-10, January 2018 22-24, June 2018 2-4

\subsection{VATT Observations}
Identification spectroscopy in the northern hemisphere was carried out with the VATTSpec spectrograph on the Vatican Advanced Technology Telescope (VATT). Using the $300\,$g/mm grating  blazed at $5000\,\text{\AA}$ in first order, we achieved a resolution of $R\sim1000$ ($1\farcs5$ slit) and a coverage of ${\sim4000}\,\text{\AA}$ around our chosen central wavelength of $\sim5850\,\text{\AA}$. Targeted PS-ELQS observations were conducted in 2017 November 7-12, 2018 March 19-21, 2018 May 17-18. Depending on the source and the conditions the exposure times varied between 15 and 30 minutes.

\subsection{MMT Observations}
We followed up our newly discovered quasars with the Red Channel Spectrograph on the MMT. We used the 300\,g/mm grating blazed at 1st/$4800\text{\AA}$ with central wavelengths of $5655\text{\AA}$, $5570\text{\AA}$ and $5900\text{\AA}$. The grating has an approximate coverage of $3310\text{\AA}$ and achieves a resolution of $R\approx400$ to $300$ for the $1\farcs25$ and the $1\farcs5$ slits. Depending on the source and the weather conditions we chose exposure times of $\sim3-15\,\text{min}$ per spectrum. Observations were taken in 2017 October 20-21, 2017 November 16, and 2018 May 14.

\subsection{Data reduction}
We reduced the spectra with the long-slit reduction methods within the IRAF software package \citep{Tody1986, Tody1993}. 
Standard bias subtraction, flat field correction and sky subtraction were applied.  Sky subtraction and spectral extraction were done using the \texttt{apall} routine with optimal extraction (weights=variance) and cosmic ray reduction. The resulting low to medium signal-to-noise spectra allowed quasars to be easily identified by their broad emission lines. 
The wavelength calibration was based on internal lamps and spectral fluxes were initially calibrated using at least one spectrophotometric standard star per night. However, changing weather conditions did not allow for absolute flux calibration and we therefore scale the fluxes to match the observed PS1 r-band magnitudes. 
The spectra have not been corrected for telluric absorption features.

\section{The PS-ELQS quasar catalog}\label{sec_qso_catalog}

\begin{figure}[t]
\centering
\includegraphics[width=0.5\textwidth]{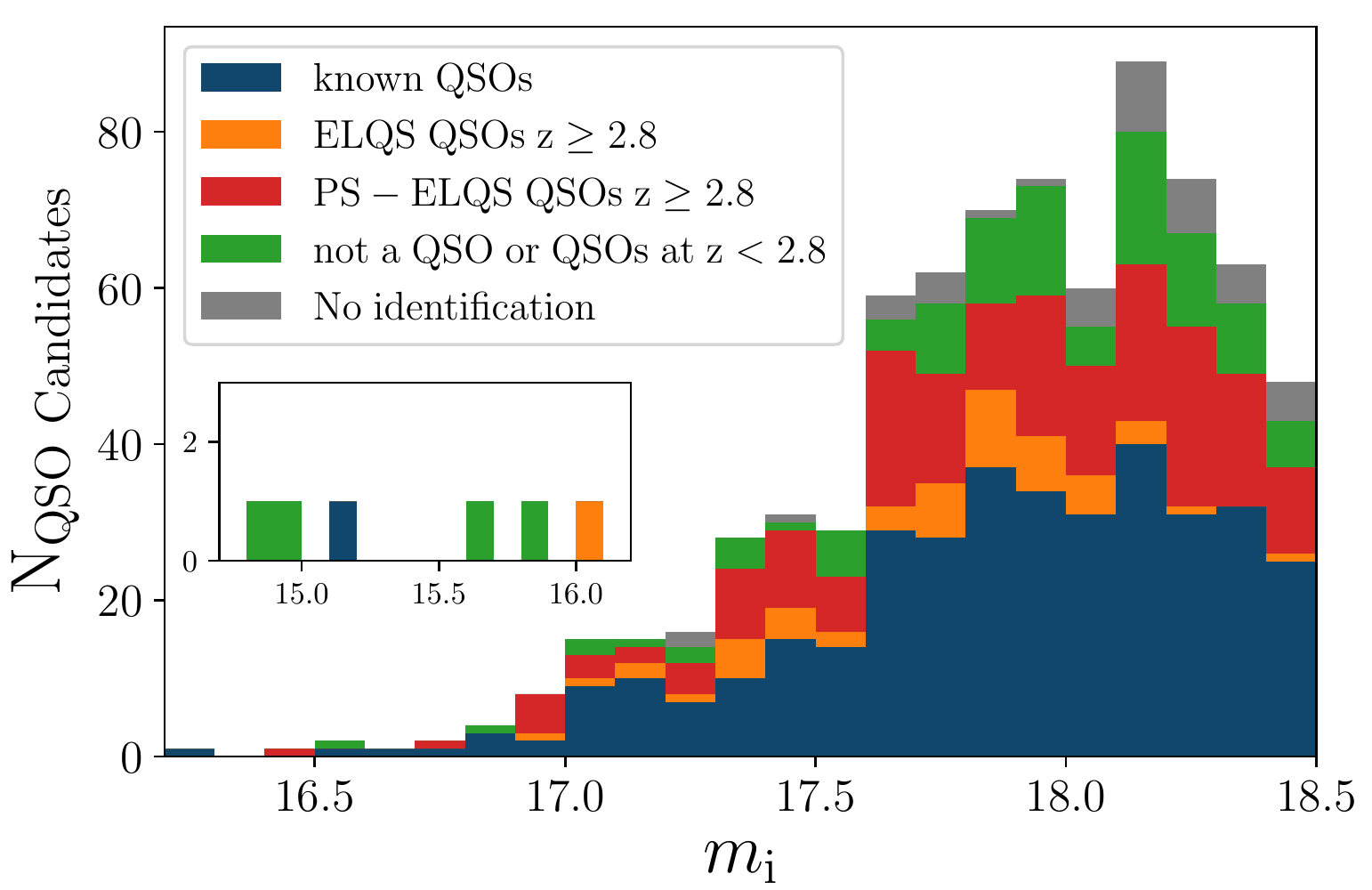}
\caption{Histogram of the spectroscopic completeness of all good PS-ELQS candidates as a function of their dereddened PS1 i-band magnitude $m_{i}$.  Dark blue and orange colors indicate all candidates from the general literature and the ELQS survey at $z\ge2.8$. All newly identified candidates in this work are highlighted in red (QSOs at $z\ge2.8$) and green (QSOs at $z<2.8$ or not a QSO). All candidates, which remain unidentified, are depicted in grey.}
\label{fig_spec_completeness}
\end{figure}

We conducted spectroscopic follow-up observations for 290 of our 334 good PS-ELQS candidates. We discovered a total of 218 new quasars, of which 190 are at $z\ge2.8$. 
%Our contaminants are mostly K- and M-stars (36/71) and galaxies (14/71).
The resulting PS-ELQS quasar catalog includes a total of 592 quasars at $z\ge2.8$:

\begin{itemize}
\item 285 quasars from DR14Q
\item 190 newly discovered quasars (PS-ELQS)
\item 54 quasars from ELQS 
\item 47 quasars from MQC
\item 13 quasars from Yang, J. et al. (in preparation)
\item 3 quasars from DR7Q
\end{itemize}

Excluding the 40 candidates with unreliable photometry we could identify 592 quasars at $z\ge2.8$ out of 766 good candidates ($= 806-40$), of which 44 have not been observed.  Therefore we calculate a minimum selection efficiency $77\%$ ($592/766$) for the PS-ELQS quasar selection. For quasars at all redshifts our selection efficiency reaches $\sim85\%$ (650/766). 
Out of the 72 objects, which were identified not to be quasars at $z\ge2.8$, we classified 36 as stars (K:18, M:7, G:2, F:2, 1 cataclysmic variable, and 6 unidentified stars) and 14 as galaxies. The remaining 22 objects are not quasars at $z\ge2.8$, but have often too low signal-to-noise to classify them with certainty as a star, galaxy or low-redshift quasar.
The random forest classification results on the validation set (see Figure\,\ref{fig_cnf_mat}) suggest a much higher efficiency of $99\%$ for quasars at any redshift.
The discrepancy with the much lower observed efficiency of $85\%$ is likely to originate in differences between the training set and the photometric sample. As we do not apply the JKW2 color cut to the stars training set, because it results in an undersized sample size (314 objects), the distribution of stellar types in the training set is different than from the photometric candidate sample. For example, $42\%$ of stars in the training set are K and M stars, while this ratio increases to $69\%$ once the JKW2 color cut is applied to the training set. Therefore, the photometric candidate sample has likely a higher fraction of contaminants that can mimic quasars at $z>2.8$, decreasing our overall efficiency.

The full PS-ELQS quasar catalog is published alongside this work in digital form. It also includes information on matches to GALEX, ROSAT 2RXS, and XMMSL2 (see Section\,\ref{sec_cross_matches} for details). Table\,\ref{tab_pselqs_cat_cols} provides an overview over all the columns.

Figure\,\ref{fig_spec_completeness} shows a histogram of all good candidates in the PS-ELQS sample as a function of their dereddened PS1 i-band magnitude, $m_{i}$. Known quasars at $z\ge2.8$ in the literature are colored dark blue, excluding the known objects published as part of the ELQS (orange). All 190 newly identified quasars at $z\ge2.8$ are shown in red, while all lower redshift quasars and non-quasars are shown in green. The remaining 44 good candidates are highlighted in grey. 

\begin{figure*}[htp]
\centering
\includegraphics[width=1.0\textwidth]{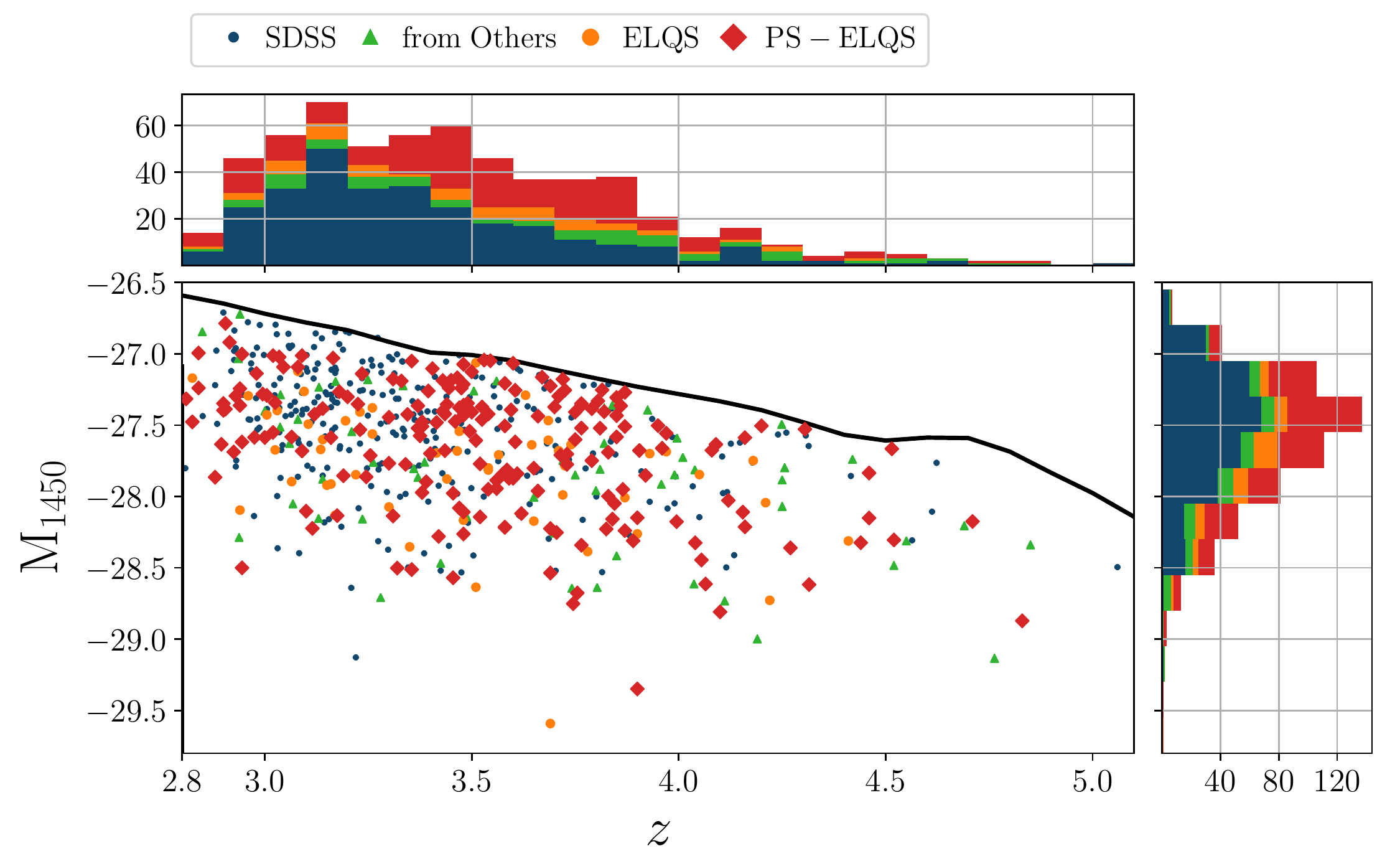}
\caption{All quasars in the PS-ELQS quasar catalog in the absolute magnitude ($M_{1450}$) and redshift ($z$) plane. Quasars identified in the literature are shown in dark blue dots (SDSS DR7Q/DR14Q), green triangles (other sources) and small orange circles (ELQS). Newly identified quasars by the PS-ELQS are depicted as larger red diamonds. The solid black line shows our limit on the apparent PS1 i-band magnitude ($m_{i}=18.5$). Histograms summarize the distribution of sources along their respective axis.}
\label{fig_M_z}
\end{figure*}

We present the discovery spectra of all newly identified quasars at $z\ge2.8$ in Figure\,\ref{fig_new_qso_spectra} in Appendix\,\ref{app_newqsos}. Additional identification spectra for new lower redshift quasars are shown in Figure\,\ref{fig_new_qso_lowz}.
All discovery spectra are sorted by redshift, beginning with the lowest redshift spectrum. A dark blue, orange and red colored bar at the top of each spectrum indicates the broad Ly$\alpha$, \ion{Si}{4} and \ion{C}{4} emission lines at $1215\,\rm{\AA}$, $1400\,\rm{\AA}$, $1549\,\rm{\AA}$ (rest-frame). Strong artifacts, like cosmic rays, were removed using an iterative sigma-clipping procedure. The spectra are not smoothed and spectral fluxes are scaled according to the PS1 r-band photometric measurement.
Redshifts for all newly discovered quasars were measured by visual comparison to a redshifted quasar template spectrum \citep{Vandenberk2001}. We estimate this method to have a redshift uncertainty of $\Delta z \approx 0.02$. 

We calculate absolute magnitudes at $1450\,\rm{\AA}$ rest-frame, $M_{1450}$ from the dereddened PS1 i-band magnitudes for all quasars. This transformation includes a k-correction term that we estimate using a large grid of simulated quasar spectra. The simulated quasar spectra and synthetic PS1 magnitudes are generated by the \texttt{simqso}\footnote{\url{https://github.com/imcgreer/simqso}} package \citep{McGreer2013}. 
The code begins by building a quasar continuum from a number of specified power-law slopes. Broad and narrow emission lines are then added onto the continuum as well as Fe emission and an IGM absorption model for the Ly$\alpha$-forest. 
Our model does not account for intrinsic extinction from the quasar host galaxy, as the survey is targeted at the unobscured UV-bright quasar population.
We adopt the values of \citet{Schindler2018} to build the spectral model and calculate a large grid of quasars with 28 cells over $m_{i} = 14-18$ and 53 cells across $z=0.2-5.5$ with a total of 200 model spectra per cell. K-corrections are then calculated for each grid cell and interpolated to retrieve individual k-corrections for all newly identified quasars.

Figure\,\ref{fig_M_z} shows all 592 quasars of the PS-ELQS quasar catalog as a function of their absolute magnitudes, $M_{1450}$, and redshifts, $z$. Known quasars from the literature are divided into objects from the SDSS quasar catalogs (dark blue dots), sources from the MQC and Yang, J. et al. (in preparation, green triangles), and the ELQS quasar catalog (orange circles). Newly discovered quasars identified with this work are highlighted as red diamonds. The black solid line shows the faint end magnitude limit of $m_{i}=18.5$ converted into $M_{1450}$. Histograms depict the distributions along each axis.

We provide a list of all newly discovered quasars within the PS-ELQS quasar catalog ($z\ge2.8$) in Table\,\ref{tab_pselqs_newqsos}. It contains the position in equatorial coordinates, the PS1 i-band magnitude, the absolute magnitude at $1450\,\rm{\AA}$, the determined spectroscopic redshift, near- and far-UV magnitudes from GALEX GR 6/7,  and further notes. We provide the same information for all newly discovered quasars at $z<2.8$ in Table\,\ref{tab_pselqs_lowzqsos}.  

\subsection{Quasar discoveries at low declinations}
The PS1 $3\pi$ survey covers $3\pi$ steradian of the northern sky, including the entire SDSS footprint. Therefore the PS-ELQS extends the efforts of the SDSS quasar surveys to a new region of $\sim9600\,\rm{deg}^2$. The majority of that area ($\sim5600\,\rm{deg}^2$) is at lower declinations ($\rm{Decl.}={-}30-0$). The PS-ELQS quasar catalog contains 207 quasars at lower declinations, of which only 70 were known before. Therefore we approximately triple the known population of luminous intermediate redshift quasars in this region.

\subsection{Notes on broad absorption line quasars} \label{sec_bal_quasars}
As part of the PS-ELQS we discovered a range of broad absorption line (BAL) quasars. Some of these objects have low-ionization BALs (LoBALs), showing broad absorption troughs from \ion{Mg}{2} and some of them also show absorption from meta-stable \ion{Fe}{2} (FeLoBALs). We provide notes on the redshift identification and the absorption properties for some BAL quasars below, where the redshift measurement from the broad emission lines is difficult due to the absorption features. This is not an exhaustive list of all PS-ELQS quasars with BAL absorption features.

\subsubsection*{J003856.98-292224.3}
This quasar is likely a FeLoBAL  at $z\approx2.27\pm0.005$ with strong nitrogen emission. The peak at $\sim4150\text{\AA}$ is \ion{N}{5} in the observed frame and the narrower line at $5750\text{\AA}$ is \ion{N}{3}] at $1750\text{\AA}$ rest-frame. BAL troughs are seen in \ion{N}{5}, \ion{C}{2}, \ion{Si}{4}, \ion{C}{4}, \ion{Al}{2}, \ion{Al}{3} and \ion{Fe}{2}.
%I believe this is at $z=2.270 \pm 0.005$.  That would make it an FeLoBAL quasar with strong nitrogen emission: the peak at ~4150 Ang observed is N V, not Ly A, and the narrowish line at 5750 observed is N III] 1750, which is rarely this strong.  The 1900 Ang complex at 6000-6200 Ang observed has stronger Al III 1860 and Si III] 1892 than C III] 1909.  (Assuming the strong peak at 4150 Ang is LyA does not yield a believable redshift.)
%The BAL troughs are seen in N V, C II, Si IV, C IV, Al II, Al III, and Fe II.

\subsubsection*{J021119.80-195943.0}
J021119.80-195943.0 is a FeLoBAL at $z=2.45\pm0.02$. The emission redshift is uncertain, because even the $1900\text{\AA}$ complex is affected by absorption. It also displays unusually strong \ion{N}{3}] $1750\text{\AA}$ emission. The red edges of the troughs are around $z=2.365$.
%
%$z=2.45 \pm 0.02$ FeLoBAL; emission redshift is uncertain because even the 1900 Ang complex
%is affected by absorption.  There seems to be unusually strong N III] 1750 emission in this object too.
%The red edge of troughs are at z=2.365 (narrow Fe III is seen, as well as many broad troughs).

\subsubsection*{J023500.45+023829.2}
This BAL quasars has a redshift of $z=1.975 \pm 0.005$, primarily identified by the $1900\text{\AA}$ complex. There is a \ion{C}{4} mini-BAL at $z=1.96$ and a \ion{C}{4} BAL at $z=1.85$.
%$z=1.975 \pm 0.005 from the 1900 Ang complex, primarily.
%CIV mini-BAL at z=1.96, C IV BAL at z=1.85.

\subsubsection*{J033559.99-132601.8}
This LoBAL can be identified to be at $z=1.900\pm0.005$ from the narrow \ion{C}{4} and \ion{Al}{3} emission. It displays absorption in \ion{C}{4}, \ion{Al}{3} and \ion{Mg}{2}.
%$z=1.900 \pm 0.005$ from narrow C IV and Al III emission.
%This is a LoBAL with C IV, Al III, Mg II absorption.

\subsubsection*{J044756.84-230748.3}
Based on multiple absorption lines arising from a near damped Ly$\alpha$ absorber this quasar has a redshift of at least $z=2.945\pm0.005$. It displays weak broad \ion{C}{4} emission from $5700-6300\text{\AA}$ and weak broad $1900\text{\AA}$ emission around $7100-7600\text{\AA}$. It has a LoBAL with \ion{Si}{4}, \ion{C}{4}, and \ion{Al}{3} at $z=2.82$ as well as a mini-BAL in \ion{N}{5}, \ion{Si}{4}, \ion{C}{4} at $z=2.745$.
%At least $z=2.945 \pm 0.005$ from multiple metal absorption lines arising from a near-DLA (seen at 4800 Ang).  Weak broad C IV emission line from 5700-6300 Ang; weak broad 1900 Ang emission feature from 7100-7600 Ang. 
%LoBAL with N V?, Si IV, C IV, Al III mini-BAL at z=2.82, and N V, Si IV, C IV, Al III? mini-BAL at z=2.745.
%Intervening absorption at z=1.321 (Fe II, Mg II, Mg I).

\subsubsection*{J113252.86-063243.3}
J113252.86-063243.3 is likely at $z=2.41\pm0.01$ mainly identified from the $1900\text{\AA}$ emission complex.  It displays a LoBAL (\ion{Si}{4}, \ion{C}{4}, \ion{Al}{2}, \ion{Al}{3}) at $z=2.27$. It further has narrower absorption troughs (\ion{Si}{4}, \ion{C}{4}, \ion{Al}{2}, \ion{Al}{3}) at $z=2.319$, $z=2.333$, and $z=2.363$.
%$z=2.41 \pm 0.01$, mostly from the 1900 Ang emission complex.
%LoBAL with Si IV, C IV, Al II, and Al III BALs at z=2.7.
%Plus narrow Si IV, C IV, Al II, Al III at z=2.319, z=2.333, z=2.363.

\subsubsection*{J191946.08+743747.1}
This BAL quasar is at $z=1.604\pm0.001$ identified by the narrow \ion{Mg}{2} emission. It shows a FeLoBAL with \ion{Al}{2}, \ion{Al}{3}, \ion{Fe}{3}, \ion{Fe}{2} and \ion{Mg}{2} absorption. The red end of the troughs is around $z=1.55$. It further displays narrow absorption in \ion{Al}{2}, \ion{Al}{3}, \ion{Fe}{2} and \ion{Mg}{2} at $z=1.5915\pm0.0005$.
%$z=1.604 \pm 0.001$ from narrow Mg II mission.
%FeLoBAL with Al II, Al III, Fe III, Fe II, Mg II absorption; red end of troughs at z=1.55.
%Narrow Al II, Al III, Fe II, Mg II at $z=1.5915 \pm 0.0005$.

\subsubsection*{J220912.01+061920.1}
This quasar is at $z=1.91\pm0.01$ identified by the narrow \ion{C}{4} and \ion{Al}{3} lines. It displays a FeLoBAL (\ion{Si}{4}, \ion{C}{4}, \ion{Al}{2}, \ion{Al}{3}, \ion{Fe}{3}, \ion{Mg}{2}), where the blue end of the troughs is around $z=1.76$. It further has narrower \ion{Si}{4}, \ion{C}{4}, \ion{Al}{2}, \ion{Al}{3}, \ion{Fe}{2} absorption around $z=1.893$. 
%$z=1.91 \pm 0.01$ from narrow C IV and Al III.
%FeLoBAL with Si IV, C IV, Al II, Al III, Fe III, Mg II BALs; blue end of troughs at z=1.76.
%Narrow Si IV, C IV, Al II, Al III, Fe III?, Fe II at z=1.893.

\subsection{Cross-matches to GALEX, ROSAT 2RXS, and XMMSL2}\label{sec_cross_matches}

To obtain near- and far-UV photometry, we cross-matched the PS-ELQS quasar catalog with  the GALEX GR6/7 Data Release \citep{Martin2005} within $2\farcs0$. We obtained the available photometry in the far- and near-UV bands at $1350-1750\text{\AA}$ and $1750-2750\text{\AA}$, respectively.

We could identify a total of 49 matches to the full PS-ELQS catalog, of which 17 were detections in both bands, 3 only in the far-UV band, and 29 only in the near-UV band.
Of all 190 newly discovered quasars 4 have detections in both GALEX bands and another 7 are detected in the near-UV. The detections could potentially signal that the quasars' flux has not been fully absorbed by neutral hydrogen along the line of sight and thus these objects are likely promising targets for studies of He-reionization \citep[e.g.][]{Worseck2016}.

In addition, we utilize the AllWISE counterparts to X-ray detections \citep{Salvato2018} from the ROSAT \citep{Truemper1982} reprocessed 2RXS catalog \citep{Boller2016} and the XMM Newton Slew 2 Survey (XMMSL2) to match the PS-ELQS quasar catalog with X-ray sources. We have matched the AllWISE positions of our sources to the AllWISE positions of the counterpart catalogs within a  $1\farcs0$ aperture. 

There are a total of 12 matches between PS-ELQS and ROSAT 2RXS and 3 between PS-ELQS and XMMSL2. 

Of the newly discovered quasars one,  J124615.10+713923.6 at $z=3.995$ has an X-ray counterpart in ROSAT 2RXS with $f_{0.1-2.4\,\rm{keV}} = 2.26\cdot 10^{-13}\,\rm{erg}\,\rm{cm}^{-2}\,\rm{s}^{-1}$ at a distance of $62\farcs7$. As the match distance implies this AllWISE source is not the most probable match to the X-ray source
(\texttt{TRXS\_match\_flag}$=2$) with a $p_i\approx0.24$ relative probability to be the correct counterpart to the X-ray source among all possible AllWISE candidates. While we wanted to report this here for completeness, we would also like to caution to blindly associate the X-ray flux with the quasar.

Another new PS-ELQS quasar,  J095947.52-103437.7 ($z=3.165$), has a counterpart in XMMSL2. We further want to report one a XMMSL2 counterpart for the quasar J171721.32+422428.3 ($z=3.495$). The latter quasar was discovered during pilot observations of the ELQS \citep{Schindler2018}, but not selected to be in the ELQS quasar catalog.
In both cases the AllWISE counterparts constitute the best match to the X-ray source and seem reliable.  
The counterpart to J095947.52-103437.7 has a $0.24\times10^{-12}\,\rm{erg}\,\rm{cm}^{-2}\,\rm{s}^{-1}$ detection in the soft band and a flux of $0.67\times10^{-12}\,\rm{erg}\,\rm{cm}^{-2}\,\rm{s}^{-1}$ in the total band.
The X-ray counterpart to  J171721.32+422428.3 was detected in the soft band with $0.94\times10^{-12}\,\rm{erg}\,\rm{cm}^{-2}\,\rm{s}^{-1}$ and in the total band with $2.42\times10^{-12}\,\rm{erg}\,\rm{cm}^{-2}\,\rm{s}^{-1}$. 

% 
% both match_flag 1
% p_i > 0.8
%20     0.842780
%289    0.999937
%
% p_any >0.68
%  20     0.681196
%289    0.949618
%
% XMM match stuff 
% flux B8
% 20     2.416651
%289    0.672934
%flux B6 
%20     0.935114
%289    0.237490

Information on all GALEX, ROSAT 2RXS and XMMSL2 matches is included in the full PS-ELQS quasar catalog (see Table\,\ref{tab_pselqs_cat_cols}).

\section{Discussion}\label{sec_discussion}

\subsection{PS-ELQS Completeness estimates}

The goal of the PS-ELQS was to extend the novel quasar selection method of the ELQS to a much larger area. However, the final PS-ELQS quasar catalog includes only 343 quasars with $m_i\le18.0$ at $z\ge2.8$ compared to the 407 quasars in the ELQS quasar catalog. A further comparison to the DR14Q showed that we only recover  roughly $56\%$ ($71\%$, $89\%$) of the bright ($m_i\le18.0$) quasars at at $z\ge2.8$ ($z\ge3.0$, $z\ge3.5$), indicating that the PS-ELQS missed quasars at $z\le3.5$ compared to the ELQS.

In order to understand this difference and characterize the biases in our selection we calculated the PS-ELQS completeness for the random forest classification and redshift estimation. The selection function is estimated based on the sample of simulated quasars, which were used to determine the k-correction described in Section\,\ref{sec_qso_catalog}.

\begin{figure*}
\centering 
\includegraphics[width=\textwidth]{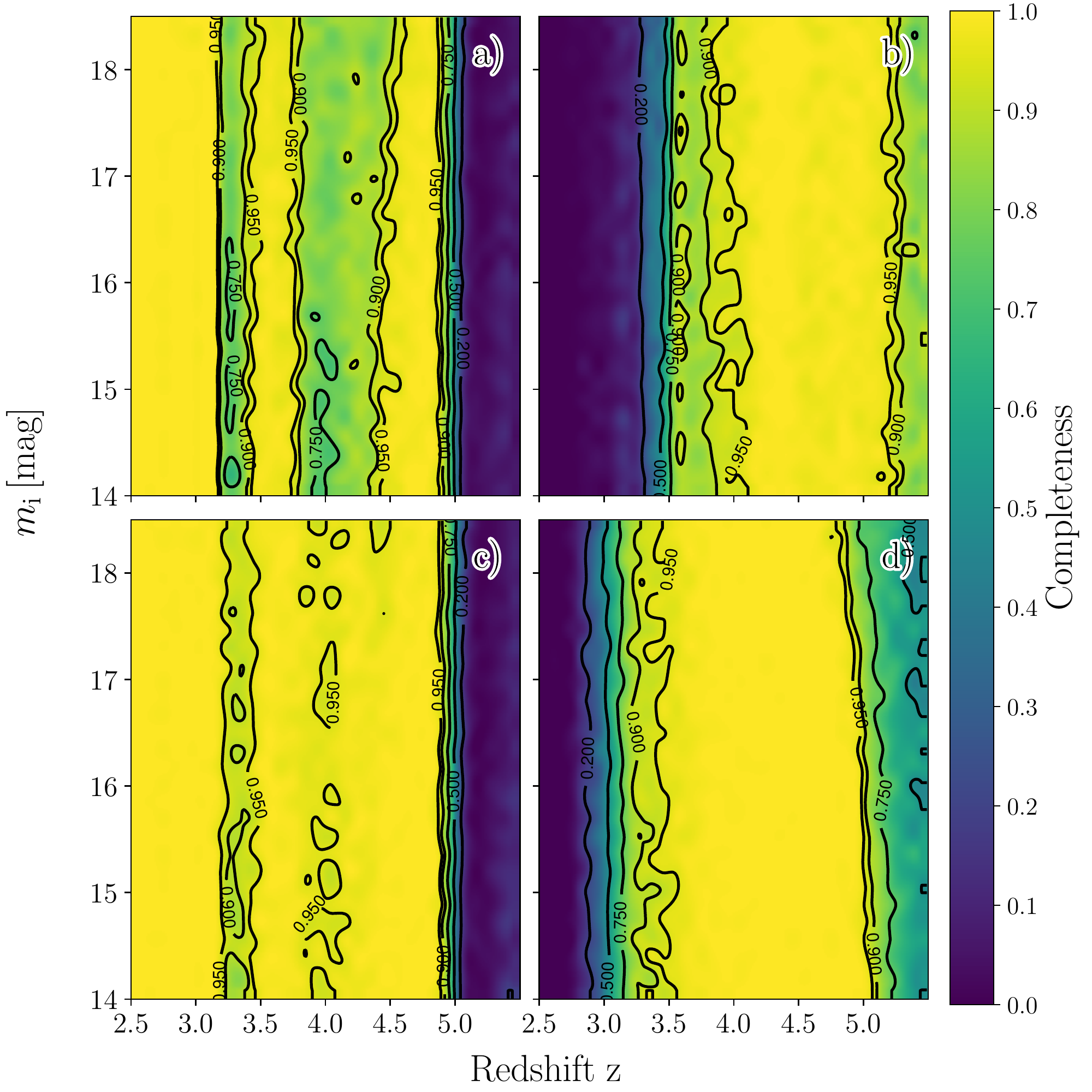}
\caption{ PS-ELQS completeness estimates as a function of redshift and \textit{i}-band magnitude. The completeness maps were determined by the fraction of simulated quasars, selected by our different criteria, to all simulated quasars: \textit{Panel a)} Completeness of the PS-ELQS classification selection  \textit{Panel (b)} Completeness of the PS-ELQS  photometric redshift selection \textit{Panel (c)} Completeness of the classification selection (including the SDSS u-band) \textit{Panel (d)} Completeness of the  photometric redshift selection (including the SDSS u-band). Contour levels are drawn with solid lines at $20\%$, $50\%$, $75\%$, $90\%$, and $95\%$ completeness.}
\label{fig_completeness}
\end{figure*}

The PS-ELQS completeness of the quasar selection based on the random forest classification is shown in panel (a) of Figure\,\ref{fig_completeness}. The results are generally independent of the i-band magnitude, but show clear features as a function of redshift. There are two redshift ranges, $z\approx3.2-3.5$ and $z\approx3.7-4.0$, where the completeness  drops to $75\%-85\%$. In the absence of u-band photometry these redshift ranges can be associated with stellar confusion. At $z\ge5$ our completeness drops steeply, analogous to \citetalias{Schindler2018}. The absence of quasars at these redshifts and magnitudes in the training set results in incorrect classifications.

Panel (b) of Figure\,\ref{fig_completeness} shows the selection function resulting from the redshift selection based on the random forest regression (photometric redshift estimation). While we should select all quasars with $z_{\rm{reg}}\ge2.8$, we miss the majority ($>50\%$) of sources between $z=2.8$ and $z=3.5$.  This effect clearly explains why the PS-ELQS missed many known quasars at $z\le3.5$, which were selected by the original ELQS. 

As the main difference of the PS-ELQS to the ELQS is the optical photometry, we suspect that our decreased completeness at lower redshifts is a result of the missing u-band photometry in PS1. In order to test this, we add the SDSS u-band photometry to the number of features for the random forest and rerun our completeness calculation. 
The selection functions for the re-run classification and redshift selections are displayed in panels (c) and (d) of Figure\,\ref{fig_completeness}. The results clearly confirm our suspicion. Comparing panels (a) and (c) highlights how adding the u-band visibly improves the classification selection at $z\approx3.2-3.5$ and $z\approx3.7-4.0$. Furthermore, panels (b) and (d) emphasize that the missing u-band in PS1 is the cause for the low completeness at $z\le3.5$ for PS-ELQS. 
A second look at the comparison between photometric redshifts and spectroscopic values in Figure\,\ref{fig_reg_zz} shows that the quasars at $z\ge2.5$ are predominantly found below the $z_{\rm{spec}} = z_{\rm{reg}}$ diagonal line. 
This translates into a bias on our photometric redshift estimates towards lower values, which in turn causes the low completeness in the redshift selection between $z=2.8$ and $z=3.5$.

% 
%QSO     10628
%STAR       37
%
%Completeness study higher than a certain redshift in DR14 and rf_emp_photoz
%z>=2.8 (513, 252)
%Missed: (222, 252)
%Retrieved: (285, 252)
%z>=3.0 (358, 252)
%Missed: (180, 252)
%Retrieved: (174, 252)
%z>=3.5 (90, 252)
%Missed: (27, 252)
%Retrieved: (60, 252)
%--------------------------------
%
%
%--------------------------------
%Completeness study higher than a certain redshift in DR14 and rf_emp_photoz>=2.8
%z>=2.8 (513, 252)
%Missed: (222, 252)
%Retrieved: (285, 252)
%z>=3.0 (358, 252)
%Missed: (98, 252)
%Retrieved: (254, 252)
%z>=3.5 (90, 252)
%Missed: (6, 252)
%Retrieved: (80, 252)
%--------------------------------
%
%QSO     151
%STAR      2
%
%Missed objects identified by ELQS
%z>=2.8 (112, 252)
%Missed: (67, 252)
%Retrieved: (44, 252)
%z>=3.0 (79, 252)
%Missed: (38, 252)
%Retrieved: (40, 252)
%z>=3.5 (25, 252)
%Missed: (7, 252)
%Retrieved: (18, 252)

\subsection{Applications for the PS-ELQS quasar sample}

In spite of the shortcomings of the PS-ELQS selection due to the missing u-band photometry of PS1, the PS-ELQS provides the most complete sample of extremely luminous quasars at $z=3.5-5.0$. Our survey covered around $21,486\,\rm{deg}^2$, making it the largest homogeneous spectroscopic quasar survey at these redshifts. We have successfully discovered $190$ new quasars at $z\ge2.8$, tripling the known quasars at $\rm{Decl.}\le0$ in the PS-ELQS quasar catalog.

Our area coverage and the high completeness at $z=3.5-5.0$ makes the PS-ELQS quasar sample uniquely equipped to constrain the volume density of extremely luminous quasars at intermediate redshifts. This is our first science goal, once the remaining 44 quasar candidates are observed to guarantee spectroscopic completeness. 

Four of our newly discovered quasars at $z=2.900, 2.905, 2.905$ and $3.320$ have far- and near-UV fluxes as measured by GALEX. We have visually confirmed these detections in GALEX photometry. These objects provide promising targets to investigate the He-reionization of the universe.
In general, the PS-ELQS quasars' strong flux facilitates efficient high resolution spectroscopy to characterize the Lyman-$\alpha$ forest and intervening absorption systems. 

From the PS-ELQS quasar catalog we have already identified five potential quasar pairs with angular separations of $\le30\,\rm{arcmin}$ and proper distances of $\le20\,h^{-1}\rm{Mpc}$ as well as an association of three quasars with larger angular distances but similar redshifts. We provide details for all of these objects in Table\,\ref{tab_qso_pairs}. The relative distances are always measured with respect to the first object in the pair or association.

These quasar associations demonstrate the value of the PS-ELQS sample for quasar clustering measurements. Previous measurements on quasar clustering \citep[e.g.][]{Myers2006, Myers2007, Shen2007, Angela2008, Ross2009, White2012, Eftekharzadeh2015, RodriguezTorres2017, Timlin2018} have shown that these populations show a high level of clustering and that quasars are inferred to reside in dark matter haloes of $10^{12}\,h^{-1}M_{\odot}$ at most redshifts. 
Due to its size the PS-ELQS quasar sample will only allow for a very sparse measurement, however, the number of quasars should allow to constrain the minimum dark matter halo mass for this extremely luminous population at $z\sim3.8$. In combination with the measurement of the ELQS quasar luminosity function \citep{Schindler2019a}, one will be able to place constraints on the duty cycle of these rare objects.

\begin{table*}[htb]
\centering 
\caption{Possible quasar associations identified from the PS-ELQS sample}
\label{tab_qso_pairs}
\begin{tabular}{ccccccc}
 \tableline
 AllWISE designation & R.A.(J2000) & Decl.(J2000) & $z$ & separation angle & proper 3D distance  \\
 &  [hh:mm:ss.sss] & [dd:mm:ss.ss] &   & [$\rm{arcmin}$]&  [$h^{-1}\rm{Mpc}$]  \\
\tableline
%J020825.26+170548.8 & 02:08:25.254 & +17:05:48.91 & 3.300 & &   \\ 
%J020838.40+170652.3 & 02:08:38.403 & +17:06:52.36 & 3.510 & 3.31 & 28.54 \\ 
%\tableline
J030341.04-002321.8 & 03:03:41.045 & -00:23:21.87 & 3.227 & &   \\ 
J030449.85-000813.4 & 03:04:49.859 & -00:08:13.54 & 3.295 & 22.92 & 12.30 \\ 
\tableline
J091647.60-113009.9 & 09:16:47.616 & -11:30:09.91 & 3.870 & &   \\ 
J091746.54-115331.9 & 09:17:46.542 & -11:53:31.89 & 3.920 & 27.46 & 9.63 \\
\tableline 
%J092456.66+305354.7 & 09:24:56.669 & +30:53:54.77 & 3.457 & &   \\ 
%J092636.33+305505.0 & 09:26:36.331 & +30:55:04.93 & 4.190 & 21.41 & 84.61 \\ 
%\tableline
J102000.80-121151.4 & 10:20:00.800 & -12:11:51.45 & 3.715 & &   \\ 
J102126.15-115621.4 & 10:21:26.131 & -11:56:22.39 & 3.670 & 25.98 & 9.30 \\
\tableline 
%J105254.50+254303.9 & 10:52:54.494 & +25:43:03.87 & 3.078 & &   \\ 
%J105353.49+253115.4 & 10:53:53.499 & +25:31:15.50 & 3.500 & 17.78 & 63.17 \\ 
%\tableline
%J115542.90+552329.8 & 11:55:42.921 & +55:23:29.81 & 3.301 & &   \\ 
%J115659.59+551308.2 & 11:56:59.602 & +55:13:08.23 & 3.123 & 15.05 & 26.17 \\ 
%\tableline
J172237.85+385951.9 & 17:22:37.854 & +38:59:51.87 & 3.359 & &   \\ 
J172338.84+392621.4 & 17:23:38.809 & +39:26:21.38 & 3.465 & 29.00 & 16.84 \\ 
\tableline
J235054.64+200938.6 & 23:50:54.634 & +20:09:38.62 & 3.170 & &   \\ 
J235201.30+200902.3 & 23:52:01.307 & +20:09:02.47 & 3.087 & 15.66 & 13.70 \\ 
\tableline
\tableline
J230432.31-124819.6 & 23:04:32.312 & -12:48:19.64 & 3.850 &   &   \\ 
J230827.04-133256.2 & 23:08:27.042 & -13:32:56.21 & 3.830 & 72.49 & 21.57 \\ 
J230959.29-122603.2 & 23:09:59.293 & -12:26:03.16 & 3.730 & 82.82 & 27.54 \\ 
\tableline
\end{tabular}
\end{table*}

\section{Conclusions}\label{sec_conclusion}

We present the results of the PS-ELQS, a spectroscopic quasar survey covering $\sim21486\,\rm{deg}^2$ of the $3\pi$ PS1 footprint. 
We apply the successful quasar selection strategy of the ELQS survey in the SDSS footprint to this larger area using the PS1 photometry. The candidates are selected based on a JKW2 color cut using 2MASS $J$ and $K_s$, and AllWISE W2 photometry. Random forest photometric redshift estimation and classification on all Pan-STARRS PS1 and AllWISE W1 and W2 photometric bands lead to a highly efficient ($\ge77\%$) quasar selection.

We select a total of 806 bright ($m_{i}\le18.5$) quasar candidates with regression redshifts of $z_{\rm{reg}}\ge2.8$. After the exclusion of 428 sources known in the literature, of which 70 were identified with the ELQS survey, we have rejected 40 candidates during visual inspection of their photometry. 

We have selected a total of 334 good candidates for spectroscopic follow-up observations and were able to observe 290 of the objects with the SOAR telescope, the MMT and the VATT. We have discovered a total of 218 new quasars, of which 190 are at $z\ge2.8$. 
Based on our quasar selection we have constructed the PS-ELQS quasar catalog with a total of 592 quasars, including the 190 newly discovered quasars at $z\ge2.8$.

Estimates of the PS-ELQS completeness for the classification and photometric redshift selections, show that the missing u-band photometry of PS1 lowers our completeness to select quasars at $z=2.8-3.5$. This effect explains the low numbers of $z<3.5$ quasars compared to the original ELQS selection. 
However, not accounting for the photometric completeness of PS1, the PS-ELQS general completeness at $z=3.5-5.0$ is consistently above $70\%$. As a result, the PS-ELQS provides the most complete sample of extremely luminous quasars at $z=3.5-5.0$.

A range of scientific applications will be able to leverage the information of this sample. For example, it is uniquely equipped to constrain the volume density of intermediate redshift extremely luminous quasars, to provide insight into quasar clustering of this rare population, and to facilitate studies of the intergalactic medium.

\section{Acknowledgements} 

We would like to the thank the anonymous referee for a thorough report with many helpful suggestions, which helped to improve this paper.

J.-T.S., X.F., I.D.M., and J.Y. acknowledge support from NSF grant AST-1515115 and NASA ADAP grant NNX17AF28G.

K.C.L. acknowledges support from the U.S. National Science Foundation (NSF) under awards AST1715213 and AST-1312950 and through award SOSPA4-007 from the National Radio Astronomy Observatory (NRAO).

Based on observations obtained at the Southern Astrophysical Research (SOAR) telescope, which is a joint project of the Minist\'{e}rio da Ci\^{e}ncia, Tecnologia, Inova\c{c}\~{o}es e Comunica\c{c}\~{o}es (MCTIC) do Brasil, the U.S. National Optical Astronomy Observatory (NOAO), the University of North Carolina at Chapel Hill (UNC), and Michigan State University (MSU).

Based on observations with the VATT: the Alice P. Lennon Telescope and the Thomas J. Bannan Astrophysics Facility.

Observations reported here were obtained at the MMT Observatory, a joint facility of the University of Arizona and the Smithsonian Institution.

This publication makes use of data products from the Two Micron All Sky Survey, which is a joint project of the University of Massachusetts and the Infrared Processing and Analysis Center/California Institute of Technology, funded by the National Aeronautics and Space Administration and the National Science Foundation.

This publication makes use of data products from the Wide-field Infrared Survey Explorer, which is a joint project of the University of California, Los Angeles, and the Jet Propulsion Laboratory/California Institute of Technology, funded by the National Aeronautics and Space Administration.

The Pan-STARRS1 Surveys (PS1) and the PS1 public science archive have been made possible through contributions by the Institute for Astronomy, the University of Hawaii, the Pan-STARRS Project Office, the Max-Planck Society and its participating institutes, the Max Planck Institute for Astronomy, Heidelberg and the Max Planck Institute for Extraterrestrial Physics, Garching, The Johns Hopkins University, Durham University, the University of Edinburgh, the Queen's University Belfast, the Harvard-Smithsonian Center for Astrophysics, the Las Cumbres Observatory Global Telescope Network Incorporated, the National Central University of Taiwan, the Space Telescope Science Institute, the National Aeronautics and Space Administration under Grant No. NNX08AR22G issued through the Planetary Science Division of the NASA Science Mission Directorate, the National Science Foundation Grant No. AST-1238877, the University of Maryland, Eotvos Lorand University (ELTE), the Los Alamos National Laboratory, and the Gordon and Betty Moore Foundation.
 
Funding for the Sloan Digital Sky Survey IV has been provided by the Alfred P. Sloan Foundation, the U.S. Department of Energy Office of Science, and the Participating Institutions. SDSS acknowledges support and resources from the Center for High-Performance Computing at the University of Utah. The SDSS web site is \url{www.sdss.org}.

SDSS is managed by the Astrophysical Research Consortium for the Participating Institutions of the SDSS Collaboration including the Brazilian Participation Group, the Carnegie Institution for Science, Carnegie Mellon University, the Chilean Participation Group, the French Participation Group, Harvard-Smithsonian Center for Astrophysics, Instituto de Astrofísica de Canarias, The Johns Hopkins University, Kavli Institute for the Physics and Mathematics of the Universe (IPMU) / University of Tokyo, Lawrence Berkeley National Laboratory, Leibniz Institut f\"ur Astrophysik Potsdam (AIP), Max-Planck-Institut f\"ur Astronomie (MPIA Heidelberg), Max-Planck-Institut f\"ur Astrophysik (MPA Garching), Max-Planck-Institut f\"ur Extraterrestrische Physik (MPE), National Astronomical Observatories of China, New Mexico State University, New York University, University of Notre Dame, Observatório Nacional / MCTI, The Ohio State University, Pennsylvania State University, Shanghai Astronomical Observatory, United Kingdom Participation Group, Universidad Nacional Autónoma de México, University of Arizona, University of Colorado Boulder, University of Oxford, University of Portsmouth, University of Utah, University of Virginia, University of Washington, University of Wisconsin, Vanderbilt University, and Yale University.

This research has made use of the NASA/IPAC Extragalactic Database (NED) which is operated by the Jet Propulsion Laboratory, California Institute of Technology, under contract with the National Aeronautics and Space Administration.

This research made use of Astropy, a community-developed core Python package for Astronomy (Astropy Collaboration, 2013, \url{http://www.astropy.org}). In addition, python routines from scikit-learn\citep{scikit-learn}, SciPy \citep{scipy}, Matplotlib \citep{matplotlib}, and Pandas \citep{pandas} were used in the quasar selection, data analysis and creation of the figures.

\bibliography{all.bib}

\begin{thebibliography}{101}
\expandafter\ifx\csname natexlab\endcsname\relax\def\natexlab#1{#1}\fi

\bibitem[{{Abazajian} {et~al.}(2009){Abazajian}, {Adelman-McCarthy},
  {Ag{\"u}eros}, {Allam}, {Allende Prieto}, {An}, {Anderson}, {Anderson},
  {Annis}, {Bahcall}, \& et~al.}]{Abazajian2009}
{Abazajian}, K.~N., {Adelman-McCarthy}, J.~K., {Ag{\"u}eros}, M.~A., {et~al.}
  2009, \apjs, 182, 543

\bibitem[{{Aihara} {et~al.}(2018){Aihara}, {Arimoto}, {Armstrong}, {Arnouts},
  {Bahcall}, {Bickerton}, {Bosch}, {Bundy}, {Capak}, {Chan}, {Chiba}, {Coupon},
  {Egami}, {Enoki}, {Finet}, {Fujimori}, {Fujimoto}, {Furusawa}, {Furusawa},
  {Goto}, {Goulding}, {Greco}, {Greene}, {Gunn}, {Hamana}, {Harikane},
  {Hashimoto}, {Hattori}, {Hayashi}, {Hayashi}, {He{\l}miniak}, {Higuchi},
  {Hikage}, {Ho}, {Hsieh}, {Huang}, {Huang}, {Ikeda}, {Imanishi}, {Inoue},
  {Iwasawa}, {Iwata}, {Jaelani}, {Jian}, {Kamata}, {Karoji}, {Kashikawa},
  {Katayama}, {Kawanomoto}, {Kayo}, {Koda}, {Koike}, {Kojima}, {Komiyama},
  {Konno}, {Koshida}, {Koyama}, {Kusakabe}, {Leauthaud}, {Lee}, {Lin}, {Lin},
  {Lupton}, {Mandelbaum}, {Matsuoka}, {Medezinski}, {Mineo}, {Miyama},
  {Miyatake}, {Miyazaki}, {Momose}, {More}, {More}, {Moritani}, {Moriya},
  {Morokuma}, {Mukae}, {Murata}, {Murayama}, {Nagao}, {Nakata}, {Niida},
  {Niikura}, {Nishizawa}, {Obuchi}, {Oguri}, {Oishi}, {Okabe}, {Okamoto},
  {Okura}, {Ono}, {Onodera}, {Onoue}, {Osato}, {Ouchi}, {Price}, {Pyo}, {Sako},
  {Sawicki}, {Shibuya}, {Shimasaku}, {Shimono}, {Shirasaki}, {Silverman},
  {Simet}, {Speagle}, {Spergel}, {Strauss}, {Sugahara}, {Sugiyama}, {Suto},
  {Suyu}, {Suzuki}, {Tait}, {Takada}, {Takata}, {Tamura}, {Tanaka}, {Tanaka},
  {Tanaka}, {Tanaka}, {Terai}, {Terashima}, {Toba}, {Tominaga}, {Toshikawa},
  {Turner}, {Uchida}, {Uchiyama}, {Umetsu}, {Uraguchi}, {Urata}, {Usuda},
  {Utsumi}, {Wang}, {Wang}, {Wong}, {Yabe}, {Yamada}, {Yamanoi}, {Yasuda},
  {Yeh}, {Yonehara}, \& {Yuma}}]{Aihara2018}
{Aihara}, H., {Arimoto}, N., {Armstrong}, R., {et~al.} 2018, \pasj, 70, S4

\bibitem[{{Akiyama} {et~al.}(2018){Akiyama}, {He}, {Ikeda}, {Niida}, {Nagao},
  {Bosch}, {Coupon}, {Enoki}, {Imanishi}, {Kashikawa}, {Kawaguchi}, {Komiyama},
  {Lee}, {Matsuoka}, {Miyazaki}, {Nishizawa}, {Oguri}, {Ono}, {Onoue}, {Ouchi},
  {Schulze}, {Silverman}, {Tanaka}, {Tanaka}, {Terashima}, {Toba}, \&
  {Ueda}}]{Akiyama2018}
{Akiyama}, M., {He}, W., {Ikeda}, H., {et~al.} 2018, \pasj, 70, S34

\bibitem[{{Ba{\~n}ados} {et~al.}(2014){Ba{\~n}ados}, {Venemans}, {Morganson},
  {Decarli}, {Walter}, {Chambers}, {Rix}, {Farina}, {Fan}, {Jiang}, {McGreer},
  {De Rosa}, {Simcoe}, {Wei{\ss}}, {Price}, {Morgan}, {Burgett}, {Greiner},
  {Kaiser}, {Kudritzki}, {Magnier}, {Metcalfe}, {Stubbs}, {Sweeney}, {Tonry},
  {Wainscoat}, \& {Waters}}]{Banados2014}
{Ba{\~n}ados}, E., {Venemans}, B.~P., {Morganson}, E., {et~al.} 2014, \aj, 148,
  14

\bibitem[{{Ba{\~n}ados} {et~al.}(2016){Ba{\~n}ados}, {Venemans}, {Decarli},
  {Farina}, {Mazzucchelli}, {Walter}, {Fan}, {Stern}, {Schlafly}, {Chambers},
  {Rix}, {Jiang}, {McGreer}, {Simcoe}, {Wang}, {Yang}, {Morganson}, {De Rosa},
  {Greiner}, {Balokovi{\'c}}, {Burgett}, {Cooper}, {Draper}, {Flewelling},
  {Hodapp}, {Jun}, {Kaiser}, {Kudritzki}, {Magnier}, {Metcalfe}, {Miller},
  {Schindler}, {Tonry}, {Wainscoat}, {Waters}, \& {Yang}}]{Banados2016}
{Ba{\~n}ados}, E., {Venemans}, B.~P., {Decarli}, R., {et~al.} 2016, \apjs, 227,
  11

\bibitem[{{Ba{\~n}ados} {et~al.}(2018){Ba{\~n}ados}, {Venemans},
  {Mazzucchelli}, {Farina}, {Walter}, {Wang}, {Decarli}, {Stern}, {Fan},
  {Davies}, {Hennawi}, {Simcoe}, {Turner}, {Rix}, {Yang}, {Kelson}, {Rudie}, \&
  {Winters}}]{Banados2018}
{Ba{\~n}ados}, E., {Venemans}, B.~P., {Mazzucchelli}, C., {et~al.} 2018, \nat,
  553, 473

\bibitem[{Bishop(2006)}]{Bishop2006}
Bishop, C.~M. 2006, Pattern recognition and machine learning (Springer)

\bibitem[{{Boller} {et~al.}(2016){Boller}, {Freyberg}, {Tr{\"u}mper}, {Haberl},
  {Voges}, \& {Nandra}}]{Boller2016}
{Boller}, T., {Freyberg}, M.~J., {Tr{\"u}mper}, J., {et~al.} 2016, \aap, 588,
  A103

\bibitem[{{Bovy} {et~al.}(2011){Bovy}, {Hennawi}, {Hogg}, {Myers},
  {Kirkpatrick}, {Schlegel}, {Ross}, {Sheldon}, {McGreer}, {Schneider}, \&
  {Weaver}}]{Bovy2011}
{Bovy}, J., {Hennawi}, J.~F., {Hogg}, D.~W., {et~al.} 2011, \apj, 729, 141

\bibitem[{Breiman(2001)}]{Breiman2001}
Breiman, L. 2001, Machine learning, 45, 5

\bibitem[{{Carliles} {et~al.}(2010){Carliles}, {Budav{\'a}ri}, {Heinis},
  {Priebe}, \& {Szalay}}]{Carliles2010}
{Carliles}, S., {Budav{\'a}ri}, T., {Heinis}, S., {Priebe}, C., \& {Szalay},
  A.~S. 2010, \apj, 712, 511

\bibitem[{{Carnall} {et~al.}(2015){Carnall}, {Shanks}, {Chehade}, {Fumagalli},
  {Rauch}, {Irwin}, {Gonzalez-Solares}, {Findlay}, \& {Metcalfe}}]{Carnall2015}
{Carnall}, A.~C., {Shanks}, T., {Chehade}, B., {et~al.} 2015, \mnras, 451, L16

\bibitem[{{Carrasco} {et~al.}(2015){Carrasco}, {Barrientos}, {Pichara},
  {Anguita}, {Murphy}, {Gilbank}, {Gladders}, {Yee}, {Hsieh}, \&
  {L{\'o}pez}}]{Carrasco2015}
{Carrasco}, D., {Barrientos}, L.~F., {Pichara}, K., {et~al.} 2015, \aap, 584,
  A44

\bibitem[{{Carrasco Kind} \& {Brunner}(2013)}]{CarrascoKind2013}
{Carrasco Kind}, M., \& {Brunner}, R.~J. 2013, \mnras, 432, 1483

\bibitem[{{Chambers} {et~al.}(2016){Chambers}, {Magnier}, {Metcalfe},
  {Flewelling}, {Huber}, {Waters}, {Denneau}, {Draper}, {Farrow}, {Finkbeiner},
  {Holmberg}, {Koppenhoefer}, {Price}, {Saglia}, {Schlafly}, {Smartt},
  {Sweeney}, {Wainscoat}, {Burgett}, {Grav}, {Heasley}, {Hodapp}, {Jedicke},
  {Kaiser}, {Kudritzki}, {Luppino}, {Lupton}, {Monet}, {Morgan}, {Onaka},
  {Stubbs}, {Tonry}, {Banados}, {Bell}, {Bender}, {Bernard}, {Botticella},
  {Casertano}, {Chastel}, {Chen}, {Chen}, {Cole}, {Deacon}, {Frenk},
  {Fitzsimmons}, {Gezari}, {Goessl}, {Goggia}, {Goldman}, {Grebel}, {Hambly},
  {Hasinger}, {Heavens}, {Heckman}, {Henderson}, {Henning}, {Holman}, {Hopp},
  {Ip}, {Isani}, {Keyes}, {Koekemoer}, {Kotak}, {Long}, {Lucey}, {Liu},
  {Martin}, {McLean}, {Morganson}, {Murphy}, {Nieto-Santisteban}, {Norberg},
  {Peacock}, {Pier}, {Postman}, {Primak}, {Rae}, {Rest}, {Riess}, {Riffeser},
  {Rix}, {Roser}, {Schilbach}, {Schultz}, {Scolnic}, {Szalay}, {Seitz},
  {Shiao}, {Small}, {Smith}, {Soderblom}, {Taylor}, {Thakar}, {Thiel},
  {Thilker}, {Urata}, {Valenti}, {Walter}, {Watters}, {Werner}, {White},
  {Wood-Vasey}, \& {Wyse}}]{Chambers2016}
{Chambers}, K.~C., {Magnier}, E.~A., {Metcalfe}, N., {et~al.} 2016, arXiv
  e-prints

\bibitem[{{Chehade} {et~al.}(2018){Chehade}, {Carnall}, {Shanks}, {Diener},
  {Fumagalli}, {Findlay}, {Metcalfe}, {Hennawi}, {Leibler}, {Murphy},
  {Prochaska}, {Irwin}, \& {Gonzalez-Solares}}]{Chehade2018}
{Chehade}, B., {Carnall}, A.~C., {Shanks}, T., {et~al.} 2018, \mnras, 478, 1649

\bibitem[{{Clemens} {et~al.}(2004){Clemens}, {Crain}, \&
  {Anderson}}]{Clemens2004}
{Clemens}, J.~C., {Crain}, J.~A., \& {Anderson}, R. 2004, in \procspie, Vol.
  5492, Ground-based Instrumentation for Astronomy, ed. A.~F.~M. {Moorwood} \&
  M.~{Iye}, 331--340

\bibitem[{{da {\^A}ngela} {et~al.}(2008){da {\^A}ngela}, {Shanks}, {Croom},
  {Weilbacher}, {Brunner}, {Couch}, {Miller}, {Myers}, {Nichol}, {Pimbblet},
  {de Propris}, {Richards}, {Ross}, {Schneider}, \& {Wake}}]{Angela2008}
{da {\^A}ngela}, J., {Shanks}, T., {Croom}, S.~M., {et~al.} 2008, \mnras, 383,
  565

\bibitem[{{Dawson} {et~al.}(2013){Dawson}, {Schlegel}, {Ahn}, {Anderson},
  {Aubourg}, {Bailey}, {Barkhouser}, {Bautista}, {Beifiori}, {Berlind},
  {Bhardwaj}, {Bizyaev}, {Blake}, {Blanton}, {Blomqvist}, {Bolton}, {Borde},
  {Bovy}, {Brandt}, {Brewington}, {Brinkmann}, {Brown}, {Brownstein}, {Bundy},
  {Busca}, {Carithers}, {Carnero}, {Carr}, {Chen}, {Comparat}, {Connolly},
  {Cope}, {Croft}, {Cuesta}, {da Costa}, {Davenport}, {Delubac}, {de Putter},
  {Dhital}, {Ealet}, {Ebelke}, {Eisenstein}, {Escoffier}, {Fan}, {Filiz Ak},
  {Finley}, {Font-Ribera}, {G{\'e}nova-Santos}, {Gunn}, {Guo}, {Haggard},
  {Hall}, {Hamilton}, {Harris}, {Harris}, {Ho}, {Hogg}, {Holder}, {Honscheid},
  {Huehnerhoff}, {Jordan}, {Jordan}, {Kauffmann}, {Kazin}, {Kirkby}, {Klaene},
  {Kneib}, {Le Goff}, {Lee}, {Long}, {Loomis}, {Lundgren}, {Lupton}, {Maia},
  {Makler}, {Malanushenko}, {Malanushenko}, {Mandelbaum}, {Manera}, {Maraston},
  {Margala}, {Masters}, {McBride}, {McDonald}, {McGreer}, {McMahon}, {Mena},
  {Miralda-Escud{\'e}}, {Montero-Dorta}, {Montesano}, {Muna}, {Myers},
  {Naugle}, {Nichol}, {Noterdaeme}, {Nuza}, {Olmstead}, {Oravetz}, {Oravetz},
  {Owen}, {Padmanabhan}, {Palanque-Delabrouille}, {Pan}, {Parejko},
  {P{\^a}ris}, {Percival}, {P{\'e}rez-Fournon}, {P{\'e}rez-R{\`a}fols},
  {Petitjean}, {Pfaffenberger}, {Pforr}, {Pieri}, {Prada}, {Price-Whelan},
  {Raddick}, {Rebolo}, {Rich}, {Richards}, {Rockosi}, {Roe}, {Ross}, {Ross},
  {Rossi}, {Rubi{\~n}o-Martin}, {Samushia}, {S{\'a}nchez}, {Sayres}, {Schmidt},
  {Schneider}, {Sc{\'o}ccola}, {Seo}, {Shelden}, {Sheldon}, {Shen}, {Shu},
  {Slosar}, {Smee}, {Snedden}, {Stauffer}, {Steele}, {Strauss}, {Streblyanska},
  {Suzuki}, {Swanson}, {Tal}, {Tanaka}, {Thomas}, {Tinker}, {Tojeiro},
  {Tremonti}, {Vargas Maga{\~n}a}, {Verde}, {Viel}, {Wake}, {Watson}, {Weaver},
  {Weinberg}, {Weiner}, {West}, {White}, {Wood-Vasey}, {Yeche}, {Zehavi},
  {Zhao}, \& {Zheng}}]{Dawson2013}
{Dawson}, K.~S., {Schlegel}, D.~J., {Ahn}, C.~P., {et~al.} 2013, \aj, 145, 10

\bibitem[{{Dawson} {et~al.}(2016){Dawson}, {Kneib}, {Percival}, {Alam},
  {Albareti}, {Anderson}, {Armengaud}, {Aubourg}, {Bailey}, {Bautista},
  {Berlind}, {Bershady}, {Beutler}, {Bizyaev}, {Blanton}, {Blomqvist},
  {Bolton}, {Bovy}, {Brandt}, {Brinkmann}, {Brownstein}, {Burtin}, {Busca},
  {Cai}, {Chuang}, {Clerc}, {Comparat}, {Cope}, {Croft}, {Cruz-Gonzalez}, {da
  Costa}, {Cousinou}, {Darling}, {de la Macorra}, {de la Torre}, {Delubac}, {du
  Mas des Bourboux}, {Dwelly}, {Ealet}, {Eisenstein}, {Eracleous}, {Escoffier},
  {Fan}, {Finoguenov}, {Font-Ribera}, {Frinchaboy}, {Gaulme}, {Georgakakis},
  {Green}, {Guo}, {Guy}, {Ho}, {Holder}, {Huehnerhoff}, {Hutchinson}, {Jing},
  {Jullo}, {Kamble}, {Kinemuchi}, {Kirkby}, {Kitaura}, {Klaene}, {Laher},
  {Lang}, {Laurent}, {Le Goff}, {Li}, {Liang}, {Lima}, {Lin}, {Lin}, {Lin},
  {Long}, {Lundgren}, {MacDonald}, {Geimba Maia}, {Malanushenko},
  {Malanushenko}, {Mariappan}, {McBride}, {McGreer}, {M{\'e}nard}, {Merloni},
  {Meza}, {Montero-Dorta}, {Muna}, {Myers}, {Nandra}, {Naugle}, {Newman},
  {Noterdaeme}, {Nugent}, {Ogando}, {Olmstead}, {Oravetz}, {Oravetz},
  {Padmanabhan}, {Palanque-Delabrouille}, {Pan}, {Parejko}, {P{\^a}ris},
  {Peacock}, {Petitjean}, {Pieri}, {Pisani}, {Prada}, {Prakash}, {Raichoor},
  {Reid}, {Rich}, {Ridl}, {Rodriguez-Torres}, {Carnero Rosell}, {Ross},
  {Rossi}, {Ruan}, {Salvato}, {Sayres}, {Schneider}, {Schlegel}, {Seljak},
  {Seo}, {Sesar}, {Shandera}, {Shu}, {Slosar}, {Sobreira}, {Streblyanska},
  {Suzuki}, {Taylor}, {Tao}, {Tinker}, {Tojeiro}, {Vargas-Maga{\~n}a}, {Wang},
  {Weaver}, {Weinberg}, {White}, {Wood-Vasey}, {Yeche}, {Zhai}, {Zhao}, {Zhao},
  {Zheng}, {Ben Zhu}, \& {Zou}}]{Dawson2016}
{Dawson}, K.~S., {Kneib}, J.-P., {Percival}, W.~J., {et~al.} 2016, \aj, 151, 44

\bibitem[{{D'Isanto} \& {Polsterer}(2018)}]{DIsanto2018}
{D'Isanto}, A., \& {Polsterer}, K.~L. 2018, \aap, 609, A111

\bibitem[{{Dubath} {et~al.}(2011){Dubath}, {Rimoldini}, {S{\"u}veges},
  {Blomme}, {L{\'o}pez}, {Sarro}, {De Ridder}, {Cuypers}, {Guy}, {Lecoeur},
  {Nienartowicz}, {Jan}, {Beck}, {Mowlavi}, {De Cat}, {Lebzelter}, \&
  {Eyer}}]{Dubath2011}
{Dubath}, P., {Rimoldini}, L., {S{\"u}veges}, M., {et~al.} 2011, \mnras, 414,
  2602

\bibitem[{{Eftekharzadeh} {et~al.}(2015){Eftekharzadeh}, {Myers}, {White},
  {Weinberg}, {Schneider}, {Shen}, {Font-Ribera}, {Ross}, {Paris}, \&
  {Streblyanska}}]{Eftekharzadeh2015}
{Eftekharzadeh}, S., {Myers}, A.~D., {White}, M., {et~al.} 2015, \mnras, 453,
  2779

\bibitem[{{Eisenstein} {et~al.}(2011){Eisenstein}, {Weinberg}, {Agol},
  {Aihara}, {Allende Prieto}, {Anderson}, {Arns}, {Aubourg}, {Bailey},
  {Balbinot}, \& et~al.}]{Eisenstein2011}
{Eisenstein}, D.~J., {Weinberg}, D.~H., {Agol}, E., {et~al.} 2011, \aj, 142, 72

\bibitem[{{Fan} {et~al.}(2000){Fan}, {White}, {Davis}, {Becker}, {Strauss},
  {Haiman}, {Schneider}, {Gregg}, {Gunn}, {Knapp}, {Lupton}, {Anderson},
  {Anderson}, {Annis}, {Bahcall}, {Boroski}, {Brunner}, {Chen}, {Connolly},
  {Csabai}, {Doi}, {Fukugita}, {Hennessy}, {Hindsley}, {Ichikawa},
  {Ivezi{\'c}}, {Loveday}, {Meiksin}, {McKay}, {Munn}, {Newberg}, {Nichol},
  {Okamura}, {Pier}, {Sekiguchi}, {Shimasaku}, {Stoughton}, {Szalay},
  {Szokoly}, {Thakar}, {Vogeley}, \& {York}}]{Fan2000}
{Fan}, X., {White}, R.~L., {Davis}, M., {et~al.} 2000, \aj, 120, 1167

\bibitem[{{Fan} {et~al.}(2001){Fan}, {Strauss}, {Schneider}, {Gunn}, {Lupton},
  {Becker}, {Davis}, {Newman}, {Richards}, {White}, {Anderson}, {Annis},
  {Bahcall}, {Brunner}, {Csabai}, {Hennessy}, {Hindsley}, {Fukugita}, {Kunszt},
  {Ivezi{\'c}}, {Knapp}, {McKay}, {Munn}, {Pier}, {Szalay}, \&
  {York}}]{Fan2001b}
{Fan}, X., {Strauss}, M.~A., {Schneider}, D.~P., {et~al.} 2001, \aj, 121, 54

\bibitem[{{Fan} {et~al.}(2003){Fan}, {Strauss}, {Schneider}, {Becker}, {White},
  {Haiman}, {Gregg}, {Pentericci}, {Grebel}, {Narayanan}, {Loh}, {Richards},
  {Gunn}, {Lupton}, {Knapp}, {Ivezi{\'c}}, {Brandt}, {Collinge}, {Hao},
  {Harbeck}, {Prada}, {Schaye}, {Strateva}, {Zakamska}, {Anderson},
  {Brinkmann}, {Bahcall}, {Lamb}, {Okamura}, {Szalay}, \& {York}}]{Fan2003}
---. 2003, \aj, 125, 1649

\bibitem[{{Fan} {et~al.}(2004){Fan}, {Hennawi}, {Richards}, {Strauss},
  {Schneider}, {Donley}, {Young}, {Annis}, {Lin}, {Lampeitl}, {Lupton}, {Gunn},
  {Knapp}, {Brandt}, {Anderson}, {Bahcall}, {Brinkmann}, {Brunner}, {Fukugita},
  {Szalay}, {Szokoly}, \& {York}}]{Fan2004}
{Fan}, X., {Hennawi}, J.~F., {Richards}, G.~T., {et~al.} 2004, \aj, 128, 515

\bibitem[{{Fan} {et~al.}(2006){Fan}, {Strauss}, {Richards}, {Hennawi},
  {Becker}, {White}, {Diamond-Stanic}, {Donley}, {Jiang}, {Kim}, {Vestergaard},
  {Young}, {Gunn}, {Lupton}, {Knapp}, {Schneider}, {Brandt}, {Bahcall},
  {Barentine}, {Brinkmann}, {Brewington}, {Fukugita}, {Harvanek}, {Kleinman},
  {Krzesinski}, {Long}, {Neilsen}, {Nitta}, {Snedden}, \& {Voges}}]{Fan2006}
{Fan}, X., {Strauss}, M.~A., {Richards}, G.~T., {et~al.} 2006, \aj, 131, 1203

\bibitem[{{Flesch}(2015)}]{Flesch2015}
{Flesch}, E.~W. 2015, \pasa, 32, e010

\bibitem[{{G{\'o}rski} {et~al.}(2005){G{\'o}rski}, {Hivon}, {Banday},
  {Wandelt}, {Hansen}, {Reinecke}, \& {Bartelmann}}]{Gorski2005}
{G{\'o}rski}, K.~M., {Hivon}, E., {Banday}, A.~J., {et~al.} 2005, \apj, 622,
  759

\bibitem[{{Green} {et~al.}(2018){Green}, {Schlafly}, {Finkbeiner}, {Rix},
  {Martin}, {Burgett}, {Draper}, {Flewelling}, {Hodapp}, {Kaiser}, {Kudritzki},
  {Magnier}, {Metcalfe}, {Tonry}, {Wainscoat}, \& {Waters}}]{Green2018}
{Green}, G.~M., {Schlafly}, E.~F., {Finkbeiner}, D., {et~al.} 2018, \mnras,
  478, 651

\bibitem[{Hunter(2007)}]{matplotlib}
Hunter, J.~D. 2007, Computing In Science \& Engineering, 9, 90

\bibitem[{{Jiang} {et~al.}(2008){Jiang}, {Fan}, {Annis}, {Becker}, {White},
  {Chiu}, {Lin}, {Lupton}, {Richards}, {Strauss}, {Jester}, \&
  {Schneider}}]{Jiang2008}
{Jiang}, L., {Fan}, X., {Annis}, J., {et~al.} 2008, \aj, 135, 1057

\bibitem[{{Jiang} {et~al.}(2009){Jiang}, {Fan}, {Bian}, {Annis}, {Chiu},
  {Jester}, {Lin}, {Lupton}, {Richards}, {Strauss}, {Malanushenko},
  {Malanushenko}, \& {Schneider}}]{Jiang2009}
{Jiang}, L., {Fan}, X., {Bian}, F., {et~al.} 2009, \aj, 138, 305

\bibitem[{{Jiang} {et~al.}(2016){Jiang}, {McGreer}, {Fan}, {Strauss},
  {Ba{\~n}ados}, {Becker}, {Bian}, {Farnsworth}, {Shen}, {Wang}, {Wang},
  {Wang}, {White}, {Wu}, {Wu}, {Yang}, \& {Yang}}]{Jiang2016}
{Jiang}, L., {McGreer}, I.~D., {Fan}, X., {et~al.} 2016, \apj, 833, 222

\bibitem[{Jones {et~al.}(2001--)Jones, Oliphant, Peterson, {et~al.}}]{scipy}
Jones, E., Oliphant, T., Peterson, P., {et~al.} 2001--, {SciPy}: Open source
  scientific tools for {Python}

\bibitem[{{Kaiser} {et~al.}(2002){Kaiser}, {Aussel}, {Burke}, {Boesgaard},
  {Chambers}, {Chun}, {Heasley}, {Hodapp}, {Hunt}, {Jedicke}, {Jewitt},
  {Kudritzki}, {Luppino}, {Maberry}, {Magnier}, {Monet}, {Onaka}, {Pickles},
  {Rhoads}, {Simon}, {Szalay}, {Szapudi}, {Tholen}, {Tonry}, {Waterson}, \&
  {Wick}}]{Kaiser2002}
{Kaiser}, N., {Aussel}, H., {Burke}, B.~E., {et~al.} 2002, in \procspie, Vol.
  4836, Survey and Other Telescope Technologies and Discoveries, ed. J.~A.
  {Tyson} \& S.~{Wolff}, 154--164

\bibitem[{{Kaiser} {et~al.}(2010){Kaiser}, {Burgett}, {Chambers}, {Denneau},
  {Heasley}, {Jedicke}, {Magnier}, {Morgan}, {Onaka}, \& {Tonry}}]{Kaiser2010}
{Kaiser}, N., {Burgett}, W., {Chambers}, K., {et~al.} 2010, in \procspie, Vol.
  7733, Ground-based and Airborne Telescopes III, 77330E

\bibitem[{{Kashikawa} {et~al.}(2015){Kashikawa}, {Ishizaki}, {Willott},
  {Onoue}, {Im}, {Furusawa}, {Toshikawa}, {Ishikawa}, {Niino}, {Shimasaku},
  {Ouchi}, \& {Hibon}}]{Kashikawa2015}
{Kashikawa}, N., {Ishizaki}, Y., {Willott}, C.~J., {et~al.} 2015, \apj, 798, 28

\bibitem[{{Magnier}(2006)}]{Magnier2006}
{Magnier}, E. 2006, in The Advanced Maui Optical and Space Surveillance
  Technologies Conference, E50

\bibitem[{{Magnier}(2007)}]{Magnier2007}
{Magnier}, E. 2007, in Astronomical Society of the Pacific Conference Series,
  Vol. 364, The Future of Photometric, Spectrophotometric and Polarimetric
  Standardization, ed. C.~{Sterken}, 153

\bibitem[{{Mainzer} {et~al.}(2011){Mainzer}, {Bauer}, {Grav}, {Masiero},
  {Cutri}, {Dailey}, {Eisenhardt}, {McMillan}, {Wright}, {Walker}, {Jedicke},
  {Spahr}, {Tholen}, {Alles}, {Beck}, {Brandenburg}, {Conrow}, {Evans},
  {Fowler}, {Jarrett}, {Marsh}, {Masci}, {McCallon}, {Wheelock}, {Wittman},
  {Wyatt}, {DeBaun}, {Elliott}, {Elsbury}, {Gautier}, {Gomillion}, {Leisawitz},
  {Maleszewski}, {Micheli}, \& {Wilkins}}]{Mainzer2011}
{Mainzer}, A., {Bauer}, J., {Grav}, T., {et~al.} 2011, \apj, 731, 53

\bibitem[{{Martin} {et~al.}(2005){Martin}, {Fanson}, {Schiminovich},
  {Morrissey}, {Friedman}, {Barlow}, {Conrow}, {Grange}, {Jelinsky},
  {Milliard}, {Siegmund}, {Bianchi}, {Byun}, {Donas}, {Forster}, {Heckman},
  {Lee}, {Madore}, {Malina}, {Neff}, {Rich}, {Small}, {Surber}, {Szalay},
  {Welsh}, \& {Wyder}}]{Martin2005}
{Martin}, D.~C., {Fanson}, J., {Schiminovich}, D., {et~al.} 2005, \apjl, 619,
  L1

\bibitem[{{Matsuoka} {et~al.}(2016){Matsuoka}, {Onoue}, {Kashikawa}, {Iwasawa},
  {Strauss}, {Nagao}, {Imanishi}, {Niida}, {Toba}, {Akiyama}, {Asami}, {Bosch},
  {Foucaud}, {Furusawa}, {Goto}, {Gunn}, {Harikane}, {Ikeda}, {Kawaguchi},
  {Kikuta}, {Komiyama}, {Lupton}, {Minezaki}, {Miyazaki}, {Morokuma},
  {Murayama}, {Nishizawa}, {Ono}, {Ouchi}, {Price}, {Sameshima}, {Silverman},
  {Sugiyama}, {Tait}, {Takada}, {Takata}, {Tanaka}, {Tang}, \&
  {Utsumi}}]{Matsuoka2016}
{Matsuoka}, Y., {Onoue}, M., {Kashikawa}, N., {et~al.} 2016, \apj, 828, 26

\bibitem[{{Matsuoka} {et~al.}(2018{\natexlab{a}}){Matsuoka}, {Onoue},
  {Kashikawa}, {Iwasawa}, {Strauss}, {Nagao}, {Imanishi}, {Lee}, {Akiyama},
  {Asami}, {Bosch}, {Foucaud}, {Furusawa}, {Goto}, {Gunn}, {Harikane}, {Ikeda},
  {Izumi}, {Kawaguchi}, {Kikuta}, {Kohno}, {Komiyama}, {Lupton}, {Minezaki},
  {Miyazaki}, {Morokuma}, {Murayama}, {Niida}, {Nishizawa}, {Oguri}, {Ono},
  {Ouchi}, {Price}, {Sameshima}, {Schulze}, {Shirakata}, {Silverman},
  {Sugiyama}, {Tait}, {Takada}, {Takata}, {Tanaka}, {Tang}, {Toba}, {Utsumi},
  \& {Wang}}]{Matsuoka2018}
---. 2018{\natexlab{a}}, \pasj, 70, S35

\bibitem[{{Matsuoka} {et~al.}(2018{\natexlab{b}}){Matsuoka}, {Iwasawa},
  {Onoue}, {Kashikawa}, {Strauss}, {Lee}, {Imanishi}, {Nagao}, {Akiyama},
  {Asami}, {Bosch}, {Furusawa}, {Goto}, {Gunn}, {Harikane}, {Ikeda}, {Izumi},
  {Kawaguchi}, {Kato}, {Kikuta}, {Kohno}, {Komiyama}, {Lupton}, {Minezaki},
  {Miyazaki}, {Morokuma}, {Murayama}, {Niida}, {Nishizawa}, {Oguri}, {Ono},
  {Ouchi}, {Price}, {Sameshima}, {Schulze}, {Shirakata}, {Silverman},
  {Sugiyama}, {Tait}, {Takada}, {Takata}, {Tanaka}, {Tang}, {Toba}, {Utsumi},
  {Wang}, \& {Yamashita}}]{Matsuoka2018b}
{Matsuoka}, Y., {Iwasawa}, K., {Onoue}, M., {et~al.} 2018{\natexlab{b}}, \apjs,
  237, 5

\bibitem[{{Matsuoka} {et~al.}(2019){Matsuoka}, {Onoue}, {Kashikawa}, {Strauss},
  {Iwasawa}, {Lee}, {Imanishi}, {Nagao}, {Akiyama}, {Asami}, {Bosch},
  {Furusawa}, {Goto}, {Gunn}, {Harikane}, {Ikeda}, {Izumi}, {Kawaguchi},
  {Kato}, {Kikuta}, {Kohno}, {Komiyama}, {Koyama}, {Lupton}, {Minezaki},
  {Miyazaki}, {Murayama}, {Niida}, {Nishizawa}, {Noboriguchi}, {Oguri}, {Ono},
  {Ouchi}, {Price}, {Sameshima}, {Schulze}, {Shirakata}, {Silverman},
  {Sugiyama}, {Tait}, {Takada}, {Takata}, {Tanaka}, {Tang}, {Toba}, {Utsumi},
  {Wang}, \& {Yamashita}}]{Matsuoka2019}
{Matsuoka}, Y., {Onoue}, M., {Kashikawa}, N., {et~al.} 2019, \apj, 872, L2

\bibitem[{{Mazzucchelli} {et~al.}(2017){Mazzucchelli}, {Ba{\~n}ados},
  {Venemans}, {Decarli}, {Farina}, {Walter}, {Eilers}, {Rix}, {Simcoe},
  {Stern}, {Fan}, {Schlafly}, {De Rosa}, {Hennawi}, {Chambers}, {Greiner},
  {Burgett}, {Draper}, {Kaiser}, {Kudritzki}, {Magnier}, {Metcalfe}, {Waters},
  \& {Wainscoat}}]{Mazzucchelli2017}
{Mazzucchelli}, C., {Ba{\~n}ados}, E., {Venemans}, B.~P., {et~al.} 2017, \apj,
  849, 91

\bibitem[{{McGreer} {et~al.}(2013){McGreer}, {Jiang}, {Fan}, {Richards},
  {Strauss}, {Ross}, {White}, {Shen}, {Schneider}, {Myers}, {Brandt}, {DeGraf},
  {Glikman}, {Ge}, \& {Streblyanska}}]{McGreer2013}
{McGreer}, I.~D., {Jiang}, L., {Fan}, X., {et~al.} 2013, \apj, 768, 105

\bibitem[{McKinney(2011)}]{pandas}
McKinney, W. 2011, presented at PyHPC2011

\bibitem[{{Morganson} {et~al.}(2012){Morganson}, {De Rosa}, {Decarli},
  {Walter}, {Chambers}, {McGreer}, {Fan}, {Burgett}, {Flewelling}, {Greiner},
  {Hodapp}, {Kaiser}, {Magnier}, {Price}, {Rix}, {Sweeney}, \&
  {Waters}}]{Morganson2012}
{Morganson}, E., {De Rosa}, G., {Decarli}, R., {et~al.} 2012, \aj, 143, 142

\bibitem[{{Mortlock} {et~al.}(2011){Mortlock}, {Warren}, {Venemans}, {Patel},
  {Hewett}, {McMahon}, {Simpson}, {Theuns}, {Gonz{\'a}les-Solares}, {Adamson},
  {Dye}, {Hambly}, {Hirst}, {Irwin}, {Kuiper}, {Lawrence}, \&
  {R{\"o}ttgering}}]{Mortlock2011}
{Mortlock}, D.~J., {Warren}, S.~J., {Venemans}, B.~P., {et~al.} 2011, \nat,
  474, 616

\bibitem[{{Myers} {et~al.}(2007){Myers}, {Brunner}, {Nichol}, {Richards},
  {Schneider}, \& {Bahcall}}]{Myers2007}
{Myers}, A.~D., {Brunner}, R.~J., {Nichol}, R.~C., {et~al.} 2007, \apj, 658, 85

\bibitem[{{Myers} {et~al.}(2006){Myers}, {Brunner}, {Richards}, {Nichol},
  {Schneider}, {Vanden Berk}, {Scranton}, {Gray}, \& {Brinkmann}}]{Myers2006}
{Myers}, A.~D., {Brunner}, R.~J., {Richards}, G.~T., {et~al.} 2006, \apj, 638,
  622

\bibitem[{{Myers} {et~al.}(2015){Myers}, {Palanque-Delabrouille}, {Prakash},
  {P{\^a}ris}, {Yeche}, {Dawson}, {Bovy}, {Lang}, {Schlegel}, {Newman},
  {Petitjean}, {Kneib}, {Laurent}, {Percival}, {Ross}, {Seo}, {Tinker},
  {Armengaud}, {Brownstein}, {Burtin}, {Cai}, {Comparat}, {Kasliwal},
  {Kulkarni}, {Laher}, {Levitan}, {McBride}, {McGreer}, {Miller}, {Nugent},
  {Ofek}, {Rossi}, {Ruan}, {Schneider}, {Sesar}, {Streblyanska}, \&
  {Surace}}]{Myers2015}
{Myers}, A.~D., {Palanque-Delabrouille}, N., {Prakash}, A., {et~al.} 2015,
  \apjs, 221, 27

\bibitem[{{Oke} \& {Gunn}(1983)}]{Oke1983}
{Oke}, J.~B., \& {Gunn}, J.~E. 1983, \apj, 266, 713

\bibitem[{{P{\^a}ris} {et~al.}(2012){P{\^a}ris}, {Petitjean}, {Aubourg},
  {Bailey}, {Ross}, {Myers}, {Strauss}, {Anderson}, {Arnau}, {Bautista},
  {Bizyaev}, {Bolton}, {Bovy}, {Brandt}, {Brewington}, {Browstein}, {Busca},
  {Capellupo}, {Carithers}, {Croft}, {Dawson}, {Delubac}, {Ebelke},
  {Eisenstein}, {Engelke}, {Fan}, {Filiz Ak}, {Finley}, {Font-Ribera}, {Ge},
  {Gibson}, {Hall}, {Hamann}, {Hennawi}, {Ho}, {Hogg}, {Ivezi{\'c}}, {Jiang},
  {Kimball}, {Kirkby}, {Kirkpatrick}, {Lee}, {Le Goff}, {Lundgren}, {MacLeod},
  {Malanushenko}, {Malanushenko}, {Maraston}, {McGreer}, {McMahon},
  {Miralda-Escud{\'e}}, {Muna}, {Noterdaeme}, {Oravetz},
  {Palanque-Delabrouille}, {Pan}, {Perez-Fournon}, {Pieri}, {Richards},
  {Rollinde}, {Sheldon}, {Schlegel}, {Schneider}, {Slosar}, {Shelden}, {Shen},
  {Simmons}, {Snedden}, {Suzuki}, {Tinker}, {Viel}, {Weaver}, {Weinberg},
  {White}, {Wood-Vasey}, \& {Y{\`e}che}}]{Paris2012}
{P{\^a}ris}, I., {Petitjean}, P., {Aubourg}, {\'E}., {et~al.} 2012, \aap, 548,
  A66

\bibitem[{{P{\^a}ris} {et~al.}(2018){P{\^a}ris}, {Petitjean}, {Aubourg},
  {Myers}, {Streblyanska}, {Lyke}, {Anderson}, {Armengaud}, {Bautista},
  {Blanton}, {Blomqvist}, {Brinkmann}, {Brownstein}, {Brandt}, {Burtin},
  {Dawson}, {de la Torre}, {Georgakakis}, {Gil-Mar{\'{\i}}n}, {Green}, {Hall},
  {Kneib}, {LaMassa}, {Le Goff}, {MacLeod}, {Mariappan}, {McGreer}, {Merloni},
  {Noterdaeme}, {Palanque-Delabrouille}, {Percival}, {Ross}, {Rossi},
  {Schneider}, {Seo}, {Tojeiro}, {Weaver}, {Weijmans}, {Y{\`e}che}, {Zarrouk},
  \& {Zhao}}]{Paris2018}
---. 2018, \aap, 613, A51

\bibitem[{Pedregosa {et~al.}(2011)Pedregosa, Varoquaux, Gramfort, Michel,
  Thirion, Grisel, Blondel, Prettenhofer, Weiss, Dubourg, Vanderplas, Passos,
  Cournapeau, Brucher, Perrot, \& Duchesnay}]{scikit-learn}
Pedregosa, F., Varoquaux, G., Gramfort, A., {et~al.} 2011, Journal of Machine
  Learning Research, 12, 2825

\bibitem[{{Planck Collaboration} {et~al.}(2016){Planck Collaboration}, {Ade},
  {Aghanim}, {Arnaud}, {Ashdown}, {Aumont}, {Baccigalupi}, {Banday},
  {Barreiro}, {Bartlett}, \& et~al.}]{PlanckCollaboration2016}
{Planck Collaboration}, {Ade}, P.~A.~R., {Aghanim}, N., {et~al.} 2016, \aap,
  594, A13

\bibitem[{{Pons} {et~al.}(2019){Pons}, {McMahon}, {Simcoe}, {Banerji},
  {Hewett}, \& {Reed}}]{Pons2019}
{Pons}, E., {McMahon}, R.~G., {Simcoe}, R.~A., {et~al.} 2019, \mnras, 484, 5142

\bibitem[{{Prochaska} {et~al.}(2005){Prochaska}, {Herbert-Fort}, \&
  {Wolfe}}]{Prochaska2005}
{Prochaska}, J.~X., {Herbert-Fort}, S., \& {Wolfe}, A.~M. 2005, \apj, 635, 123

\bibitem[{{Reed} {et~al.}(2015){Reed}, {McMahon}, {Banerji}, {Becker},
  {Gonzalez-Solares}, {Martini}, {Ostrovski}, {Rauch}, {Abbott}, {Abdalla},
  {Allam}, {Benoit-Levy}, {Bertin}, {Buckley-Geer}, {Burke}, {Carnero Rosell},
  {da Costa}, {D'Andrea}, {DePoy}, {Desai}, {Diehl}, {Doel}, {Cunha},
  {Estrada}, {Evrard}, {Fausti Neto}, {Finley}, {Fosalba}, {Frieman}, {Gruen},
  {Honscheid}, {James}, {Kent}, {Kuehn}, {Kuropatkin}, {Lahav}, {Maia},
  {Makler}, {Marshall}, {Merritt}, {Miquel}, {Mohr}, {Nord}, {Ogando},
  {Plazas}, {Romer}, {Roodman}, {Rykoff}, {Sako}, {Sanchez}, {Santiago},
  {Schubnell}, {Sevilla}, {Smith}, {Soares-Santos}, {Suchyta}, {Swanson},
  {Tarle}, {Thomas}, {Tucker}, {Walker}, \& {Wechsler}}]{Reed2015}
{Reed}, S.~L., {McMahon}, R.~G., {Banerji}, M., {et~al.} 2015, \mnras, 454,
  3952

\bibitem[{{Reed} {et~al.}(2017){Reed}, {McMahon}, {Martini}, {Banerji},
  {Auger}, {Hewett}, {Koposov}, {Gibbons}, {Gonzalez-Solares}, {Ostrovski},
  {Tie}, {Abdalla}, {Allam}, {Benoit-L{\'e}vy}, {Bertin}, {Brooks},
  {Buckley-Geer}, {Burke}, {Carnero Rosell}, {Carrasco Kind}, {Carretero}, {da
  Costa}, {DePoy}, {Desai}, {Diehl}, {Doel}, {Evrard}, {Finley}, {Flaugher},
  {Fosalba}, {Frieman}, {Garc{\'{\i}}a-Bellido}, {Gaztanaga}, {Goldstein},
  {Gruen}, {Gruendl}, {Gutierrez}, {James}, {Kuehn}, {Kuropatkin}, {Lahav},
  {Lima}, {Maia}, {Marshall}, {Melchior}, {Miller}, {Miquel}, {Nord}, {Ogando},
  {Plazas}, {Romer}, {Sanchez}, {Scarpine}, {Schubnell}, {Sevilla-Noarbe},
  {Smith}, {Sobreira}, {Suchyta}, {Swanson}, {Tarle}, {Tucker}, {Walker}, \&
  {Wester}}]{Reed2017}
{Reed}, S.~L., {McMahon}, R.~G., {Martini}, P., {et~al.} 2017, \mnras, 468,
  4702

\bibitem[{{Reed} {et~al.}(2019){Reed}, {Banerji}, {Becker}, {Hewett},
  {Martini}, {McMahon}, {Pons}, {Rauch}, {Abbott}, {Allam}, {Annis}, {Avila},
  {Bertin}, {Brooks}, {Buckley-Geer}, {Carnero Rosell}, {Carrasco Kind},
  {Carretero}, {Castander}, {Cunha}, {D'Andrea}, {da Costa}, {De Vicente},
  {Desai}, {Diehl}, {Doel}, {Evrard}, {Flaugher}, {Frieman}, {Garcia-Bellido},
  {Gaztanaga}, {Gruen}, {Gschwend}, {Gutierrez}, {Hollowood}, {Honscheid},
  {Hoyle}, {James}, {Kuehn}, {Lahav}, {Lima}, {Maia}, {Marshall}, {Miquel},
  {Ogand o}, {Plazas}, {Roodman}, {Sanchez}, {Scarpine}, {Schubnell},
  {Serrano}, {Sevilla-Noarbe}, {Smith}, {Smith}, {Sobreira}, {Suchyta},
  {Swanson}, {Tarle}, {Thomas}, {Tucker}, \& {Vikram}}]{Reed2019}
{Reed}, S.~L., {Banerji}, M., {Becker}, G.~D., {et~al.} 2019, arXiv e-prints,
  arXiv:1901.07456

\bibitem[{{Richards} {et~al.}(2002){Richards}, {Fan}, {Newberg}, {Strauss},
  {Vanden Berk}, {Schneider}, {Yanny}, {Boucher}, {Burles}, {Frieman}, {Gunn},
  {Hall}, {Ivezi{\'c}}, {Kent}, {Loveday}, {Lupton}, {Rockosi}, {Schlegel},
  {Stoughton}, {SubbaRao}, \& {York}}]{Richards2002}
{Richards}, G.~T., {Fan}, X., {Newberg}, H.~J., {et~al.} 2002, \aj, 123, 2945

\bibitem[{{Richards} {et~al.}(2011){Richards}, {Kruczek}, {Gallagher}, {Hall},
  {Hewett}, {Leighly}, {Deo}, {Kratzer}, \& {Shen}}]{Richards2011}
{Richards}, G.~T., {Kruczek}, N.~E., {Gallagher}, S.~C., {et~al.} 2011, \aj,
  141, 167

\bibitem[{{Richards} {et~al.}(2015){Richards}, {Myers}, {Peters}, {Krawczyk},
  {Chase}, {Ross}, {Fan}, {Jiang}, {Lacy}, {McGreer}, {Trump}, \&
  {Riegel}}]{Richards2015}
{Richards}, G.~T., {Myers}, A.~D., {Peters}, C.~M., {et~al.} 2015, \apjs, 219,
  39

\bibitem[{{Rodr{\'{\i}}guez-Torres} {et~al.}(2017){Rodr{\'{\i}}guez-Torres},
  {Comparat}, {Prada}, {Yepes}, {Burtin}, {Zarrouk}, {Laurent}, {Hahn},
  {Behroozi}, {Klypin}, {Ross}, {Tojeiro}, \& {Zhao}}]{RodriguezTorres2017}
{Rodr{\'{\i}}guez-Torres}, S.~A., {Comparat}, J., {Prada}, F., {et~al.} 2017,
  \mnras, 468, 728

\bibitem[{{Ross} {et~al.}(2009){Ross}, {Shen}, {Strauss}, {Vanden Berk},
  {Connolly}, {Richards}, {Schneider}, {Weinberg}, {Hall}, {Bahcall}, \&
  {Brunner}}]{Ross2009}
{Ross}, N.~P., {Shen}, Y., {Strauss}, M.~A., {et~al.} 2009, \apj, 697, 1634

\bibitem[{{Ross} {et~al.}(2013){Ross}, {McGreer}, {White}, {Richards}, {Myers},
  {Palanque-Delabrouille}, {Strauss}, {Anderson}, {Shen}, {Brandt},
  {Y{\`e}che}, {Swanson}, {Aubourg}, {Bailey}, {Bizyaev}, {Bovy}, {Brewington},
  {Brinkmann}, {DeGraf}, {Di Matteo}, {Ebelke}, {Fan}, {Ge}, {Malanushenko},
  {Malanushenko}, {Mandelbaum}, {Maraston}, {Muna}, {Oravetz}, {Pan},
  {P{\^a}ris}, {Petitjean}, {Schawinski}, {Schlegel}, {Schneider}, {Silverman},
  {Simmons}, {Snedden}, {Streblyanska}, {Suzuki}, {Weinberg}, \&
  {York}}]{Ross2013}
{Ross}, N.~P., {McGreer}, I.~D., {White}, M., {et~al.} 2013, \apj, 773, 14

\bibitem[{{Salvato} {et~al.}(2018){Salvato}, {Buchner}, {Budav{\'a}ri},
  {Dwelly}, {Merloni}, {Brusa}, {Rau}, {Fotopoulou}, \& {Nandra}}]{Salvato2018}
{Salvato}, M., {Buchner}, J., {Budav{\'a}ri}, T., {et~al.} 2018, \mnras, 473,
  4937

\bibitem[{Schindler {et~al.}(2017)Schindler, Fan, McGreer, Yang, Wu, Jiang, \&
  Green}]{Schindler2017}
Schindler, J.-T., Fan, X., McGreer, I.~D., {et~al.} 2017, \apj, 851, 13

\bibitem[{{Schindler} {et~al.}(2018){Schindler}, {Fan}, {McGreer}, {Yang},
  {Wang}, {Green}, {Garavito-Camargo}, {Huang}, {O{\textquoteright}Donnell},
  {Patej}, {Pucha}, {Rees}, \& {Spalding}}]{Schindler2018}
{Schindler}, J.-T., {Fan}, X., {McGreer}, I.~D., {et~al.} 2018, \apj, 863, 144

\bibitem[{{Schindler} {et~al.}(2019){Schindler}, {Fan}, {McGreer}, {Yang},
  {Wang}, {Green}, {Fynbo}, {Krogager}, {Green}, {Huang}, {Kadowaki}, {Patej},
  {Wu}, \& {Yue}}]{Schindler2019a}
---. 2019, \apj, 871, 258

\bibitem[{{Schlegel} {et~al.}(1998){Schlegel}, {Finkbeiner}, \&
  {Davis}}]{Schlegel1998}
{Schlegel}, D.~J., {Finkbeiner}, D.~P., \& {Davis}, M. 1998, \apj, 500, 525

\bibitem[{{Schmidt} {et~al.}(2018){Schmidt}, {Hennawi}, {Lee}, {Lukic},
  {Onorbe}, \& {White}}]{Schmidt2018b_arxiv}
{Schmidt}, T.~M., {Hennawi}, J.~F., {Lee}, K.-G., {et~al.} 2018, arXiv e-prints

\bibitem[{{Schneider} {et~al.}(2010){Schneider}, {Richards}, {Hall}, {Strauss},
  {Anderson}, {Boroson}, {Ross}, {Shen}, {Brandt}, {Fan}, {Inada}, {Jester},
  {Knapp}, {Krawczyk}, {Thakar}, {Vanden Berk}, {Voges}, {Yanny}, {York},
  {Bahcall}, {Bizyaev}, {Blanton}, {Brewington}, {Brinkmann}, {Eisenstein},
  {Frieman}, {Fukugita}, {Gray}, {Gunn}, {Hibon}, {Ivezi{\'c}}, {Kent}, {Kron},
  {Lee}, {Lupton}, {Malanushenko}, {Malanushenko}, {Oravetz}, {Pan}, {Pier},
  {Price}, {Saxe}, {Schlegel}, {Simmons}, {Snedden}, {SubbaRao}, {Szalay}, \&
  {Weinberg}}]{Schneider2010}
{Schneider}, D.~P., {Richards}, G.~T., {Hall}, P.~B., {et~al.} 2010, \aj, 139,
  2360

\bibitem[{{Shen} {et~al.}(2007){Shen}, {Strauss}, {Oguri}, {Hennawi}, {Fan},
  {Richards}, {Hall}, {Gunn}, {Schneider}, {Szalay}, {Thakar}, {Vanden Berk},
  {Anderson}, {Bahcall}, {Connolly}, \& {Knapp}}]{Shen2007}
{Shen}, Y., {Strauss}, M.~A., {Oguri}, M., {et~al.} 2007, \aj, 133, 2222

\bibitem[{{Simcoe} {et~al.}(2004){Simcoe}, {Sargent}, \& {Rauch}}]{Simcoe2004}
{Simcoe}, R.~A., {Sargent}, W.~L.~W., \& {Rauch}, M. 2004, \apj, 606, 92

\bibitem[{{Timlin} {et~al.}(2018){Timlin}, {Ross}, {Richards}, {Myers},
  {Pellegrino}, {Bauer}, {Lacy}, {Schneider}, {Wollack}, \&
  {Zakamska}}]{Timlin2018}
{Timlin}, J.~D., {Ross}, N.~P., {Richards}, G.~T., {et~al.} 2018, \apj, 859, 20

\bibitem[{{Tody}(1986)}]{Tody1986}
{Tody}, D. 1986, in \procspie, Vol. 627, Instrumentation in astronomy VI, ed.
  D.~L. {Crawford}, 733

\bibitem[{{Tody}(1993)}]{Tody1993}
{Tody}, D. 1993, in Astronomical Society of the Pacific Conference Series,
  Vol.~52, Astronomical Data Analysis Software and Systems II, ed. R.~J.
  {Hanisch}, R.~J.~V. {Brissenden}, \& J.~{Barnes}, 173

\bibitem[{{Truemper}(1982)}]{Truemper1982}
{Truemper}, J. 1982, Advances in Space Research, 2, 241

\bibitem[{{Vanden Berk} {et~al.}(2001){Vanden Berk}, {Richards}, {Bauer},
  {Strauss}, {Schneider}, {Heckman}, {York}, {Hall}, {Fan}, {Knapp},
  {Anderson}, {Annis}, {Bahcall}, {Bernardi}, {Briggs}, {Brinkmann}, {Brunner},
  {Burles}, {Carey}, {Castander}, {Connolly}, {Crocker}, {Csabai}, {Doi},
  {Finkbeiner}, {Friedman}, {Frieman}, {Fukugita}, {Gunn}, {Hennessy},
  {Ivezi{\'c}}, {Kent}, {Kunszt}, {Lamb}, {Leger}, {Long}, {Loveday}, {Lupton},
  {Meiksin}, {Merelli}, {Munn}, {Newberg}, {Newcomb}, {Nichol}, {Owen}, {Pier},
  {Pope}, {Rockosi}, {Schlegel}, {Siegmund}, {Smee}, {Snir}, {Stoughton},
  {Stubbs}, {SubbaRao}, {Szalay}, {Szokoly}, {Tremonti}, {Uomoto}, {Waddell},
  {Yanny}, \& {Zheng}}]{Vandenberk2001}
{Vanden Berk}, D.~E., {Richards}, G.~T., {Bauer}, A., {et~al.} 2001, \aj, 122,
  549

\bibitem[{{Venemans} {et~al.}(2007){Venemans}, {McMahon}, {Warren},
  {Gonzalez-Solares}, {Hewett}, {Mortlock}, {Dye}, \& {Sharp}}]{Venemans2007}
{Venemans}, B.~P., {McMahon}, R.~G., {Warren}, S.~J., {et~al.} 2007, \mnras,
  376, L76

\bibitem[{{Venemans} {et~al.}(2013){Venemans}, {Findlay}, {Sutherland}, {De
  Rosa}, {McMahon}, {Simcoe}, {Gonz{\'a}lez-Solares}, {Kuijken}, \&
  {Lewis}}]{Venemans2013}
{Venemans}, B.~P., {Findlay}, J.~R., {Sutherland}, W.~J., {et~al.} 2013, \apj,
  779, 24

\bibitem[{{Volonteri}(2012)}]{Volonteri2012}
{Volonteri}, M. 2012, Science, 337, 544

\bibitem[{{Wang} {et~al.}(2018{\natexlab{a}}){Wang}, {Yang}, {Fan}, {Wu},
  {Yue}, {Li}, {Bian}, {Jiang}, {Ba{\~n}ados}, {Schindler}, {Findlay},
  {Davies}, {Decarli}, {Farina}, {Green}, {Hennawi}, {Huang}, {Mazzuccheli},
  {McGreer}, {Venemans}, {Walter}, {Dye}, {Lyke}, {Myers}, \& {Haze
  Nunez}}]{Wang2018b_arxiv}
{Wang}, F., {Yang}, J., {Fan}, X., {et~al.} 2018{\natexlab{a}}, arXiv e-prints

\bibitem[{{Wang} {et~al.}(2018{\natexlab{b}}){Wang}, {Yang}, {Fan}, {Yue},
  {Wu}, {Schindler}, {Bian}, {Li}, {Farina}, {Ba{\~n}ados}, {Davies},
  {Decarli}, {Green}, {Jiang}, {Hennawi}, {Huang}, {Mazzucchelli}, {McGreer},
  {Venemans}, {Walter}, \& {Beletsky}}]{Wang2018}
---. 2018{\natexlab{b}}, \apjl, 869, L9

\bibitem[{{White} {et~al.}(2012){White}, {Myers}, {Ross}, {Schlegel},
  {Hennawi}, {Shen}, {McGreer}, {Strauss}, {Bolton}, {Bovy}, {Fan},
  {Miralda-Escude}, {Palanque-Delabrouille}, {Paris}, {Petitjean}, {Schneider},
  {Viel}, {Weinberg}, {Yeche}, {Zehavi}, {Pan}, {Snedden}, {Bizyaev},
  {Brewington}, {Brinkmann}, {Malanushenko}, {Malanushenko}, {Oravetz},
  {Simmons}, {Sheldon}, \& {Weaver}}]{White2012}
{White}, M., {Myers}, A.~D., {Ross}, N.~P., {et~al.} 2012, \mnras, 424, 933

\bibitem[{{Willott} {et~al.}(2007){Willott}, {Delorme}, {Omont}, {Bergeron},
  {Delfosse}, {Forveille}, {Albert}, {Reyl{\'e}}, {Hill}, {Gully-Santiago},
  {Vinten}, {Crampton}, {Hutchings}, {Schade}, {Simard}, {Sawicki}, {Beelen},
  \& {Cox}}]{Willott2007}
{Willott}, C.~J., {Delorme}, P., {Omont}, A., {et~al.} 2007, \aj, 134, 2435

\bibitem[{{Willott} {et~al.}(2010){Willott}, {Delorme}, {Reyl{\'e}}, {Albert},
  {Bergeron}, {Crampton}, {Delfosse}, {Forveille}, {Hutchings}, {McLure},
  {Omont}, \& {Schade}}]{Willott2010a}
{Willott}, C.~J., {Delorme}, P., {Reyl{\'e}}, C., {et~al.} 2010, \aj, 139, 906

\bibitem[{{Worseck} \& {Prochaska}(2011)}]{Worseck2011}
{Worseck}, G., \& {Prochaska}, J.~X. 2011, \apj, 728, 23

\bibitem[{{Worseck} {et~al.}(2016){Worseck}, {Prochaska}, {Hennawi}, \&
  {McQuinn}}]{Worseck2016}
{Worseck}, G., {Prochaska}, J.~X., {Hennawi}, J.~F., \& {McQuinn}, M. 2016,
  \apj, 825, 144

\bibitem[{{Wu} {et~al.}(2012){Wu}, {Hao}, {Jia}, {Zhang}, \& {Peng}}]{Wu2012}
{Wu}, X.-B., {Hao}, G., {Jia}, Z., {Zhang}, Y., \& {Peng}, N. 2012, \aj, 144,
  49

\bibitem[{{Wu} \& {Jia}(2010)}]{Wu2010}
{Wu}, X.-B., \& {Jia}, Z. 2010, \mnras, 406, 1583

\bibitem[{{Yang} {et~al.}(2018{\natexlab{a}}){Yang}, {Wang}, {Fan}, {Yue},
  {Wu}, {Li}, {Bian}, {Jiang}, {Ba{\~n}ados}, \& {Beletsky}}]{Yang2018c_arxiv}
{Yang}, J., {Wang}, F., {Fan}, X., {et~al.} 2018{\natexlab{a}}, arXiv e-prints

\bibitem[{{Yang} {et~al.}(2018{\natexlab{b}}){Yang}, {Wang}, {Fan}, {Wu},
  {Bian}, {Ba{\~n}ados}, {Yue}, {Schindler}, {Yang}, {Jiang}, {McGreer},
  {Green}, \& {Dye}}]{Yang2018b_arxiv}
---. 2018{\natexlab{b}}, arXiv e-prints

\bibitem[{{York} {et~al.}(2000){York}, {Adelman}, {Anderson}, {Anderson},
  {Annis}, {Bahcall}, {Bakken}, {Barkhouser}, {Bastian}, {Berman}, {Boroski},
  {Bracker}, {Briegel}, {Briggs}, {Brinkmann}, {Brunner}, {Burles}, {Carey},
  {Carr}, {Castander}, {Chen}, {Colestock}, {Connolly}, {Crocker}, {Csabai},
  {Czarapata}, {Davis}, {Doi}, {Dombeck}, {Eisenstein}, {Ellman}, {Elms},
  {Evans}, {Fan}, {Federwitz}, {Fiscelli}, {Friedman}, {Frieman}, {Fukugita},
  {Gillespie}, {Gunn}, {Gurbani}, {de Haas}, {Haldeman}, {Harris}, {Hayes},
  {Heckman}, {Hennessy}, {Hindsley}, {Holm}, {Holmgren}, {Huang}, {Hull},
  {Husby}, {Ichikawa}, {Ichikawa}, {Ivezi{\'c}}, {Kent}, {Kim}, {Kinney},
  {Klaene}, {Kleinman}, {Kleinman}, {Knapp}, {Korienek}, {Kron}, {Kunszt},
  {Lamb}, {Lee}, {Leger}, {Limmongkol}, {Lindenmeyer}, {Long}, {Loomis},
  {Loveday}, {Lucinio}, {Lupton}, {MacKinnon}, {Mannery}, {Mantsch}, {Margon},
  {McGehee}, {McKay}, {Meiksin}, {Merelli}, {Monet}, {Munn}, {Narayanan},
  {Nash}, {Neilsen}, {Neswold}, {Newberg}, {Nichol}, {Nicinski}, {Nonino},
  {Okada}, {Okamura}, {Ostriker}, {Owen}, {Pauls}, {Peoples}, {Peterson},
  {Petravick}, {Pier}, {Pope}, {Pordes}, {Prosapio}, {Rechenmacher}, {Quinn},
  {Richards}, {Richmond}, {Rivetta}, {Rockosi}, {Ruthmansdorfer}, {Sandford},
  {Schlegel}, {Schneider}, {Sekiguchi}, {Sergey}, {Shimasaku}, {Siegmund},
  {Smee}, {Smith}, {Snedden}, {Stone}, {Stoughton}, {Strauss}, {Stubbs},
  {SubbaRao}, {Szalay}, {Szapudi}, {Szokoly}, {Thakar}, {Tremonti}, {Tucker},
  {Uomoto}, {Vanden Berk}, {Vogeley}, {Waddell}, {Wang}, {Watanabe},
  {Weinberg}, {Yanny}, {Yasuda}, \& {SDSS Collaboration}}]{York2000}
{York}, D.~G., {Adelman}, J., {Anderson}, Jr., J.~E., {et~al.} 2000, \aj, 120,
  1579

\end{thebibliography}
\bibliographystyle{apj}

\appendix

\section{The PS-ELQS Quasar Catalog} \label{app_qso_catalog}
The PS-ELQS quasar catalog is available as a machine readable table on-line. It has 51 columns, detailed in Table\,\ref{tab_pselqs_cat_cols}.

\begin{table*}[htp]
 \centering 
 \caption{Description of the full PS-ELQS quasar catalog table}
 \label{tab_pselqs_cat_cols}
 \begin{tabular}{cccp{8cm}}
  \tableline
  Column & Column Name & Unit & Description \\
  \tableline
  1 & wise\_designation &  - & Designation of the WISE AllWISE survey\\
  2 & ps\_ra & deg & Right ascension from PS1 \\
  3 & ps\_dec & deg & Declination from PS1 \\
  4 & ps\_ra\_hms & hh:mm:ss.sss & Right ascension from PS1 \\
  5 & ps\_dec\_dms & dd:mm:ss.ss & Declination from the PS1 \\
  6 & wise\_ra & deg & Right ascension from the AllWISE \\
  7 & wise\_dec & deg & Declination from the AllWISE \\
  8 & reference & - & Reference to the quasar classification \\
  9 & reference\_z & - & Best redshift of the quasar according to the reference \\
  10 & M\_1450 & mag & Absolute magnitude at $1450\text{\AA}$ calculated using the k-correction determined for this work\\
%  11 & sel\_prob & - & Selection probability according to our completeness calculation \\
  11-20 & [survey]\_mag\_[band] & mag & Dereddened AB magnitudes of the PS1 grizy, 2MASS jh$\rm{k}_{\rm{s}}$ and WISE W1W2 bands (surveys = [PS,TMASS,WISE]; bands = [g,r,i,z,y],[j,h,k],[w1,w2]). \\
  21-30 & [survey]\_magerr\_[band] & mag & $1\sigma$ errors on the AB magnitudes. \\
  31 & EBV & mag & E(B-V) \\
%  33 & extinction\_i & mag & Extinction in the SDSS i-band\\ 
%  34 & FIRST\_match & True/False & Boolean to indicate successful matches with the FIRST catalog \\
%  35 & FIRST\_distance & arcsec & Distance of the FIRST source relative to the SDSS position \\
%  36 & FIRST\_peak\_flux\_mJy/bm & mJy/bm & FIRST peak flux\\
%  37 & FIRST\_RMS\_mJy/bm & mJy/bm & RMS error on the FIRST flux\\
  32 & GALEX\_match & True/NaN &  Boolean to indicate successful matches with the GALEX GR6/7 catalog\\
  33 & GALEX\_distance & arcsec & Distance of the GALEX GR6/7 match relative to the SDSS position \\
  34 & GALEX\_nuv\_mag & mag & GALEX near-UV flux in magnitudes \\
  35 & GALEX\_nuv\_magErr & mag & Error on the GALEX near-UV flux \\
  36 & GALEX\_fuv\_mag & mag & GALEX far-UV flux in magnitudes \\
  37 & GALEX\_fuv\_magErr & mag & Error on the GALEX far-UV flux \\
  38 & TRXS\_match & True/False & Boolean to indicate successful matches to the ROSAT 2RXS AllWISE counterparts \\
  39 & TRXS\_distance & arcsec & Match distance between the ELQS AllWISE position to the ROSAT 2RXS AllWISE position. The distance values are often 0 or otherwise extremely small, because the positions match to numerical accuracy.\\
  40 & TRXS\_match\_flag & - & A flag indicating the most probable AllWISE ROSAT 2RXS cross-match with 1. This is the case for all matched objects.\\
  41 & TRXS\_2RXS\_SRC\_FLUX & $\rm{erg}\,\rm{cm}^{-2}\,\rm{s}^{-1}$ &  2RXS flux \\
  42 & TRXS\_2RXS\_SRC\_FLUX\_ERR &  $\rm{erg}\,\rm{cm}^{-2}\,\rm{s}^{-1}$ & 2RXS flux error \\
   43 & XMM\_match & True/False & Boolean to indicate successful matches to the XMMSL2 AllWISE counterparts \\ 
  44 & XMM\_distance & arcsec & Match distance between the ELQS AllWISE position to the XMMSL2 AllWISE counterparts\\
  45 & XMM\_match\_flag & - &  A flag indicating the most probable AllWISE XMMSL2 cross-match with 1. This is the case for all matched objects.  \\
   46 & XMM\_XMMSL2\_FLUX\_B8 &   $10^{-12}\,\rm{erg}\,\rm{cm}^{-2}\,\rm{s}^{-1}$& Total band ($0.2-12.0\,\rm{keV}$) flux \\
   47 & XMM\_XMMSL2\_FLUX\_B7 &  $10^{-12}\,\rm{erg}\,\rm{cm}^{-2}\,\rm{s}^{-1}$& Hard band ($2.0-12.0\,\rm{keV}$) flux\\
   48 & XMM\_XMMSL2\_FLUX\_B6 &   $10^{-12}\,\rm{erg}\,\rm{cm}^{-2}\,\rm{s}^{-1}$& Soft band ($0.2-2.0\,\rm{keV}$) flux \\ 
   49 & XMM\_XMMSL2\_FLUX\_B8\_ERR & $10^{-12}\,\rm{erg}\,\rm{cm}^{-2}\,\rm{s}^{-1}$ & Total band ($0.2-12.0\,\rm{keV}$) flux error\\ 
   50 & XMM\_XMMSL2\_FLUX\_B7\_ERR &  $10^{-12}\,\rm{erg}\,\rm{cm}^{-2}\,\rm{s}^{-1}$& Hard band ($2.0-12.0\,\rm{keV}$) flux error\\ 
   51 & XMM\_XMMSL2\_FLUX\_B6\_ERR &  $10^{-12}\,\rm{erg}\,\rm{cm}^{-2}\,\rm{s}^{-1}$& Soft band ($0.2-2.0\,\rm{keV}$) flux error\\

%  49 & BAL\_VI & 1/0/-1 & A flag indicating whether the object is visually identified as a broad absorption line (BAL) quasar ($1=$ BAL quasars, $0=$ quasars, $-1=$ no visual classification).\\
%  50 & qlf\_sample & True/False & Boolean to indicate whether the quasar is included in the estimation of the quasar luminosity function (Section\,\ref{sec_QLF}).\\
  \tableline
 \end{tabular}

\end{table*}

\clearpage

\section {Newly discovered QSOs at $z\ge2.8$}\label{app_newqsos}

We present general properties of the 190 newly discovered PS-ELQS quasars in Table\,\ref{tab_pselqs_newqsos}. Their discovery spectra are shown in Figure\,\ref{fig_new_qso_spectra}.

% Change the footnote style to lowercase letters
\renewcommand{\thefootnote}{\alph{footnote}}

\begin{longtable}{cccccccc}
 \caption{Newly discovered quasars at $z\geq2.8$ in the PS-ELQS sample} \label{tab_pselqs_newqsos}\\
 \tableline
 R.A.(J2000) & Decl.(J2000) & $m_{i}$ &  $M_{1450}$ & Spectroscopic & near UV\tablenotemark{a} & far UV\tablenotemark{a}  & Notes\tablenotemark{b} \\
 
 [hh:mm:ss.sss] & [dd:mm:ss.ss] & [mag]  & [mag] &  Redshift & [mag] & [mag]  & \\
  \tableline
  \tableline
\endfirsthead
% \begin{tabular}{cccccccc}
 \tableline
 R.A.(J2000) & Decl.(J2000) & $m_{i}$ &  $M_{1450}$ & Spectroscopic & near UV\tablenotemark{a} & far UV\tablenotemark{a}  & Notes\tablenotemark{b} \\
 
 [hh:mm:ss.sss] & [dd:mm:ss.ss] & [mag]  & [mag] &  Redshift & [mag] & [mag]  & \\
  \tableline
  \tableline
 \endhead
 \tableline\multicolumn{8}{r}{{Continued on the next page}} \\\tableline
\endfoot
\\[-1.8ex] \tableline\tableline
\endlastfoot
00:20:27.082 & -18:44:00.97 & $18.31\pm0.01$ & -27.35 & 3.765 & - &  -& 171008 \\
00:27:25.651 & -26:44:32.00 & $17.93\pm0.01$ & -27.29 & 3.005 & -&  -& 171008 \\
00:43:46.841 & -11:17:02.06 & $17.41\pm0.00$ & -28.11 & 3.480 & -&  -& 171007 \\
00:55:15.845 & -14:59:15.50 & $18.39\pm0.01$ & -27.50 & 4.200 & -&  -& 171008 \\
01:02:48.769 & -20:07:28.70 & $18.32\pm0.01$ & -27.30 & 3.710 & -&  -& 171010 \\
01:03:05.501 & -24:49:25.25 & $17.77\pm0.01$ & -27.95 & 3.865 & -&  -& 180122 \\
01:03:18.075 & -13:05:10.19 & $17.22\pm0.00$ & -28.61 & 4.065 & -&  -& 171007 \\
01:09:33.398 & +38:20:15.82 & $18.45\pm0.01$ & -27.18 & 3.720 & -&  -& 171020 \\
01:18:52.261 & -09:40:16.07 & $17.98\pm0.00$ & -27.54 & 3.495 & -&  -& 171008 \\
01:28:18.883 & -09:57:00.44 & $18.14\pm0.01$ & -27.52 & 3.765 & -&  -& 171008 \\
01:29:48.978 & -04:21:49.49 & $18.48\pm0.00$ & -27.07 & 3.600 & -&  -& 171010 \\
01:39:11.231 & -02:31:33.65 & $18.38\pm0.01$ & -27.23 & 3.690 & -&  -& 171010 \\
01:40:46.361 & +36:41:30.22 & $18.25\pm0.02$ & -27.18 & 3.310 & -&  -& 171021 \\
01:50:41.591 & -25:08:46.35 & $17.70\pm0.01$ & -27.87 & 3.600 & -&  -& 171008 \\
01:51:06.839 & -28:39:33.76 & $17.87\pm0.00$ & -27.78 & 3.730 & -&  -& 171007 \\
02:01:58.777 & +37:17:45.47 & $18.16\pm0.01$ & -27.68 & 4.080 & -&  -& 171020 \\
02:08:25.254 & +17:05:48.91 & $17.66\pm0.01$ & -27.77 & 3.300 & -&  -& 171010 \\
02:12:20.417 & +09:17:49.15 & $17.64\pm0.01$ & -27.59 & 3.000 & -&  -& 171008 \\
02:14:21.635 & +09:04:07.05 & $17.61\pm0.01$ & -27.94 & 3.560 & -&  -& 171009 \\
02:19:48.831 & +34:47:19.63 & $17.74\pm0.01$ & -27.59 & 3.160 & -&  -& 171020 \\
02:21:23.915 & -14:16:54.82 & $17.80\pm0.01$ & -27.80 & 3.650 & -&  -& 180124 \\
02:21:26.889 & -28:22:51.31 & $18.31\pm0.01$ & -27.21 & 3.480 & -&  -& 171008 \\
02:23:25.100 & +22:12:11.77 & $18.43\pm0.01$ & -27.25 & 3.815 & -&  -& 171021 \\
02:28:40.587 & +35:26:17.59 & $17.52\pm0.01$ & -27.97 & 3.380 & -&  -& 171020 \\
02:31:49.748 & -11:15:20.81 & $18.32\pm0.00$ & -27.19 & 3.430 & -&  -& 171010 \\
02:35:51.443 & -17:57:25.67 & $18.16\pm0.00$ & -27.39 & 3.595 & -&  -& 171009 \\
02:45:26.449 & +37:10:07.34 & $17.67\pm0.01$ & -27.89 & 3.560 & -&  -& 171111 \\
02:49:32.661 & +27:59:25.10 & $18.22\pm0.01$ & -27.01 & 3.020 & -&  -& 171116 \\
02:55:29.671 & +12:28:26.46 & $17.50\pm0.01$ & -28.24 & 3.870 & -&  -& 171010 \\
02:57:21.095 & +15:33:23.09 & $17.30\pm0.01$ & -28.14 & 3.310 & -&  -& 171010 \\
03:01:51.627 & +12:12:04.58 & $17.10\pm0.00$ & -28.54 & 3.690 & -&  -& 171010 \\
03:05:17.924 & -20:56:28.22 & $18.11\pm0.01$ & -27.66 & 3.960 & -&  -& 171007 \\
03:05:59.775 & +24:25:07.32 & $18.17\pm0.01$ & -27.52 & 3.810 & -&  -& 171109 \\
03:25:09.436 & +27:12:00.36 & $18.34\pm0.01$ & -27.21 & 3.580 & -&  -& 171111 \\
03:29:06.257 & +20:24:57.82 & $17.82\pm0.01$ & -27.53 & 3.230 & -&  -& 171111 \\
03:31:36.931 & +21:29:32.29 & $17.48\pm0.01$ & -27.85 & 3.190 & -&  -& 171109 \\
03:39:08.180 & -15:38:21.18 & $17.93\pm0.00$ & -27.75 & 3.790 & -&  -& 171006 \\
03:41:18.143 & +02:24:37.30 & $17.60\pm0.01$ & -27.68 & 3.090 & -&  -& 171008 \\
03:53:14.885 & -25:18:14.85 & $18.45\pm0.01$ & -27.53 & 4.305 & -&  -& 171008 \\
03:55:50.316 & -14:56:39.05 & $18.28\pm0.01$ & -27.24 & 3.475 & -&  -& 171010 \\
03:56:17.616 & -12:03:09.63 & $18.29\pm0.00$ & -27.36 & 3.765 & -&  -& 171008 \\
03:58:11.141 & +25:04:01.62 & $18.24\pm0.01$ & -27.41 & 3.750 & -&  -& 171111 \\
03:59:15.718 & -07:41:42.13 & $17.24\pm0.00$ & -28.28 & 3.420 & -&  -& 171006 \\
03:59:22.959 & -19:11:27.82 & $17.87\pm0.01$ & -27.24 & 2.840 & -&  -& 171009 \\
04:08:20.966 & -03:08:29.58 & $18.06\pm0.00$ & -27.60 & 3.750 & -&  -& 171009 \\
04:09:14.876 & -27:56:32.89 & $17.95\pm0.00$ & -28.15 & 4.460 & -&  -& 171006 \\
04:10:53.654 & -07:47:44.82 & $17.62\pm0.00$ & -27.59 & 2.975 & -&  -& 171009 \\
04:11:02.077 & -01:35:15.15 & $17.64\pm0.01$ & -27.96 & 3.660 & -&  -& 171006 \\
04:32:29.308 & -19:17:17.82 & $17.88\pm0.00$ & -27.30 & 2.930 & -&  -& 171009 \\
04:41:32.015 & -10:16:34.27 & $18.21\pm0.01$ & -27.56 & 3.970 & -&  -& 171008 \\
04:47:56.843 & -23:07:48.29 & $16.71\pm0.00$ & -28.50 & 2.945 & -&  -& 171006 \tablenotemark{c} \\
04:53:16.580 & -09:30:24.94 & $18.39\pm0.00$ & -27.10 & 3.405 & -&  -& 180124 \\
04:54:20.311 & -00:37:31.84 & $18.25\pm0.02$ & -26.92 & 2.915 & -&  -& 180124 \\
04:59:50.110 & +07:28:02.71 & $18.09\pm0.01$ & -27.42 & 3.435 & $22.67\pm0.32$ &  - & 180124 \\
05:00:15.026 & -24:39:27.24 & $17.92\pm0.00$ & -27.61 & 3.510 & -&  -& 171007 \\
05:03:54.146 & -06:08:25.04 & $18.22\pm0.01$ & -27.02 & 3.035 & -&  -& 180124 \\
05:20:01.728 & -20:14:40.59 & $18.26\pm0.00$ & -27.50 & 3.950 & -&  -& 171008 \\
05:21:36.923 & -13:39:38.79 & $17.60\pm0.00$ & -28.36 & 4.270 & -&  -& 171006 \\
05:39:46.870 & -20:08:41.86 & $18.29\pm0.01$ & -27.38 & 3.790 & -&  -& 171021 \\
08:18:24.472 & +82:06:48.47 & $17.35\pm0.01$ & -28.22 & 3.580 & -&  -& 180517 \\
08:51:03.208 & +13:02:53.32 & $18.47\pm0.01$ & -27.04 & 3.530 & -&  -& 180122 \\
09:16:47.616 & -11:30:09.91 & $18.44\pm0.01$ & -27.27 & 3.870 & -&  -& 180122 \\
09:17:46.542 & -11:53:31.89 & $17.90\pm0.01$ & -27.85 & 3.920 & -&  -& 171111 \\
09:19:23.109 & -00:52:08.00 & $17.56\pm0.00$ & -27.62 & 2.945 & -&  -& 180122 \\
09:28:05.302 & +28:27:19.72 & $17.80\pm0.01$ & -27.70 & 3.400 & $21.48\pm0.23$ &  - & 171109 \\
09:34:04.053 & -11:11:25.13 & $17.94\pm0.01$ & -27.62 & 3.605 & -&  -& 180122 \\
09:35:42.696 & -06:51:18.94 & $17.50\pm0.01$ & -28.32 & 4.040 & -&  -& 180122 \\
09:40:24.121 & -03:23:04.13 & $17.60\pm0.00$ & -28.15 & 3.900 & -&  -& 180123 \\
09:50:34.733 & -21:02:50.74 & $18.16\pm0.01$ & -27.36 & 3.480 & -&  -& 180124 \\
09:59:47.524 & -10:34:37.50 & $18.28\pm0.03$ & -27.03 & 3.165 & $21.81\pm0.28$ &  - & 180123 \\
10:14:30.281 & -04:21:40.32 & $17.44\pm0.01$ & -28.31 & 3.890 & -&  -& 180123 \\
10:15:29.367 & -12:13:14.34 & $17.06\pm0.00$ & -28.81 & 4.100 & -&  -& 180123 \\
10:15:40.799 & -03:27:47.25 & $17.67\pm0.01$ & -28.05 & 3.845 & -&  -& 180123 \\
10:15:44.118 & -11:09:22.77 & $17.56\pm0.00$ & -28.16 & 3.840 & -&  -& 180123 \\
10:20:00.800 & -12:11:51.45 & $17.92\pm0.01$ & -27.71 & 3.715 & -&  -& 180404 \\
10:21:26.131 & -11:56:22.39 & $18.43\pm0.01$ & -27.16 & 3.670 & -&  -& 180405 \\
10:31:58.288 & -21:44:07.40 & $17.69\pm0.00$ & -27.87 & 3.590 & -&  -& 180404 \\
10:41:38.997 & -09:44:37.94 & $18.22\pm0.02$ & -27.14 & 3.235 & -&  -& 180321 \\
10:46:27.942 & -23:39:17.54 & $18.04\pm0.01$ & -27.51 & 3.580 & -&  -& 180122 \\
10:47:13.545 & -06:45:38.19 & $18.42\pm0.01$ & -27.05 & 3.355 & -&  -& 180405 \\
10:51:22.689 & -06:50:47.84 & $17.34\pm0.01$ & -28.34 & 3.765 & -&  -& 180122 \\
10:53:53.499 & +25:31:15.50 & $18.39\pm0.01$ & -27.13 & 3.500 & -&  -& 180321 \\
10:54:49.678 & -17:11:07.39 & $16.92\pm0.00$ & -28.75 & 3.745 & $20.68\pm0.10$ &  - & 180122 \\
11:08:48.484 & -10:22:07.31 & $18.26\pm0.01$ & -27.84 & 4.460 & -&  -& 180123 \\
11:10:54.687 & -30:11:29.95 & $17.38\pm0.00$ & -28.87 & 4.830 & -&  -& 180122 \tablenotemark{d} \\
11:13:05.343 & -21:25:40.65 & $17.60\pm0.01$ & -27.90 & 3.390 & -&  -& 180122 \\
11:13:34.586 & -07:50:33.49 & $18.15\pm0.01$ & -27.09 & 3.045 & -&  -& 180405 \\
11:14:03.257 & -05:02:35.09 & $17.48\pm0.00$ & -28.23 & 3.825 & -&  -& 180122 \\
11:14:28.309 & -04:09:38.76 & $18.27\pm0.02$ & -27.24 & 3.445 & -&  -& 180405 \\
11:19:56.987 & -19:28:32.42 & $18.01\pm0.02$ & -27.69 & 3.830 & -&  -& 180123 \\
11:29:39.605 & -23:33:49.64 & $17.28\pm0.00$ & -27.86 & 2.880 & -&  -& 180123 \\
11:33:55.641 & -23:05:24.38 & $18.15\pm0.01$ & -27.44 & 3.660 & -&  -& 180405 \\
11:44:17.308 & -05:45:34.69 & $18.11\pm0.01$ & -27.41 & 3.500 & -&  -& 180405 \\
11:49:14.377 & -15:30:43.93 & $17.68\pm0.01$ & -28.21 & 4.160 & -&  -& 180404 \\
11:56:32.386 & -07:21:14.26 & $18.37\pm0.01$ & -26.79 & 2.905 & $21.70\pm0.27$ &  $21.73\pm0.49$ & 180405 \\
12:09:29.549 & -05:17:37.06 & $18.43\pm0.01$ & -27.07 & 3.480 & -&  -& 180405 \\
12:10:16.802 & +80:56:03.21 & $18.17\pm0.04$ & -27.35 & 3.490 & -&  -& 180514 \\
12:10:30.332 & -09:57:25.39 & $18.03\pm0.01$ & -27.30 & 3.200 & -&  -& 180405 \\
12:30:10.034 & -06:33:34.10 & $17.97\pm0.01$ & -27.44 & 3.300 & -&  -& 180404 \\
12:36:12.047 & -11:36:00.62 & $18.00\pm0.01$ & -27.35 & 3.225 & -&  -& 180405 \\
12:46:10.755 & +75:17:11.07 & $17.40\pm0.00$ & -28.14 & 3.520 & -&  -& 180517 \\
12:46:15.090 & +71:39:23.60 & $17.62\pm0.00$ & -28.18 & 3.995 & -&  -& 180517 \\
12:58:50.976 & -18:54:30.55 & $17.67\pm0.00$ & -27.71 & 3.255 & -&  -& 180124 \\
13:00:31.133 & -28:29:31.01 & $17.94\pm0.00$ & -28.18 & 4.710 & -&  -& 180124 \tablenotemark{d} \\
13:01:48.270 & -14:46:52.70 & $18.44\pm0.02$ & -27.67 & 4.515 & -&  -& 180124 \\
13:02:30.435 & -10:26:28.59 & $18.29\pm0.01$ & -27.41 & 3.820 & -&  -& 180124 \\
13:05:00.904 & -12:26:18.79 & $18.17\pm0.01$ & -27.09 & 3.080 & -&  -& 180405 \\
13:16:44.039 & -25:38:10.33 & $18.12\pm0.01$ & -27.36 & 3.370 & -&  -& 180405 \\
13:17:25.036 & -18:42:30.76 & $18.24\pm0.01$ & -27.37 & 3.700 & -&  -& 180404 \\
13:29:56.958 & -04:52:21.77 & $18.23\pm0.02$ & -27.26 & 3.395 & -&  -& 180406 \\
13:39:32.277 & +36:13:40.62 & $18.33\pm0.01$ & -27.19 & 3.450 & -&  -& 180321 \\
13:58:32.274 & -28:48:35.48 & $18.40\pm0.02$ & -27.31 & 3.850 & -&  -& 180602 \\
14:00:15.152 & -03:44:16.50 & $17.60\pm0.01$ & -27.95 & 3.540 & -&  -& 180404 \\
14:08:01.817 & -27:58:20.35 & $17.77\pm0.01$ & -28.32 & 4.440 & -&  -& 180404 \\
14:11:42.768 & -24:13:13.48 & $18.11\pm0.01$ & -27.42 & 3.540 & -&  -& 180404 \\
14:27:32.247 & -18:03:18.31 & $17.07\pm0.02$ & -28.22 & 3.115 & -&  -& 180404 \\
14:32:54.468 & -27:22:28.05 & $18.37\pm0.02$ & -27.25 & 3.725 & -&  -& 180602 \\
14:39:49.242 & -08:07:05.38 & $18.10\pm0.01$ & -27.40 & 3.430 & -&  -& 180602 \\
14:45:49.741 & -11:10:15.68 & $17.94\pm0.00$ & -27.28 & 2.995 & -&  -& 180404 \\
14:55:59.430 & -25:28:32.10 & $18.28\pm0.01$ & -27.59 & 4.160 & -&  -& 180404 \\
15:18:53.216 & -11:59:51.54 & $17.83\pm0.00$ & -27.68 & 3.435 & -&  -& 180404 \\
15:23:12.411 & -16:27:22.92 & $17.83\pm0.00$ & -28.03 & 4.120 & -&  -& 180404 \\
15:32:45.990 & -25:10:48.22 & $17.78\pm0.01$ & -28.11 & 4.155 & -&  -& 180404 \\
15:38:15.568 & +81:44:32.99 & $18.06\pm0.01$ & -27.27 & 3.180 & $21.43\pm0.21$ &  - & 180514 \\
15:49:16.840 & -22:37:46.53 & $18.06\pm0.01$ & -27.68 & 3.905 & -&  -& 180405 \\
15:56:35.483 & +60:37:26.78 & $17.69\pm0.00$ & -27.77 & 3.340 & -&  -& 180517 \\
15:58:41.854 & -04:03:53.42 & $17.68\pm0.01$ & -27.55 & 3.020 & -&  -& 180405 \\
16:00:19.482 & -12:17:02.75 & $18.07\pm0.01$ & -27.14 & 2.980 & -&  -& 180405 \\
16:01:11.971 & -16:43:41.63 & $18.12\pm0.01$ & -26.99 & 2.840 & -&  -& 180602 \\
16:06:08.999 & +48:41:37.02 & $18.04\pm0.01$ & -27.42 & 3.345 & $21.39\pm0.15$ &  - & 180517 \\
16:16:48.948 & -09:14:44.39 & $17.39\pm0.01$ & -28.45 & 4.055 & -&  -& 180405 \\
16:17:37.785 & +59:50:20.13 & $17.39\pm0.02$ & -28.62 & 4.315 & -&  -& 180517 \\
16:23:49.985 & -15:44:27.68 & $17.81\pm0.03$ & -27.35 & 2.900 & -&  -& 180406 \\
16:29:36.489 & +60:40:49.19 & $17.93\pm0.01$ & -27.24 & 2.940 & -&  -& 180517 \\
16:31:18.216 & -12:43:08.60 & $16.96\pm0.01$ & -28.57 & 3.455 & -&  -& 180406 \\
16:32:27.929 & -14:20:44.18 & $17.00\pm0.01$ & -28.68 & 3.755 & -&  -& 180406 \\
16:35:36.073 & +03:24:07.80 & $16.95\pm0.00$ & -28.50 & 3.320 & $20.54\pm0.15$ &  $21.53\pm0.35$ & 180321 \\
16:38:56.009 & +69:18:15.24 & $18.02\pm0.01$ & -27.48 & 3.415 & -&  -& 180517 \\
16:39:26.455 & +03:52:04.12 & $17.52\pm0.00$ & -27.86 & 3.245 & -&  -& 180321 \\
16:48:52.744 & +09:59:42.10 & $17.77\pm0.00$ & -27.39 & 2.905 & $20.80\pm0.09$ &  $21.28\pm0.12$ & 180321 \\
16:59:29.379 & +65:38:20.85 & $18.35\pm0.00$ & -27.33 & 3.805 & -&  -& 180514 \\
17:20:46.132 & +00:43:28.19 & $17.91\pm0.01$ & -27.57 & 3.375 & -&  -& 180320 \\
17:30:03.673 & +48:46:30.77 & $16.97\pm0.01$ & -28.51 & 3.355 & -&  -& 180518 \\
17:47:13.484 & +29:55:32.30 & $18.34\pm0.01$ & -27.17 & 3.465 & -&  -& 180321 \\
17:53:34.530 & +37:49:07.13 & $18.19\pm0.01$ & -27.00 & 2.945 & -&  -& 180518 \\
17:55:21.128 & +30:09:04.24 & $17.89\pm0.00$ & -27.34 & 3.025 & -&  -& 180321 \\
17:56:29.853 & +26:07:40.57 & $17.27\pm0.01$ & -28.26 & 3.480 & -&  -& 180320 \\
18:02:09.690 & +40:12:53.77 & $18.16\pm0.01$ & -27.37 & 3.525 & -&  -& 180518 \\
18:10:27.309 & +34:24:08.85 & $18.26\pm0.01$ & -27.19 & 3.330 & -&  -& 171109 \\
18:10:41.346 & +34:54:49.46 & $18.04\pm0.01$ & -27.47 & 3.460 & -&  -& 171109 \\
18:17:06.185 & +48:22:26.07 & $17.39\pm0.02$ & -28.25 & 3.705 & -&  -& 180518 \\
18:19:14.803 & +33:39:45.98 & $17.74\pm0.00$ & -27.84 & 3.610 & -&  -& 171020 \\
19:35:12.403 & -26:10:49.63 & $17.20\pm0.00$ & -28.13 & 3.175 & -&  -& 171007 \\
20:00:13.515 & -25:05:36.92 & $17.72\pm0.01$ & -27.84 & 3.575 & -&  -& 171006 \\
20:02:05.969 & -23:28:26.52 & $17.82\pm0.01$ & -27.36 & 2.940 & -&  -& 171007 \\
20:10:23.353 & -18:23:47.75 & $18.15\pm0.00$ & -27.36 & 3.440 & -&  -& 171010 \\
20:11:58.767 & -26:23:40.95 & $17.47\pm0.00$ & -28.12 & 3.620 & -&  -& 171008 \\
20:17:41.494 & -28:16:29.96 & $17.40\pm0.01$ & -28.22 & 3.690 & -&  -& 171008 \\
20:18:34.860 & -15:28:38.69 & $17.35\pm0.01$ & -28.22 & 3.580 & -&  -& 171007 \\
20:20:43.904 & -02:37:02.52 & $17.75\pm0.01$ & -27.81 & 3.585 & -&  -& 171008 \\
20:33:36.699 & -24:56:15.86 & $18.14\pm0.01$ & -27.38 & 3.470 & -&  -& 171010 \\
20:33:43.573 & -30:23:09.83 & $17.76\pm0.00$ & -27.77 & 3.520 & -&  -& 171007 \\
20:34:16.997 & -02:59:53.43 & $18.26\pm0.00$ & -27.01 & 3.090 & -&  -& 180602 \\
20:36:23.526 & -08:37:29.92 & $17.51\pm0.01$ & -27.63 & 2.895 & -&  -& 171008 \\
20:43:20.175 & -03:38:40.93 & $17.48\pm0.02$ & -27.69 & 2.925 & -&  -& 171009 \\
20:48:48.274 & -22:51:52.15 & $17.18\pm0.00$ & -28.10 & 3.100 & -&  -& 171009 \\
21:03:51.499 & -26:23:00.22 & $17.54\pm0.00$ & -27.98 & 3.455 & -&  -& 171006 \\
21:11:11.604 & -25:36:15.03 & $17.96\pm0.01$ & -27.52 & 3.370 & -&  -& 171007 \\
21:25:40.966 & -17:19:51.41 & $16.42\pm0.00$ & -29.35 & 3.900 & -&  -& 171010 \\
21:26:51.969 & -10:31:39.62 & $18.21\pm0.01$ & -27.51 & 3.870 & -&  -& 171008 \\
21:27:16.485 & -04:04:33.58 & $18.47\pm0.02$ & -27.05 & 3.545 & -&  -& 171010 \\
21:30:50.101 & -24:44:03.50 & $18.01\pm0.01$ & -27.48 & 3.380 & -&  -& 171009 \\
21:32:25.900 & -28:31:33.33 & $17.76\pm0.01$ & -27.32 & 2.810 & -&  -& 171009 \\
21:34:45.240 & -27:49:39.75 & $17.68\pm0.00$ & -27.58 & 3.065 & $22.89\pm0.47$ &  - & 171009 \\
22:21:52.882 & -18:26:02.94 & $17.80\pm0.01$ & -28.30 & 4.520 & -&  -& 171010 \\
22:29:59.998 & -26:21:05.68 & $17.91\pm0.01$ & -27.38 & 3.140 & -&  -& 171008 \\
22:30:49.477 & -21:54:02.09 & $18.30\pm0.01$ & -27.26 & 3.605 & -&  -& 180602 \\
22:37:34.408 & -31:07:14.09 & $17.87\pm0.01$ & -27.42 & 3.120 & -&  -& 171009 \\
22:46:19.167 & -25:17:20.76 & $17.76\pm0.01$ & -27.40 & 2.900 & $21.71\pm0.27$ &  $22.55\pm0.54$ & 171007 \\
22:56:33.174 & -12:43:59.64 & $17.62\pm0.01$ & -27.48 & 2.825 & -&  -& 171009 \\
22:58:20.943 & -28:18:55.25 & $18.07\pm0.01$ & -27.46 & 3.525 & -&  -& 171009 \\
22:59:39.043 & -22:50:35.10 & $17.43\pm0.01$ & -28.08 & 3.470 & -&  -& 171007 \\
23:00:22.023 & -14:40:31.37 & $18.28\pm0.00$ & -27.43 & 3.850 & -&  -& 171008 \\
23:04:32.312 & -12:48:19.64 & $18.13\pm0.01$ & -27.58 & 3.850 & -&  -& 171007 \\
23:08:27.042 & -13:32:56.21 & $17.71\pm0.01$ & -28.00 & 3.830 & -&  -& 171006 \\
23:09:59.293 & -12:26:03.16 & $17.94\pm0.00$ & -27.70 & 3.730 & -&  -& 171006 \\
23:23:06.892 & -24:51:13.78 & $18.35\pm0.01$ & -27.36 & 3.860 & -&  -& 171007 \\
23:39:26.346 & -11:20:50.35 & $18.20\pm0.01$ & -27.64 & 4.090 & -&  -& 171021 \\
23:42:10.117 & -23:21:53.60 & $18.09\pm0.01$ & -27.43 & 3.480 & -&  -& 171008 \\
23:42:41.975 & -12:26:55.02 & $17.98\pm0.00$ & -27.51 & 3.380 & -&  -& 171006 \\
\footnotetext[1]{The near and far UV magnitudes were obtained from cross-matches within $2\farcs0$ to the GALEX GR6/7 data release}
%\tablenotetext{2}{Visual qualitative BAL identification flag: $1=$BAL; $0=$no BAL; $-1=$ insufficient wavelength coverage or inconclusive archival data}
\footnotetext[2]{This column shows the observation date (YYMMDD) and provides further information on individual objects.}
\footnotetext[3]{This object has been classified as a BAL, LoBAL or FeLoBAL quasar. Details are discussed in Section\,\ref{sec_bal_quasars}}
\footnotetext[4]{These objects were also independently discovered by \citet{Yang2018b_arxiv}}
%\tablenotetext{5}{See also HST GO Proposal 13013 (PI: Gabor Worseck), \citet{Zheng2015} and \cite{Schmidt2017}}
% \end{tabular}
%\tablenotetext{1}{The near and far UV magnitudes were obtained from cross-matches within $2\farcs0$ to the GALEX GR6/7 data release}
%%\tablenotetext{2}{Visual qualitative BAL identification flag: $1=$BAL; $0=$no BAL; $-1=$ insufficient wavelength coverage or inconclusive archival data}
%\tablenotetext{3}{This column shows the observation date (YYMMDD) and provides further information on individual objects.}
%%\tablenotetext{4}{These objects were also independently discovered by Yang et al.}
%%\tablenotetext{5}{See also HST GO Proposal 13013 (PI: Gabor Worseck), \citet{Zheng2015} and \cite{Schmidt2017}}

\end{longtable}
% Reset the footnotes back to numbers
\renewcommand{\thefootnote}{\arabic{footnote}}

\clearpage

%\section{Spectra of the newly discovered PS-ELQS quasars}\label{app_new_qso}
%{
\begin{figure*}[htb]
 \centering
 \includegraphics[width=0.9\textwidth]{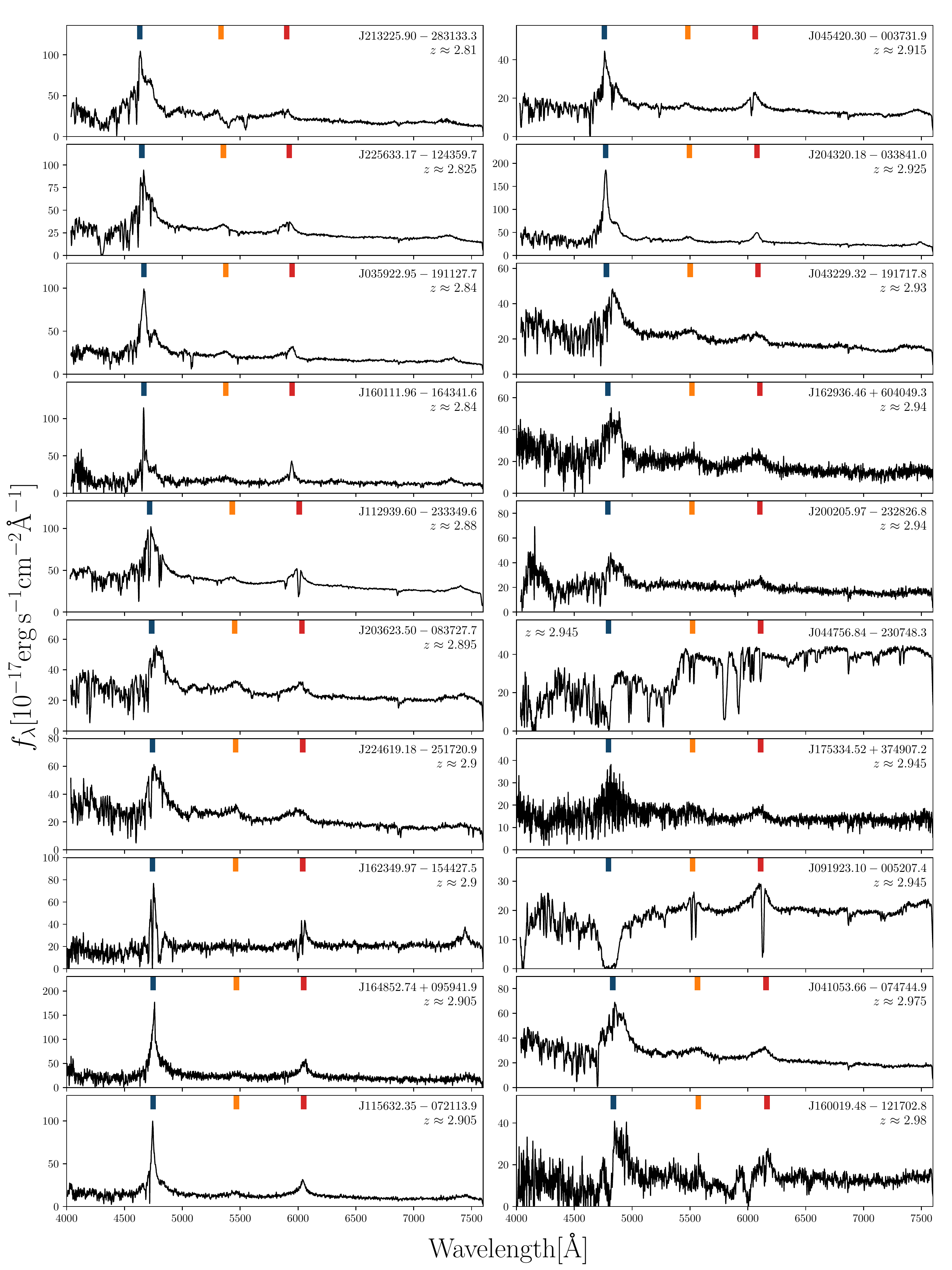}
 \caption{The discovery spectra of the newly discovered PS-ELQS quasars. The  dark blue, orange and red bars denote the center positions of the broad $\rm{Ly}\alpha$, \ion{Si}{4} and \ion{C}{4} emission lines according to the spectroscopic redshift.}
 \label{fig_new_qso_spectra}
 \ContinuedFloat
\end{figure*}
 
\begin{figure*}[htb] 
 \centering
 \includegraphics[width=0.9\textwidth]{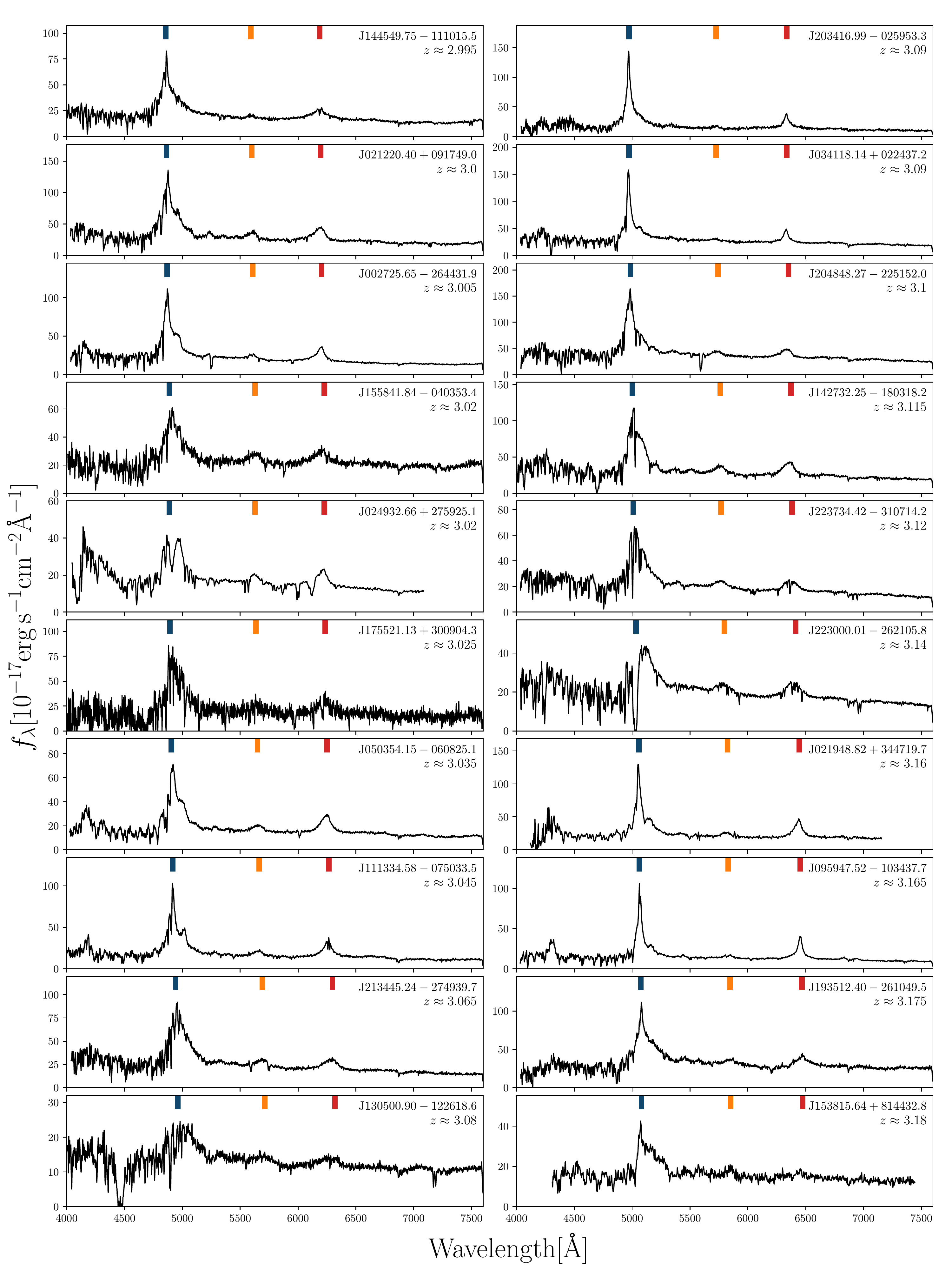}
 \caption{(continued)}
 \ContinuedFloat
\end{figure*} 

\begin{figure*}[htb] 
 \centering
 \includegraphics[width=0.9\textwidth]{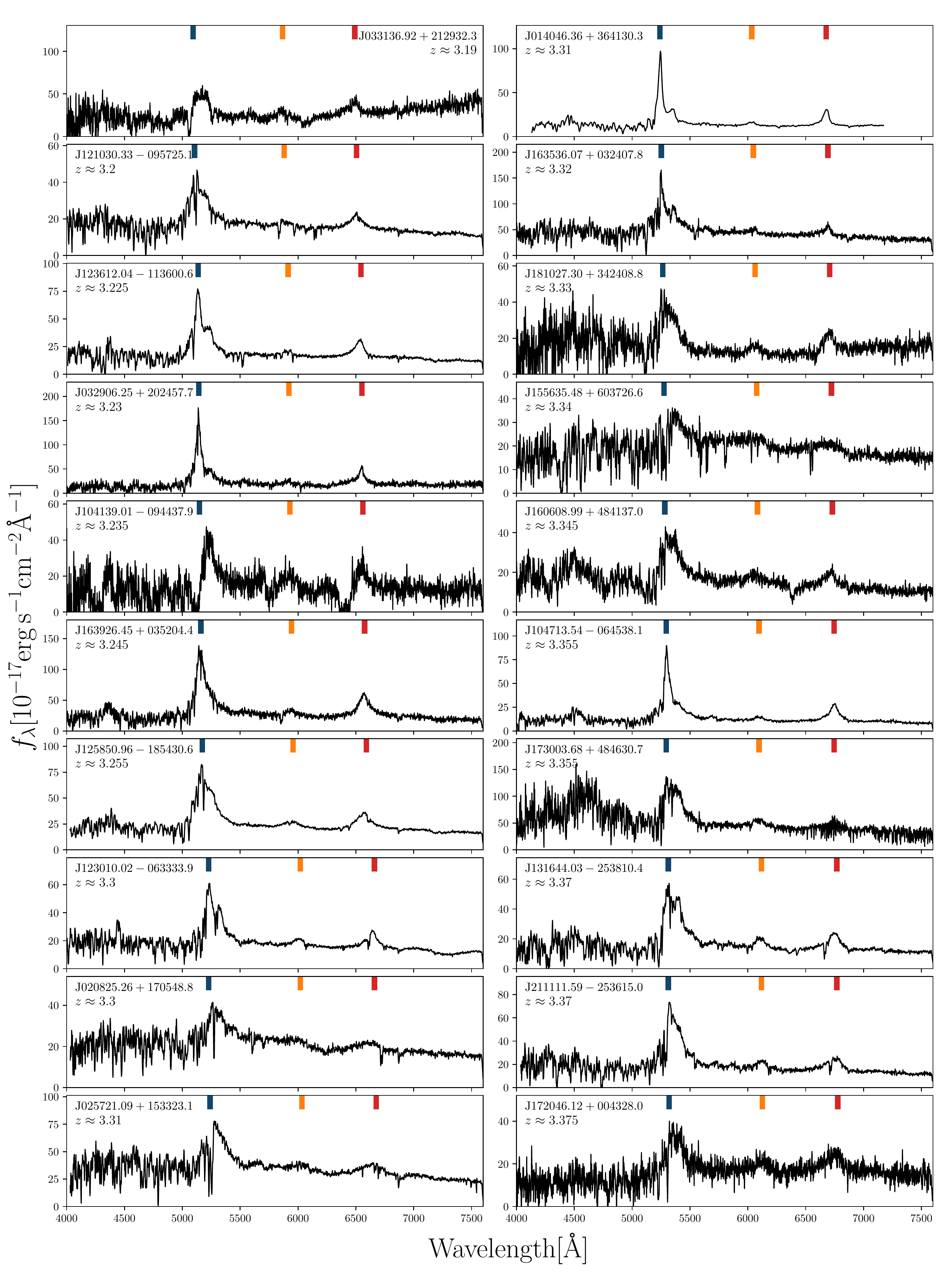}
 \caption{(continued)}
  \ContinuedFloat
\end{figure*} 

\begin{figure*}[htb] 
 \centering
 \includegraphics[width=0.9\textwidth]{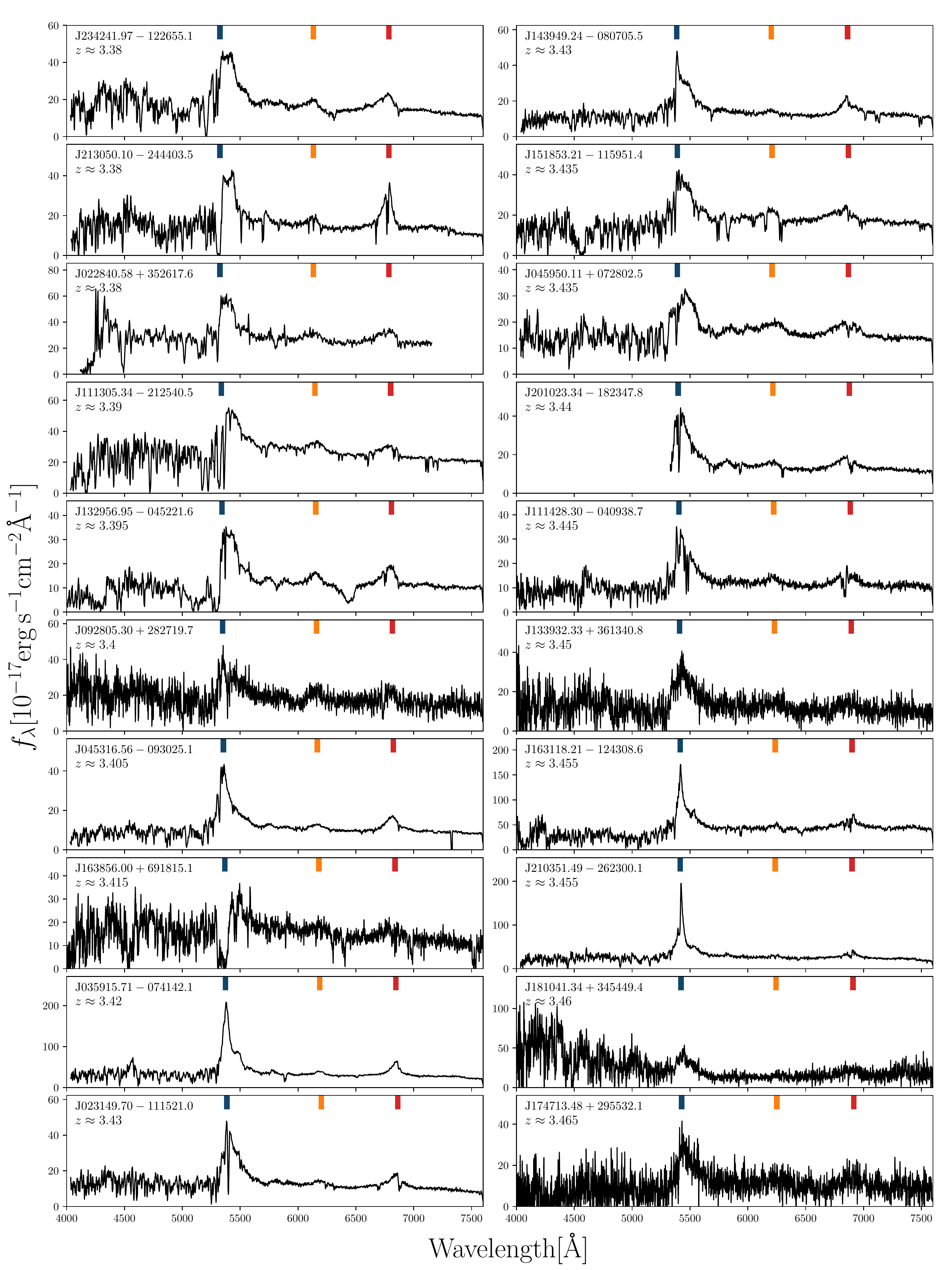}
 \caption{(continued)}
  \ContinuedFloat
\end{figure*} 

\begin{figure*}[htb] 
 \centering
 \includegraphics[width=0.9\textwidth]{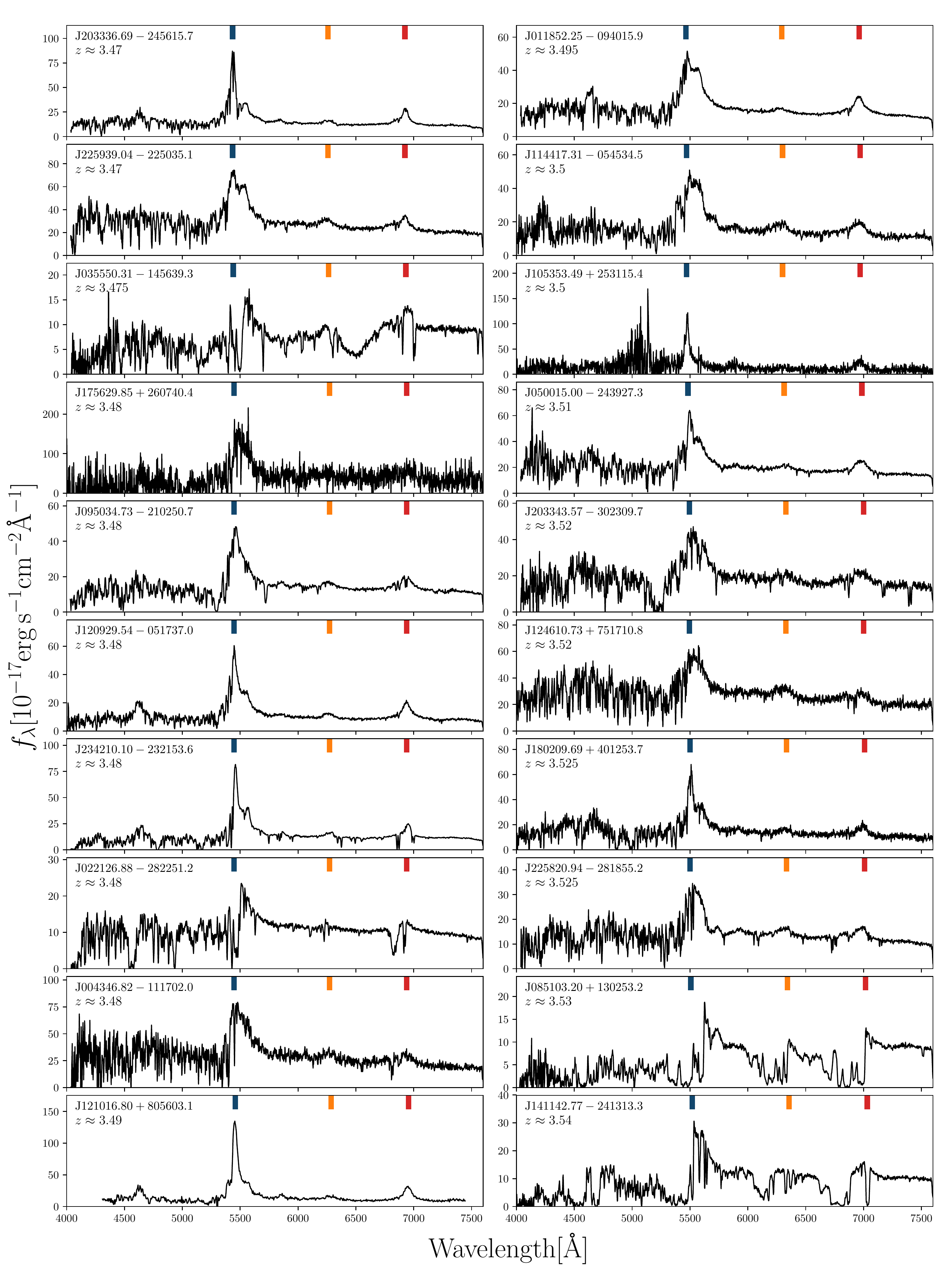}
 \caption{(continued)}
  \ContinuedFloat
\end{figure*} 

\begin{figure*}[htb] 
 \centering
 \includegraphics[width=0.9\textwidth]{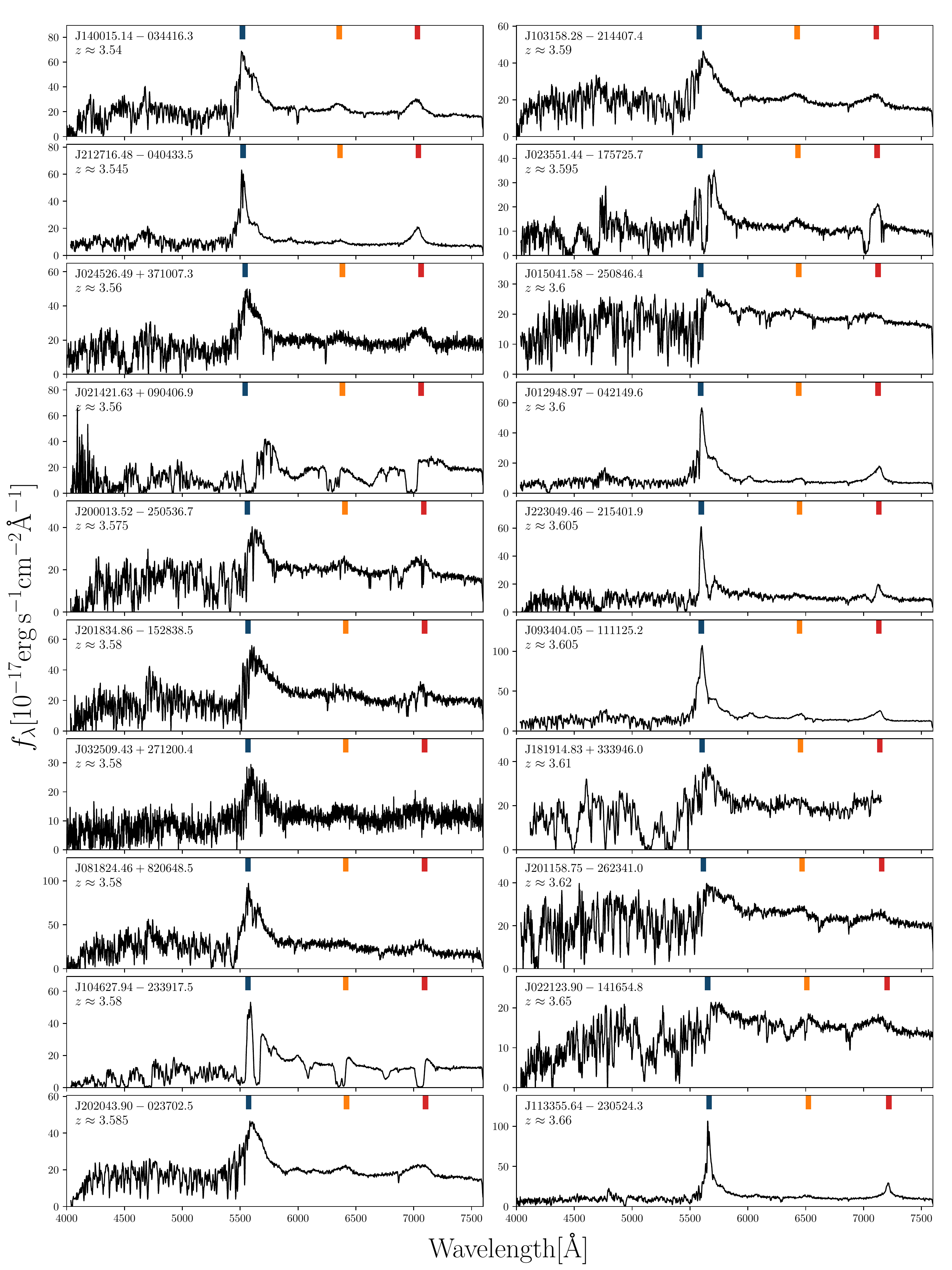}
 \caption{(continued)}
   \ContinuedFloat
\end{figure*} 

\begin{figure*}[htb] 
 \centering
 \includegraphics[width=0.9\textwidth]{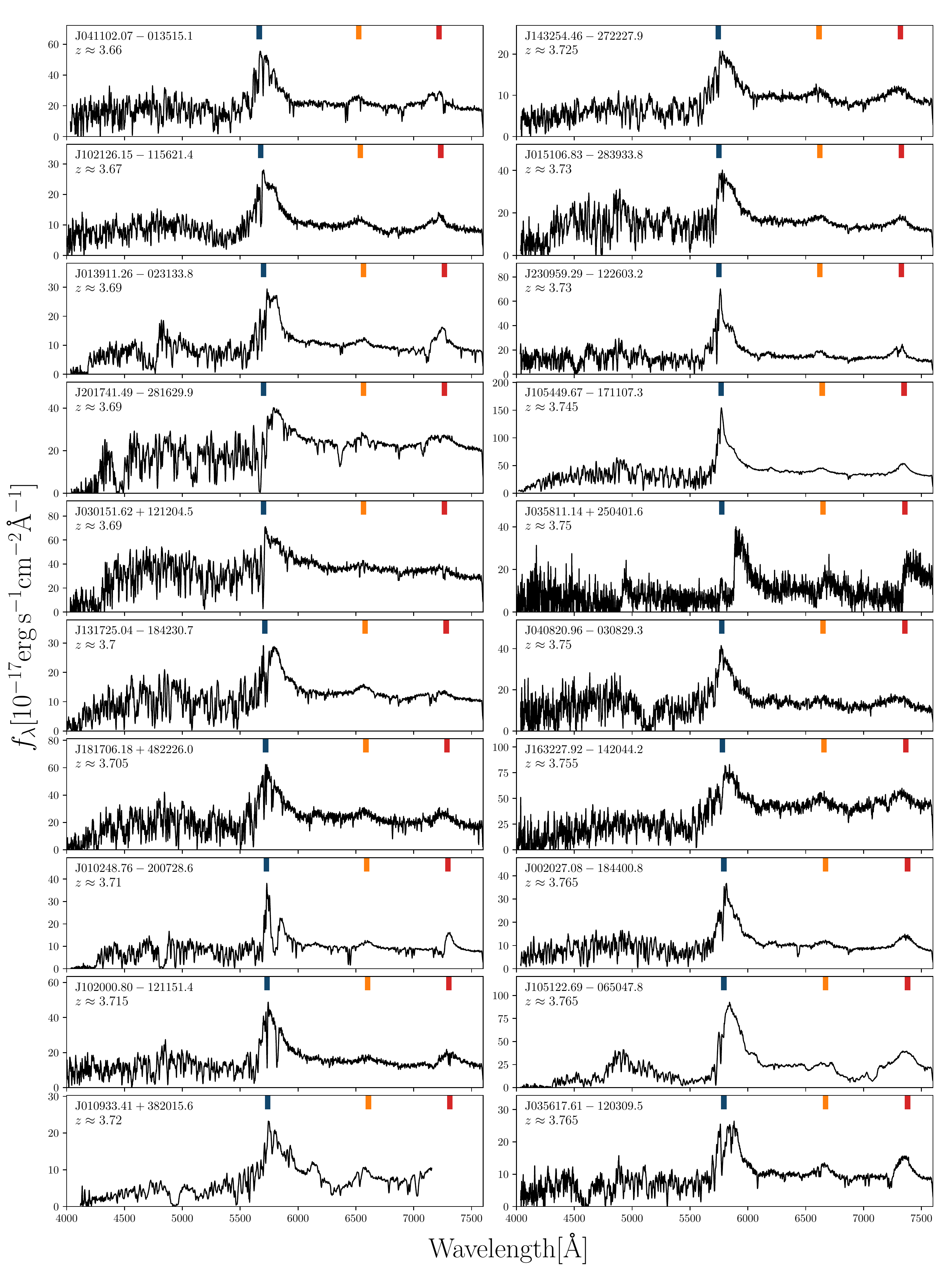}
 \caption{(continued)}
   \ContinuedFloat
\end{figure*} 

\begin{figure*}[htb] 
 \centering
 \includegraphics[width=0.9\textwidth]{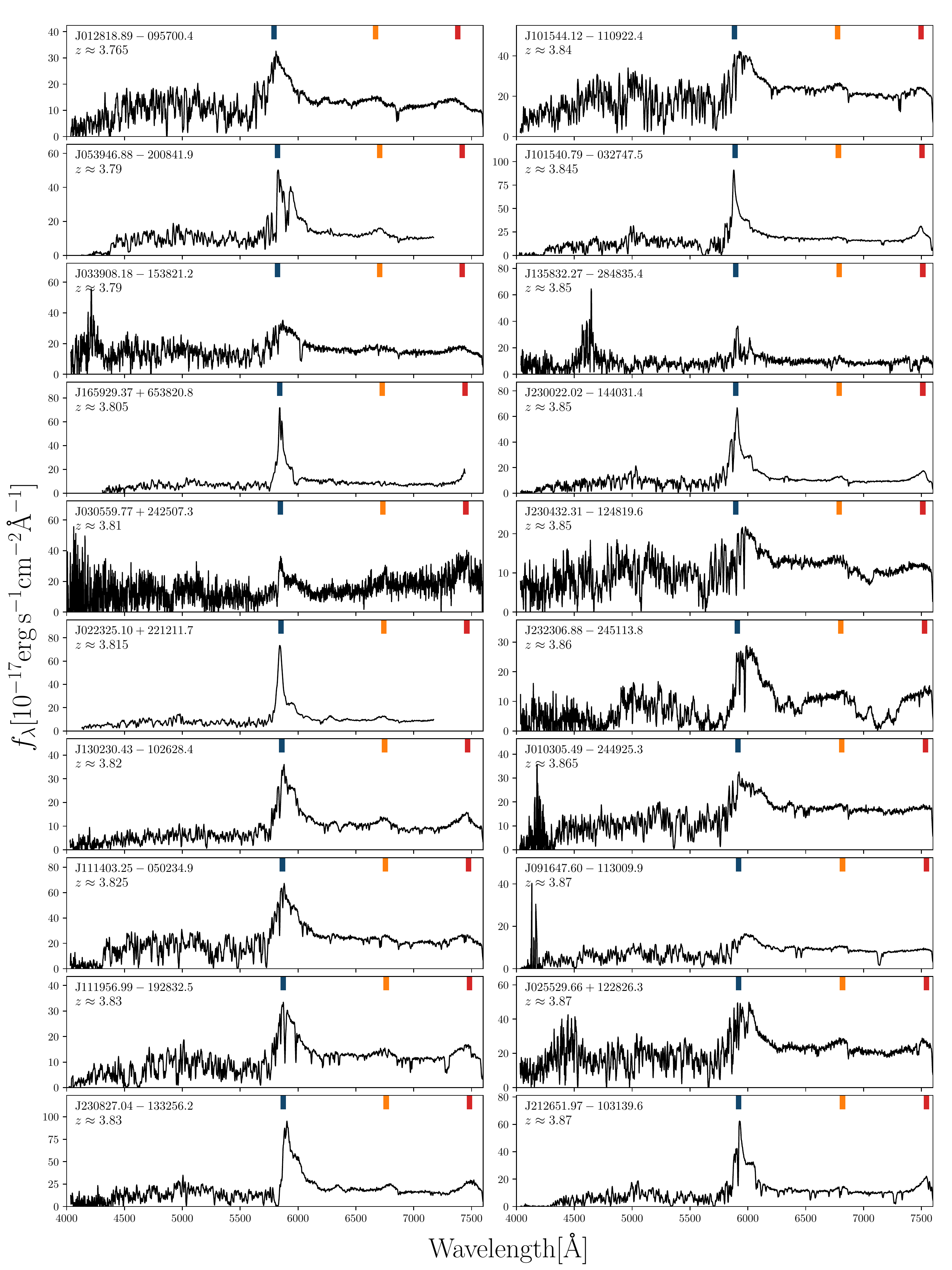}
 \caption{(continued)}
   \ContinuedFloat
\end{figure*} 

\begin{figure*}[htb] 
 \centering
 \includegraphics[width=0.9\textwidth]{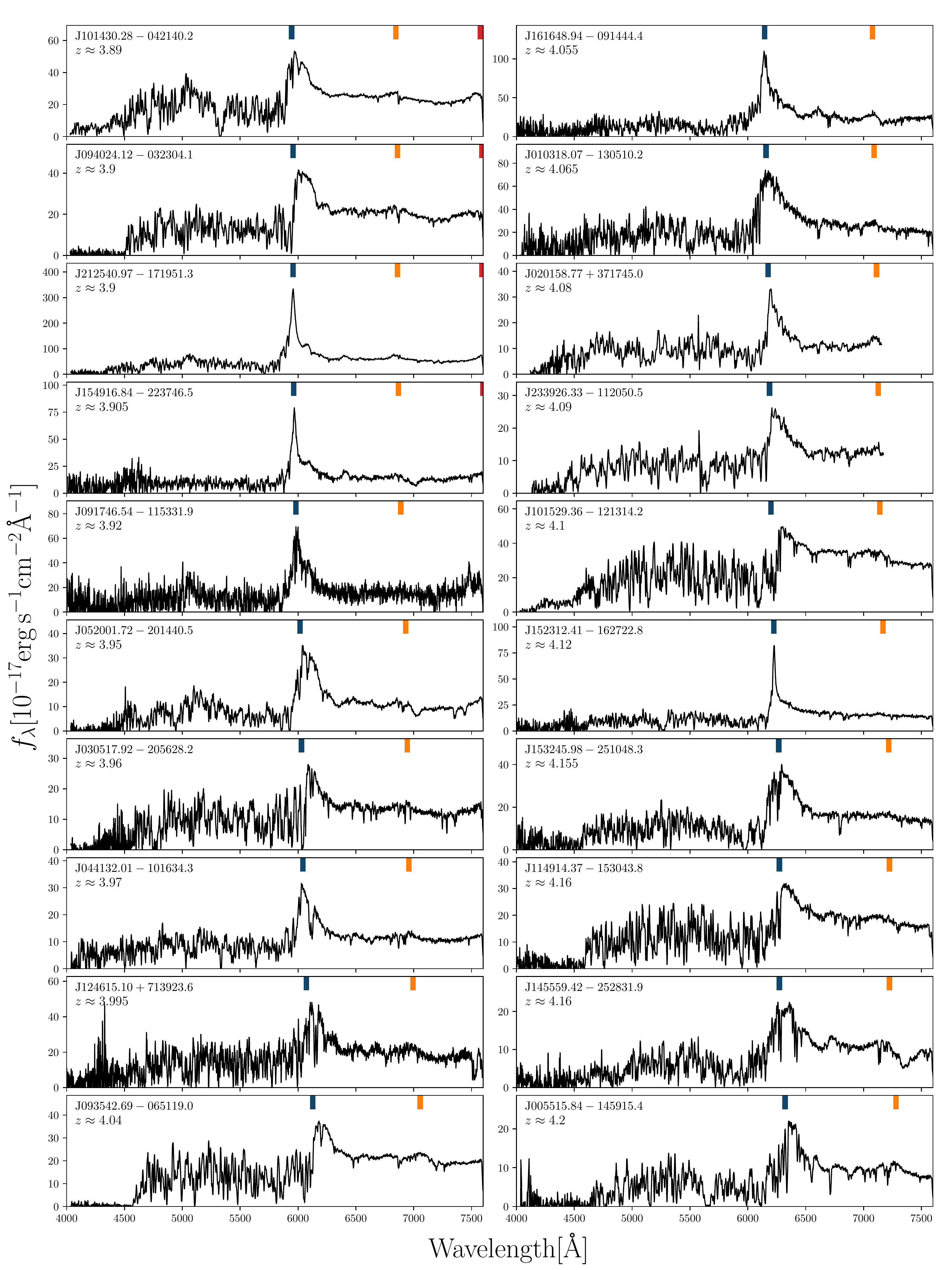}
 \caption{(continued)}
   \ContinuedFloat
\end{figure*} 

\begin{figure*}[htb] 
 \centering
 \includegraphics[width=0.9\textwidth]{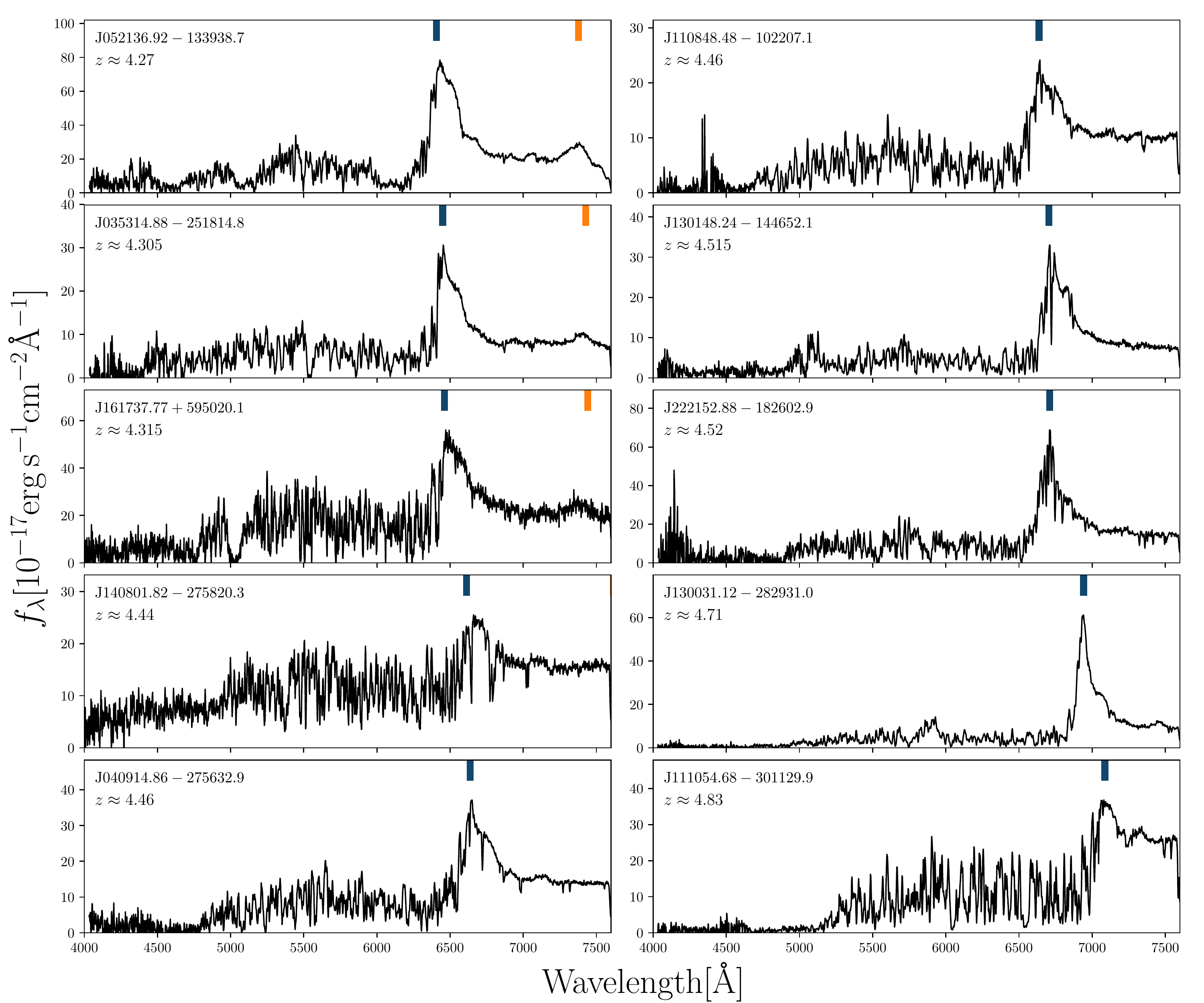}
 \caption{(continued)}
%   \ContinuedFloat
\end{figure*} 
%}

\clearpage

\section{Newly discovered QSOs at lower redshift ($z<2.8$)}\label{app_new_lowz_qsos}

In this part of the appendix we present newly discovered quasars targeted with PS-ELQS spectroscopically confirmed to be at $z<2.8$. A summary of the general properties of these objects in provided in Table\,\ref{tab_pselqs_lowzqsos}, which is also available in machine readable format on-line.  The discovery spectra are shown in Figure\,\ref{fig_new_qso_lowz}.

\begin{table*}[htb]
\centering
 \caption{Newly discovered quasars at $z<2.8$ in the PS-ELQS sample}
 \label{tab_pselqs_lowzqsos}
 \begin{tabular}{cccccccc}
  \tableline
 R.A.(J2000) & Decl.(J2000) & $m_{i}$ &  $M_{1450}$ & Spectroscopic & near UV\tablenotemark{a} & far UV\tablenotemark{a}  & Notes\tablenotemark{b} \\

 [hh:mm:ss.sss] & [dd:mm:ss.ss] & [mag]  & [mag] &  Redshift & [mag] & [mag]  & \\
  \tableline
 \tableline
00:13:10.727 & +29:18:47.74 & $17.65\pm0.01$ & -27.27 & 2.500 & -&  -& 171020 \\
00:38:56.987 & -29:22:24.43 & $17.35\pm0.00$ & -27.28 & 2.270 & -&  -& 180124 \tablenotemark{c} \\
02:11:19.800 & -19:59:43.01 & $17.94\pm0.01$ & -26.93 & 2.450 & -&  -& 171010 \tablenotemark{c}  \\
02:35:00.447 & +02:38:29.25 & $18.11\pm0.01$ & -23.73 & 0.650 & -&  -& 180124  \tablenotemark{c} \\
03:30:11.020 & -12:40:08.68 & $17.59\pm0.00$ & -26.78 & 2.075 & -&  -& 171006 \\
03:35:59.996 & -13:26:02.08 & $17.99\pm0.01$ & -26.14 & 1.900 & -&  -& 171007  \tablenotemark{c} \\
03:41:38.070 & -11:42:59.44 & $17.09\pm0.01$ & -27.98 & 2.770 & -&  -& 171008 \\
04:05:48.525 & -24:21:15.26 & $17.38\pm0.01$ & -27.68 & 2.760 & -&  -& 171007 \\
09:10:54.661 & +46:06:51.94 & $18.09\pm0.01$ & -26.81 & 2.490 & -&  -& 180514 \\
09:26:42.056 & -17:47:21.96 & $18.24\pm0.02$ & -22.39 & 0.369 & $20.10\pm0.10$ &  $20.52\pm0.17$ & 180602 \\
09:36:20.407 & +82:51:14.07 & $17.54\pm0.01$ & -27.50 & 2.715 & -&  -& 180517 \\
10:11:22.657 & -24:33:01.43 & $18.12\pm0.01$ & -26.92 & 2.720 & -&  -& 180123 \\
10:57:02.777 & +34:22:50.37 & $18.08\pm0.00$ & -26.63 & 2.320 & -&  -& 180321 \\
11:28:14.210 & +26:56:46.36 & $18.18\pm0.01$ & -26.32 & 2.170 & -&  -& 180321 \\
11:32:52.869 & -06:32:43.31 & $17.68\pm0.01$ & -27.15 & 2.410 & -&  -& 180404  \tablenotemark{c} \\
14:27:45.083 & -14:51:49.32 & $17.79\pm0.01$ & -26.96 & 2.350 & -&  -& 180404 \\
14:40:30.602 & +69:42:11.58 & $17.99\pm0.01$ & -27.04 & 2.690 & -&  -& 180514 \\
17:13:01.101 & +66:58:25.90 & $18.29\pm0.01$ & -26.67 & 2.550 & -&  -& 180518 \\
18:00:30.260 & +79:34:47.07 & $17.93\pm0.01$ & -27.12 & 2.755 & $22.06\pm0.30$ & - & 180518 \\
18:03:11.956 & +70:38:25.75 & $17.94\pm0.01$ & -27.09 & 2.715 & -&  -& 180518 \\
18:07:24.633 & +28:08:14.40 & $17.90\pm0.00$ & -26.60 & 2.150 & -&  -& 171020 \\
18:20:00.261 & +63:10:36.85 & $17.63\pm0.00$ & -26.97 & 2.235 & -&  -& 180518 \\
18:29:04.759 & +78:31:06.45 & $18.13\pm0.01$ & -26.41 & 2.200 & $22.33\pm0.35$ & - & 180514 \\
19:19:46.075 & +74:37:47.11 & $17.86\pm0.01$ & -25.92 & 1.604 & -&  -& 180518  \tablenotemark{c} \\
20:30:34.859 & -25:41:57.41 & $18.10\pm0.01$ & -26.96 & 2.743 & -&  -& 180604 \\
22:09:12.009 & +06:19:20.01 & $17.25\pm0.00$ & -26.90 & 1.910 & -&  -& 171008  \tablenotemark{c} \\
22:51:59.483 & +17:28:44.68 & $17.35\pm0.01$ & -27.36 & 2.320 & -&  -& 180518 \\
23:41:20.021 & +31:20:25.38 & $18.39\pm0.01$ & -26.22 & 2.240 & -&  -& 171021 \\
\tableline
 \end{tabular}
\tablenotetext{1}{The near and far UV magnitudes were obtained from cross-matches within $2\farcs0$ to the GALEX GR6/7 data release}
%\tablenotetext{2}{Visual qualitative BAL identification flag: $1=$BAL; $0=$no BAL; $-1=$ insufficient wavelength coverage or inconclusive archival data}
\tablenotetext{2}{This column shows the observation date (YYMMDD) and provides further information on individual objects.}
\tablenotetext{3}{This object has been classified as a BAL, LoBAL or FeLoBAL quasar. Details are discussed in Section\,\ref{sec_bal_quasars}}
%\tablenotetext{4}{These objects were also independently discovered by Yang et al.}
%\tablenotetext{5}{See also HST GO Proposal 13013 (PI: Gabor Worseck), \citet{Zheng2015} and \cite{Schmidt2017}}
\end{table*}

%\section{Spectra of the newly discovered PS-ELQS quasars at $z<2.8$}\label{app_new_qso_lowz}
\begin{figure*}[htb] 
 \centering
 \includegraphics[width=0.9\textwidth]{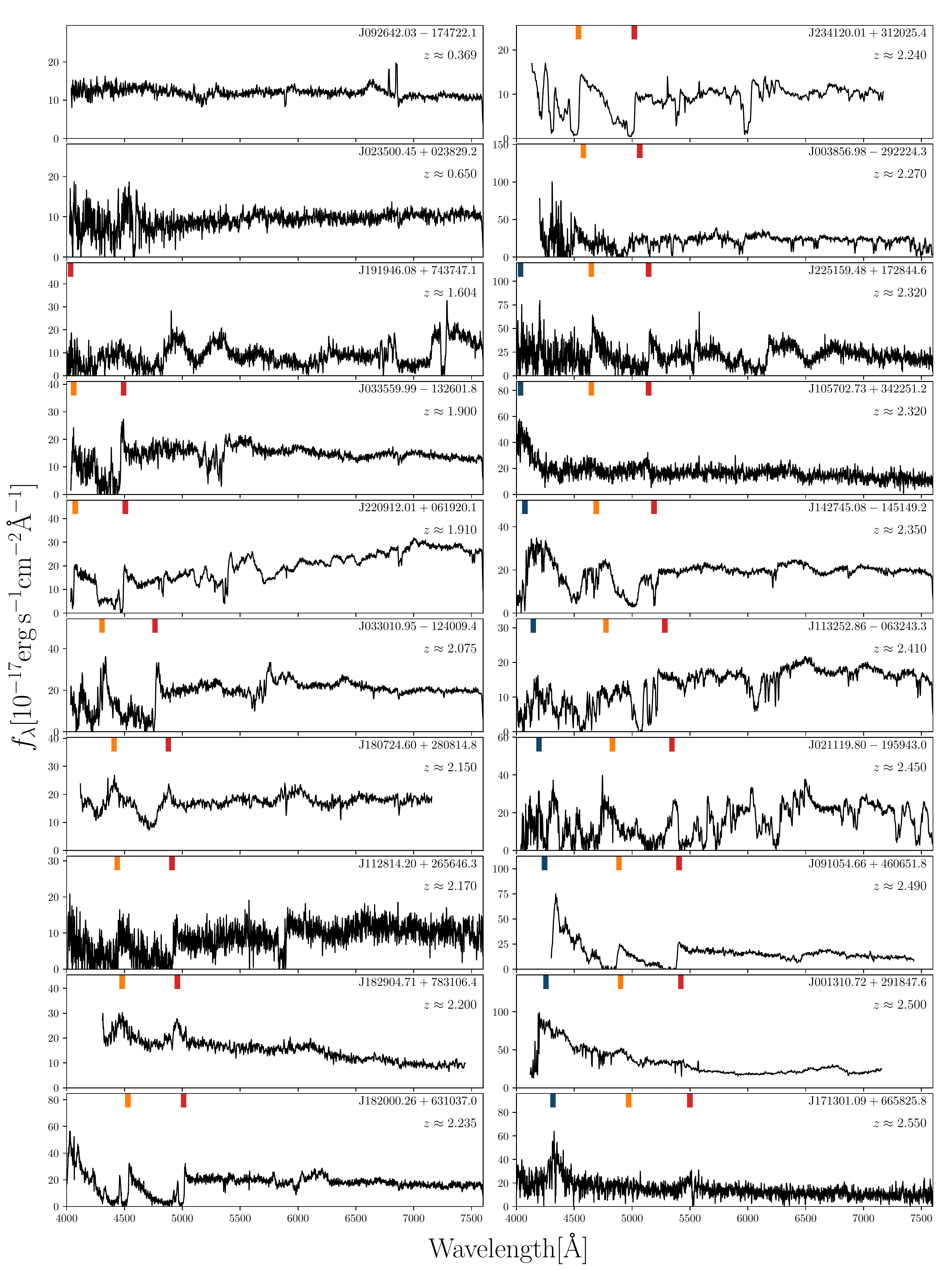}
 \caption{The discovery spectra of the newly discovered PS-ELQS quasars at $z<2.8$. The  dark blue, orange and red bars denote the center positions of the broad $\rm{Ly}\alpha$, \ion{Si}{4} and \ion{C}{4} emission lines according to the spectroscopic redshift.}
  \ContinuedFloat
 \label{fig_new_qso_lowz}
\end{figure*} 

\begin{figure*}[htb] 
 \centering
 \includegraphics[width=0.9\textwidth]{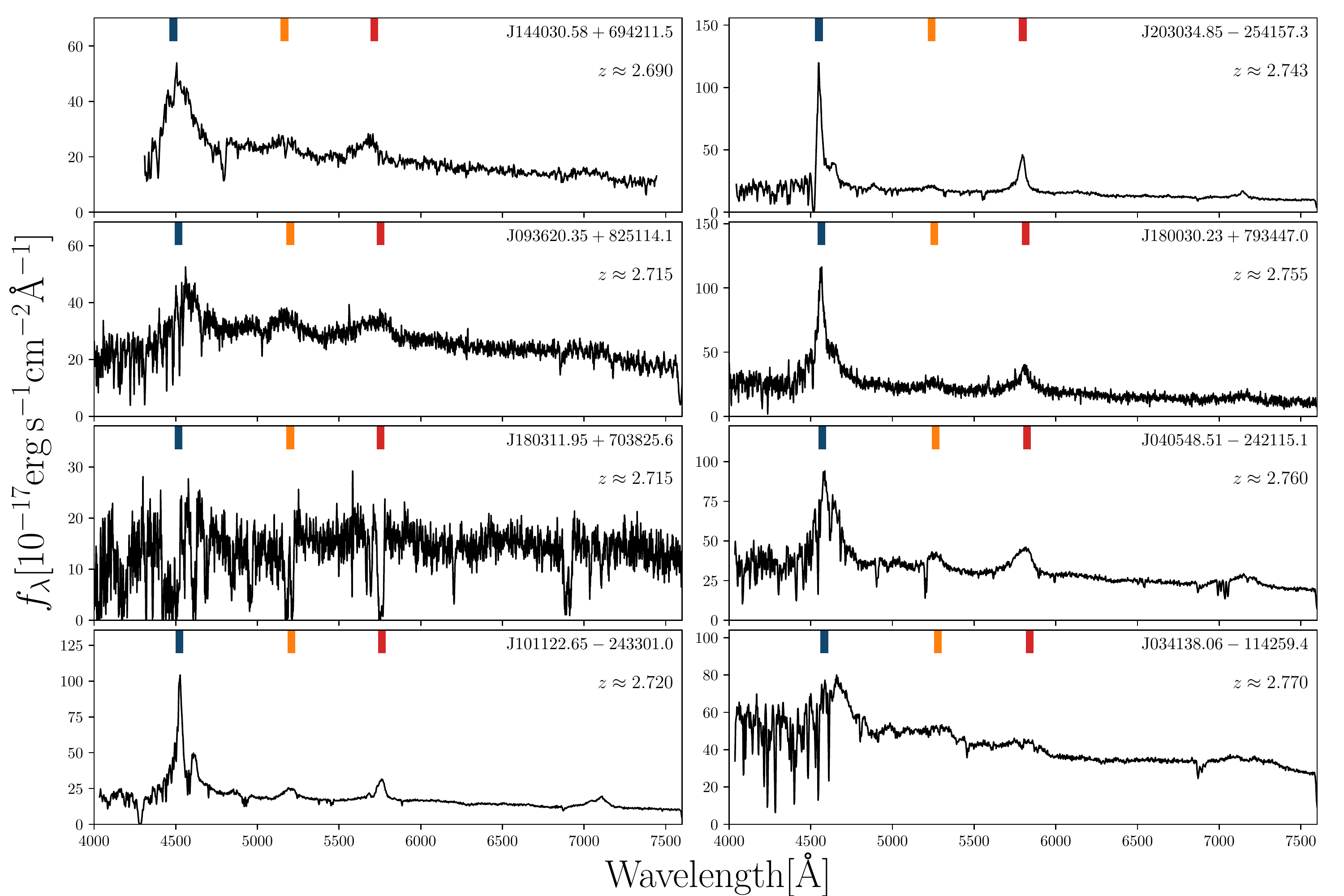}
 \caption{(continued)}
  
\end{figure*}

\clearpage

\section{Remaining good PS-ELQS candidates}\label{app_candidates}
We further present properties of the remaining good PS-ELQS candidates in Table\,\ref{tab_cand}. A machine readable version of this table is provided on-line.

\begin{table}[htb]
\centering
\caption{Properties of the remaining good PS-ELQS candidates}
\label{tab_cand}
\begin{tabular}{ccccccc}
\tableline
R.A.(J2000) & Decl.(J2000) &  $m_{i}$ &  \texttt{rf\_photoz} &  \texttt{rf\_qso\_prob} & \texttt{rf\_mult\_class\_pred} & priority\\

[hh:mm:ss.sss] & [dd:mm:ss.ss] & [mag] &  & & \\
\tableline
\tableline 
00:08:01.926 & -27:24:29.28 & $18.04\pm0.01$ & 2.83 & 1.00 & midz & 5 \\
00:25:18.462 & -15:57:50.47 & $18.12\pm0.01$ & 2.82 & 0.99 & midz & 5 \\
00:46:12.529 & +41:50:02.49 & $18.34\pm0.01$ & 3.30 & 0.84 & highz & 4 \\
00:55:09.859 & -03:49:43.89 & $18.32\pm0.01$ & 2.85 & 0.86 & midz & 5 \\
01:54:15.903 & +40:43:40.91 & $18.02\pm0.01$ & 3.59 & 0.86 & highz & 3 \\
01:58:16.432 & -01:30:38.31 & $18.18\pm0.01$ & 2.88 & 0.98 & midz & 5 \\
02:22:07.117 & -16:28:11.77 & $18.15\pm0.01$ & 2.89 & 0.99 & midz & 5 \\
02:29:43.899 & +29:33:05.50 & $18.34\pm0.01$ & 3.00 & 0.98 & midz & 5 \\
02:39:44.591 & +07:26:59.62 & $18.09\pm0.00$ & 2.92 & 0.73 & midz & 5 \\
02:56:27.355 & -18:35:49.70 & $18.11\pm0.00$ & 2.88 & 0.99 & midz & 5 \\
03:05:33.395 & +12:57:34.33 & $18.07\pm0.01$ & 2.90 & 0.99 & midz & 5 \\
03:18:29.401 & +23:34:35.21 & $18.32\pm0.01$ & 3.85 & 0.87 & highz & 3 \\
04:14:03.285 & -10:50:03.75 & $18.30\pm0.01$ & 2.83 & 0.98 & midz & 5 \\
04:23:28.876 & -27:52:23.82 & $18.14\pm0.01$ & 2.92 & 0.96 & midz & 5 \\
04:44:07.833 & +80:34:43.36 & $18.24\pm0.01$ & 3.46 & 0.60 & highz & 4 \\
04:46:49.052 & -03:54:39.44 & $18.48\pm0.01$ & 2.83 & 0.85 & midz & 5 \\
05:05:25.937 & +76:49:53.53 & $17.77\pm0.01$ & 3.58 & 0.94 & highz & 1 \\
05:34:17.420 & +75:44:13.74 & $17.27\pm0.00$ & 3.04 & 0.54 & midz & 2 \\
05:47:06.719 & +79:02:21.45 & $17.79\pm0.01$ & 3.20 & 0.89 & midz & 2 \\
05:57:01.236 & +68:30:27.86 & $18.14\pm0.01$ & 2.87 & 1.00 & midz & 5 \\
06:12:25.946 & +66:15:22.70 & $18.25\pm0.01$ & 3.15 & 0.90 & highz & 4 \\
06:34:29.752 & +56:34:42.36 & $18.46\pm0.00$ & 3.02 & 0.91 & midz & 4 \\
06:42:53.018 & +59:43:45.50 & $18.30\pm0.01$ & 3.84 & 0.96 & highz & 3 \\
06:50:56.448 & +72:53:14.65 & $18.04\pm0.00$ & 4.49 & 0.83 & highz & 3 \\
06:52:41.984 & +54:27:40.60 & $18.11\pm0.02$ & 3.65 & 0.99 & highz & 3 \\
06:57:27.418 & +57:22:11.94 & $17.49\pm0.01$ & 3.61 & 0.92 & highz & 1 \\
07:00:32.592 & +56:00:27.17 & $17.98\pm0.00$ & 3.04 & 0.76 & highz & 2 \\
07:08:02.482 & +63:15:59.67 & $17.20\pm0.00$ & 2.95 & 0.76 & midz & 3 \\
07:14:46.848 & +84:25:28.21 & $17.81\pm0.01$ & 2.93 & 0.64 & midz & 3 \\
07:15:52.373 & +42:10:06.15 & $17.62\pm0.01$ & 3.08 & 0.63 & midz & 2 \\
07:17:03.905 & +59:02:59.46 & $18.40\pm0.01$ & 3.61 & 0.95 & highz & 3 \\
07:32:57.277 & +54:52:11.55 & $17.80\pm0.01$ & 3.07 & 0.59 & midz & 2 \\
07:37:59.176 & +54:54:44.01 & $17.76\pm0.00$ & 2.88 & 0.96 & midz & 3 \\
07:42:23.031 & +68:36:31.53 & $18.49\pm0.01$ & 3.95 & 0.96 & highz & 3 \\
07:42:58.216 & +61:21:10.97 & $17.61\pm0.00$ & 3.62 & 0.94 & highz & 1 \\
07:51:07.041 & +37:11:56.34 & $18.19\pm0.01$ & 2.91 & 0.90 & midz & 5 \\
07:51:55.122 & +53:53:34.41 & $18.48\pm0.00$ & 3.50 & 0.92 & highz & 4 \\
07:52:48.270 & +70:24:33.00 & $18.16\pm0.01$ & 3.60 & 0.97 & highz & 3 \\
07:55:50.673 & +68:47:04.24 & $17.68\pm0.01$ & 2.95 & 0.47 & midz & 3 \\
08:09:10.462 & +59:01:25.22 & $18.23\pm0.01$ & 3.00 & 0.94 & highz & 5 \\
08:23:56.195 & +69:08:15.67 & $18.27\pm0.00$ & 3.21 & 0.95 & highz & 4 \\
08:32:04.867 & +57:33:15.31 & $18.49\pm0.01$ & 3.29 & 0.94 & highz & 4 \\
13:34:19.002 & +26:55:34.63 & $18.43\pm0.01$ & 3.17 & 0.88 & highz & 4 \\
23:05:05.917 & +26:47:14.04 & $18.27\pm0.01$ & 2.94 & 0.51 & midz & 5 \\

\tableline
\end{tabular}
\end{table}

\clearpage
\section{SQL query to obtain the Pan-STARRS DR1 (PS1) photometry}\label{app_ps1_sql}
\begin{verbatim}
SELECT
m.wise_designation, m.wise_ra, m.wise_dec,
o.ObjID as PS1_ObjID,x.ra as ps_ra,x.dec as ps_dec,
o.gMeanPSFMag, o.gMeanPSFMagErr, o.gMeanKronMag, o.gMeanApMag,
o.rMeanPSFMag, o.rMeanPSFMagErr, o.rMeanKronMag, o.rMeanApMag,
o.iMeanPSFMag, o.iMeanPSFMagErr, o.iMeanKronMag, o.iMeanApMag,
o.zMeanPSFMag, o.zMeanPSFMagErr, o.zMeanKronMag, o.zMeanApMag,
o.yMeanPSFMag, o.yMeanPSFMagErr, o.yMeanKronMag, o.yMeanApMag,
o.gMeanPSFmagNpt,o.rMeanPSFmagNpt,o.iMeanPSFmagNpt,o.zMeanPSFmagNpt ,o.yMeanPSFmagNpt,
o.gFlags, o.gQfPerfect,
o.rFlags, o.rQfPerfect,
o.iFlags, o.iQfPerfect,
o.zFlags, o.zQfPerfect,
o.yFlags, o.yQfPerfect,
sp.gpetRadius,sp.rpetRadius,sp.ipetRadius,sp.zpetRadius,sp.ypetRadius,
sot.iinfoFlag, sot.iinfoFlag2 


into mydb.wise_2mass_jkw2_colorcut_matched
from mydb.wise_2mass_jkw2_colorcut AS m

CROSS APPLY (SELECT * FROM dbo.fGetNearestObjEq(m.wise_ra,m.wise_dec,0.066))  AS x
JOIN MeanObject o on o.ObjID=x.ObjId 

LEFT JOIN StackPetrosian AS sp ON sp.objID = o.objID
LEFT JOIN StackObjectThin AS sot ON sot.objID = o.objID

WHERE (o.iMeanPSFMag > 0 AND o.iMeanPSFMag <= 19.0 )
AND o.zMeanPSFMag > 0
AND o.yMeanPSFMag > 0
AND o.iQfPerfect>=0.85 and o.zQfPerfect>=0.85 
--- rejects extended objects
AND (-0.3 <= iMeanPSFMag - iMeanApMag OR iMeanPSFMag - iMeanApMag <= 0.3) 
--- photometric quality criteria
AND ( sot.iinfoFlag & 0x00000008 = 0) ---FAIL
AND ( sot.iinfoFlag & 0x00000010 = 0) ---POOR / POORFIT
AND ( sot.iinfoFlag & 0x00000020 = 0) ---PAIR
AND ( sot.iinfoFlag & 0x00000080 = 0) ---SATSTAR
AND ( sot.iinfoFlag & 0x00000100 = 0) ---BLEND
AND ( sot.iinfoFlag & 0x00000400 = 0) ---BADPSF
AND ( sot.iinfoFlag & 0x00000800 = 0) ---DEFECT
AND ( sot.iinfoFlag & 0x00001000 = 0) ---SATURATED
AND ( sot.iinfoFlag & 0x00002000 = 0) ---CR_LIMIT
AND ( sot.iinfoFlag & 0x00008000 = 0) ---MOMENTS_FAILURE
AND ( sot.iinfoFlag & 0x00010000 = 0) ---SKY_FAILURE
AND ( sot.iinfoFlag & 0x00020000 = 0) ---SKYVAR_FAILURE
AND ( sot.iinfoFlag & 0x00040000 = 0) ---MOMENTS_SN
AND ( sot.iinfoFlag & 0x00400000 = 0) ---BLEND_FIT
AND ( sot.iinfoFlag & 0x10000000 = 0) ---SIZE_SKIPPED
AND ( sot.iinfoFlag & 0x20000000 = 0) ---ON_SPIKE
AND ( sot.iinfoFlag & 0x40000000 = 0) ---ON_GHOST
AND ( sot.iinfoFlag & 0x80000000 = 0) ---OFF_CHIP
AND ( sot.iinfoFlag2 & 0x00000008 = 0) ---ON_SPIKE
AND ( sot.iinfoFlag2 & 0x00000010 = 0) ---ON_STARCORE
AND ( sot.iinfoFlag2 & 0x00000020 = 0) ---ON_BURNTOOL

\end{verbatim}

\end{document}